%% file: paper.tex
\documentclass[prd,preprint,showpacs,superscriptaddress,amssymb,floatfix]{revtex4}

\usepackage{latexsym}
\usepackage{graphicx}
\usepackage{lscape}
\usepackage{epsfig}
\usepackage{bm}
\include{user_def}

\newcommand\riken{RIKEN-BNL Research Center, Brookhaven National Laboratory, Upton, NY 11973, USA}
\newcommand\bnlaf{Brookhaven National Laboratory, Upton, NY 11973, USA}
\newcommand\edinb{SUPA, School of Physics, The University of Edinburgh, Edinburgh EH9 3JZ, UK}
\newcommand\epcca{EPCC, School of Physics, The University of Edinburgh, Edinburgh EH9 3JZ, UK}
\newcommand\cuaff{Physics Department, Columbia University, New York, NY 10027, USA}
\newcommand\glasg{SUPA, Department of Physics \& Astronomy, University of Glasgow, Glasgow G12 8QQ, UK}
\newcommand\tokyo{Department of Physics, University of Tokyo, Hongo 7-3-1, Bunkyo-ku, Tokyo 113, Japan}

\begin{document}
\bibliographystyle{apsrev}
\preprint{BNL-HET-06/17, CU-TP-1159, Edinburgh 2006/39, KEK-TH-1115, RBRC-631}

\title{First results from 2+1-Flavor Domain Wall QCD:
Mass Spectrum, Topology Change and Chiral Symmetry with $L_s=8$}

\author{D.J.~Antonio}   \affiliation{\edinb}
\author{T.~Blum}        \affiliation{\riken}
\affiliation{Physics Department, University of Connecticut, Storrs CT, 06269-3046, USA}
\author{K.C.~Bowler}    \affiliation{\edinb}
\author{P.A.~Boyle}     \affiliation{\edinb}
\author{N.H.~Christ}    \affiliation{\cuaff}
\author{S.D.~Cohen}     \affiliation{\cuaff}
\author{M.A.~Clark}     \affiliation{Center for Computational Science, 3 Cummington Street, Boston University, MA 02215, USA}
\author{C.~Dawson}      \affiliation{\riken}
\author{A.~Hart}	\affiliation{\edinb}
\author{K.~Hashimoto}   \affiliation{\riken}
\affiliation{Radiation Laboratory, RIKEN, Wako, Saitama 351-0198, Japan}
\affiliation{Institute for Theoretical Physics, 
Kanazawa University, Kakuma, Kanazawa, 920-1192, Japan}
\author{T.~Izubuchi}    \affiliation{\riken}
\affiliation{Institute for Theoretical Physics, 
Kanazawa University, Kakuma, Kanazawa, 920-1192, Japan}
\author{B.~Jo\'o}       \affiliation{Jefferson Laboratory, Newport News, VA 23606, USA}
\author{C.~Jung}        \affiliation{\bnlaf}
\author{A.D.~Kennedy}   \affiliation{\edinb}
\author{R.D.~Kenway}    \affiliation{\edinb}
\author{S.~Li}	        \affiliation{\cuaff}
\author{H.W.~Lin}	\affiliation{\cuaff}
\affiliation{Jefferson Laboratory, Newport News, VA 23606, USA}
\author{M.F.~Lin}         \affiliation{\cuaff}
\author{R.D.~Mawhinney} \affiliation{\cuaff}
\author{C.M.~Maynard}   \affiliation{\epcca}
\author{J.~Noaki\footnote{Present address: Institute of Particle and Nuclear Studies, KEK, Ibaraki 305-0801, Japan}}
  \affiliation{School of Physics and Astronomy, University of Southampton,  Southampton SO17 1BJ, UK}
\author{S.~Ohta}        \affiliation{\riken}
\affiliation{Institute of Particle and Nuclear Studies, KEK, Ibaraki 305-0801, Japan}
\affiliation{Physics Department, The Graduate University for Advanced Studies (Sokendai), Tsukuba, Ibaraki 305-0801, Japan}

\author{S.~Sasaki}      \affiliation{\riken}\affiliation{\tokyo}
\author{A.~Soni}        \affiliation{\bnlaf}
\author{R.J.~Tweedie}   \affiliation{\edinb}
\author{A.~Yamaguchi}   \affiliation{\glasg}

\collaboration{RBC and UKQCD Collaborations}


\date{\today}

\begin{abstract}
We present results for the static interquark potential, light meson
and baryon masses, and light pseudoscalar meson decay constants
obtained from simulations of domain wall QCD with one dynamical
flavour approximating the $s$ quark, and two degenerate dynamical
flavours with input bare masses ranging from $m_s$ to $m_s/4$
approximating the $u$ and $d$ quarks. We compare these quantities
obtained using the Iwasaki and DBW2 improved gauge actions, and
actions with larger rectangle coefficients, on $16^3\times32$
lattices. We seek parameter values at which both the chiral symmetry
breaking residual mass due to the finite lattice extent in the fifth
dimension and the Monte Carlo time history for topological charge are
acceptable for this set of quark masses at lattice spacings above
0.1~fm. We find that the Iwasaki gauge action is best, demonstrating the
feasibility of using QCDOC to generate ensembles which are good
representations of the QCD path integral on lattices of up to 3~fm in
spatial extent with lattice spacings in the range 0.09-0.13~fm.  Despite large
residual masses and a limited number of sea quark mass values with
which to perform chiral extrapolations, our results for light hadronic
physics scale and agree with experimental measurements within our
statistical uncertainties.
\end{abstract}

\maketitle

\newpage


\section{Introduction}
\label{sec:Introduction}

Quantum Chromodynamcs (QCD), a four-dimensional, asymptotically free
gauge theory of interacting vector bosons and Dirac fermions, enjoys a
unique position due to its mathematical consistency and, as the
description of strong interactions, its relevance to current particle
physics phenomena.  However, due to the strength of the interactions
in this theory, analytic calculations are not possible for many
phenomenologically important quantities and numerical methods must be
employed in the quest to achieve first principles results.  Since
Wilson's formulation of lattice gauge theory \cite{Wilson:1974sk},
substantial progress has been made in these pursuits, although
incorporating fermions on the lattice has been a particularly
difficult enterprise.  Both Wilson's original lattice fermion
formulation and the subsequent staggered fermion approach, at
non-zero lattice spacing, break the full chiral symmetries of
continuum, massless QCD.  While these breakings vanish as the lattice
spacing goes to zero, they lead to practical difficulties with
operator mixing and a challenging extrapolation to the physical
light quark masses.

With the advent of domain wall fermions
\cite{Kaplan:1992bt,Shamir:1993zy,Furman:1994ky} and the related overlap fermions
\cite{Narayanan:1993wx,Narayanan:1993ss}, it has become possible to
perform simulations at non-zero lattice spacing which possess good chiral
and vector symmetries. When using domain wall fermions, at
sufficiently small lattice spacings, the light-quark limit should
behave as in the continuum, including the correct effects of topology,
without any infrared pathologies.  Chiral symmetry, which governs much
of the physics of low-energy QCD, is mildly broken in a controllable
way, yielding the correct number of light pseudoscalar mesons and protecting
against the mixing of operators with different chirality.  Also, as we
will see, exact numerical algorithms, based on Monte Carlo techniques,
can be employed with domain wall fermions for simulations with the
three light quarks that are relevant to low-energy QCD.

With domain wall fermions and exact algorithms for our simulations, we
now have numerical methods which nicely complement the pristine
mathematical pedigree of QCD.  These methods come at a cost. Current
simulations with domain wall fermions require more computer power at a
given lattice spacing than fermion formulations that do not fully
realise chiral and/or flavor symmetry.  This extra computational cost
may be more than recouped by better control over the light-quark limit
and potentially better scaling properties.  Also, there are many
calculations, particularly those involving nucleons and where operator
mixing is critical, such as those relevant for neutral kaon mixing,
that may only be practical with fermion formulations that have good
chiral properties.

Simulations of QCD with two flavors of domain wall fermions were first
performed some time ago on the QCDSP computer by the Columbia
group~\cite{Chen:2000zu}. More recently, simulations at finer lattice
spacings with better chiral symmetry were carried out by the RBC
collaboration~\cite{Aoki:2004ht}.  A first exploratory simulation of
three-flavor QCD with domain wall fermions was also done by the RBC
collaboration~\cite{Mawhinney:2005gn}.  This work showed that
dynamical domain wall simulations are practical, provided sufficient
computer power is available.  To increase the computer power available
for QCD simulations, the RBC and UKQCD collaborations, in
collaboration with IBM Research, the RIKEN Laboratory in Japan and the
RIKEN-BNL Research Center, have designed and built the QCDOC (QCD on a
Chip) computers \cite{Boyle:2005ibm, Boyle:2004sc,Boyle:2005gf}.  Both
collaborations have a 12,288 node QCDOC, each with a sustained speed
of 5 Tflop/s, making a serious investigation of QCD with 2+1 flavors
of domain wall fermions possible.

The primary objective of this paper is to establish a set of parameter
values for simulations of 2+1 flavor QCD on QCDOC, using domain wall
fermions and one choice from a class of improved gauge actions.  The
two light quark flavors ($u$ and $d$) are given the same mass,
$m_l$, and we seek to make $m_l$ at least as low as one quarter of the
strange quark mass, $m_s$, which is held fixed close to its physical
value.  Clearly, we need to understand how our parameter choices
affect the size of the residual chiral symmetry breaking in domain
wall fermions.  Since future simulations will require weaker couplings
to test scaling, we also pay particular attention to the rate at which
lattices decorrelate, using the evolution of topology as the most
demanding measure.  While small physical volumes will be used for the
simulations presented here, once we have determined our parameters,
volumes of 2.5~fm or larger, will be used.  Since coarser lattice
spacings make large physical volume simulations easier, we will also
explore how domain wall fermions behave for coarser lattices.

This paper is organized as follows.  In Section~\ref{sec:Details} we
give our basic definitions and notation, the precise form of the
actions used, and an overview of the simulations we have done.  The
work in this paper uses the exact Rational Hybrid Monte Carlo (RHMC)
algorithm of Clark and Kennedy~\cite{Clark:2003na,Clark:2004cp} and,
for a few parameter choices, the inexact $R$
algorithm~\cite{Gottlieb:1987mq}. We detail how we measure and extract
masses, decay constants, the string tension and other observables in
our simulations. In Section~\ref{sec:Results} we describe our results
for these quantities from simulations with a variety of gauge actions
that differ in the gauge coupling, relative size of the plaquette and
rectangle terms, and light quark masses. Varying parameters in this
space, keeping the physical lattice scale fixed, we measure and
compare the residual chiral symmetry breaking and the evolution of
topological charge.  Since we envisage large simulations at several
lattice spacings, we have also investigated the Iwasaki and DBW2 gauge
actions at weaker couplings. Our conclusions are given in Section
\ref{sec:Conclusions}.


\section{Calculation details}
\label{sec:Details}

\subsection{Lattice actions}
\label{sec:LatticeActions}

This paper reports on simulations with dynamical domain wall fermions
for a class of improved gauge actions.  Our notation, which we briefly
review here, is the same as in
\cite{Blum:2000kn,Aoki:2002vt,Aoki:2004ht}.  We denote points in four
dimensional space-time by $x$ and points in the fifth dimension of the
domain wall formulation by $s$ where $0 \leq s \leq L_s -1 $ and $L_s$
is the extent of the fifth dimension.  The partition function is given
by
\begin{equation}
Z = \int [dU] \int \prod_{i=1}^{3} [d\Psi_i d\bar{\Psi}_i]
   \int \prod_{i=1}^{3} [d\Phi^\dagger_{PV,i} d\Phi_{PV,i}]
   e^{-S}
\label{eq:Z}
\end{equation}
where the index $i$ runs over the $u$, $d$ and $s$ quark flavors.
The Pauli-Villars fields, $\Phi_{PV,i}$ are needed to cancel
the bulk infinity that would be produced by the domain wall
fermions as $L_s \to \infty$. 
In particular, the total action is
\begin{equation}
S = S_G(U) + S_F(\bar{\Psi}, \Psi, U) +
S_{PV}(\Phi^\dagger, \Phi, U).
\label{eq:total_action}
\end{equation}
The class of improved gauge actions we consider are of the form
\begin{equation}
  S_{\rm G}[U] =
   - \frac{\beta}{3} \left[
   (1-8\,c_1) \sum_{x;\mu<\nu} P[U]_{x,\mu\nu} \\
 + c_1 \sum_{x;\mu\neq\nu} R[U]_{x,\mu\nu}\right]
  \label{eq:gauge_action}
\end{equation}
where $P[U]_{x,\mu\nu}$ and $R[U]_{x,\mu\nu}$ represent the real
part of the trace of the path ordered product of link variables
around the $1\times 1$ plaquette and $1\times 2$ rectangle,
respectively, in the $\mu,\nu$ plane at the point $x$, and $\beta
\equiv 6/g^2$ with $g$ the bare quark-gluon coupling.
Different approximations to the renormalization group trajectory motivate
two common choices for $c_1$:  1) the Iwasaki action which
sets $c_{1} = -0.331$~\cite{Iwasaki:1983ck,Iwasaki:1985we,Iwasaki:1984cj}
and 2) the DBW2 action which has $c_1 = -1.4069$~\cite{Takaishi:1996xj,deForcrand:1999bi}.

For the fermion action in Eq.~(\ref{eq:total_action}),
we use the domain wall fermion formulation of Shamir
\cite{Shamir:1993zy}, and Furman and Shamir \cite{Furman:1994ky}.
In our notation, the domain wall fermion operator $D^{\rm DWF}$, for a 
fermion of mass $m_f$, is defined as
\begin{equation}
  D^{\rm DWF}_{x,s; x^\prime, s^\prime}(M_5, m_f)
   = \delta_{s,s^\prime} D^\parallel_{x,x^\prime}(M_5)
   + \delta_{x,x^\prime} D^\bot_{s,s^\prime}(m_f)
\label{eq:D_dwf}
\end{equation}
\begin{eqnarray}
D^\parallel_{x,x^\prime}(M_5) & =&
  {1\over 2} \sum_{\mu=1}^4 \left[ (1-\gamma_\mu)
  U_{x,\mu} \delta_{x+\hat\mu,x^\prime} + (1+\gamma_\mu)
  U^\dagger_{x^\prime, \mu} \delta_{x-\hat\mu,x^\prime} \right] 
  \nonumber \\
  & + & (M_5 - 4)\delta_{x,x^\prime}
\label{eq:D_parallel}
\end{eqnarray}
\begin{eqnarray}
D^\bot_{s,s^\prime}(m_f) 
     &=& {1\over 2}\Big[(1-\gamma_5)\delta_{s+1,s^\prime} 
                 + (1+\gamma_5)\delta_{s-1,s^\prime} 
                 - 2\delta_{s,s^\prime}\Big] \nonumber\\
     &-& {m_f\over 2}\Big[(1-\gamma_5) \delta_{s, L_s-1}
       \delta_{0, s^\prime}
      +  (1+\gamma_5)\delta_{s,0}\delta_{L_s-1,s^\prime}\Big].
\label{eq:D_perp}
\end{eqnarray}
$D^{\rm DWF}$ is not Hermitian, but does satisfy $\gamma_5 R_5 D_{\rm
  DWF} \gamma_5 R_5 = D_{\rm DWF}^\dagger$, where $R_5$ is a
reflection operator in the fifth dimension.  This, along with the
transfer matrix formalism of \cite{Furman:1994ky}, suffices
to show that $\det(D^{\rm DWF})$ is positive for positive mass,
allowing us rigorously to write the fermion action in a form suitable
for simulations as
\begin{equation}
S_F = - \sum_{i=1}^{3} \bar{\Psi}_i \left[ D_{\rm DWF}^\dagger(M_5, m_i)
  D_{\rm DWF}(M_5, m_i) \right]^{1/2} \Psi_i
\label{eq:fermion_action}
\end{equation}
where $m_i$ is the input bare quark mass for the $i$th light quark
flavor.  We only consider the case where all light quarks have the
same value for the five-dimensional domain wall height, $M_5$.  The
action for the Pauli-Villars fields is similar, except that the
quark mass $m_f$ is replaced by 1 to yield
\begin{equation}
S_{PV} =
  \sum_{i=1}^{3} \Phi^\dagger_i \left[ D_{\rm DWF}^\dagger(M_5,
  1) D_{\rm DWF}(M_5, 1) \right]^{1/2} \Phi_i .
\label{eq:pv_action} \end{equation}
It should be noted that this is not the precise form 
of the Pauli-Villars action density given in \cite{Furman:1994ky},
but a variant introduced in \cite{Vranas:1997da}.
For the case of 2 dynamical flavors with the same mass, $m_l$,
integrating out the fermions and Pauli-Villars fields yields
the following determinants to include in the generation of
gauge fields
\begin{equation}
\frac{
 \det \left[ D_{\rm DWF}^\dagger(M_5, m_l) D_{\rm DWF}(M_5, m_l) \right]
 }{
 \det \left[ D_{\rm DWF}^\dagger(M_5, 1) D_{\rm DWF}(M_5, 1) \right]
 }.
 \label{eq:nf2_det}
\end{equation}
This can be readily simulated by conventional Hybrid Monte Carlo by
introducing pseudofermion fields for the numerator and using
bosonic Pauli-Villars fields for the denominator, as was done in
\cite{Chen:2000zu}.   The more recent 2-flavor simulations 
in \cite{Aoki:2004ht} used a single pseudofermion field to
evaluate directly the ratio of determinants in Eq.~(\ref{eq:nf2_det}).
This reduces the stochastic noise in the molecular dynamics evolution
and speeds up the calculation, since a larger step size can be used.

For 2+1-flavor simulations, the term in Eq.~(\ref{eq:nf2_det})
has to be multiplied by
\begin{equation}
\frac{
 \det^{1/2} \left[ D_{\rm DWF}^\dagger(M_5, m_s) D_{\rm DWF}(M_5, m_s)
   \right] }{
 \det^{1/2} \left[ D_{\rm DWF}^\dagger(M_5, 1) D_{\rm DWF}(M_5, 1)
    \right]
 }
 \label{eq:nf3_det}
\end{equation}
where $m_s$ is the strange quark mass.  
In this work, we have handled the fractional power in two ways. For
most of the simulations we have used the exact RHMC
algorithm~\cite{Clark:2003na,Clark:2004cp}.  For some of our initial
simulations we used the $R$ algorithm~\cite{Gottlieb:1987mq}, an
inexact algorithm with finite step-size errors that easily handles
fractional powers of the fermion determinant.  At the time of this
work, our implementation of the RHMC algorithm used separate fields as
stochastic estimators for the numerator and denominator of
Eq.~(\ref{eq:nf3_det}).  Code to stochastically estimate the ratio was
being finished while these simulations were underway and is now in
use.

The RHMC algorithm allows us to simulate many decompositions of the
same fermionic determinant, since $\det(M) = [\det^{1/n}(M)]^n$.
If we adopt the notation ${\cal D}(m_i) = D_{\rm
  DWF}^\dagger(M_5, m_i) D_{\rm DWF}(M_5, m_i)$ and, by convention,
let every determinant appearing be represented by a separate
pseudofermion field, then our decomposition can be written as
\begin{equation}
\frac{
  \det^{1/2}[{\cal D}(m_l)] \; \det^{1/2}[{\cal D}(m_l)] \;
  \det^{1/2}[{\cal D}(m_s)]
}{
  \det[{\cal D}(1)] \; \det^{1/4}[{\cal D}(1)] \;
  \det^{1/4}[{\cal D}(1)]
}.
\label{eq:DeterminantDecomp}
\end{equation}
Thus, we used six pseudofermion fields: five associated
with the RHMC, since five of the determinants in Eq.~(\ref{eq:DeterminantDecomp})
involve fractional powers, and one associated with conventional HMC.  

\subsection{Details of the ensembles}

In our search for the optimal parameters for 2+1-flavor QCD with
domain wall fermions, we have performed a large number of simulations
for various values of $\beta$ and $c_1$ in
Eq.~(\ref{eq:gauge_action}) with two different goals in mind: 1)
to work at a fixed inverse lattice spacing ($\approx 1.6$ GeV) and see how the
residual chiral symmetry breaking of domain wall fermions varies with
$c_1$, and 2) to then explore the weaker coupling behaviour at fixed
$c_1$.

In earlier work using the QCDSP computer and the $R$ algorithm, the
RBC collaboration estimated that the inverse lattice spacing for 3-flavor QCD 
with domain wall fermions on a $16^3 \times 32 \times 8$ lattice with
the DBW2 gauge action at $\beta = 0.72$ was 1.6--1.7
GeV~\cite{Mawhinney:2005gn}.  Since only a single dynamical quark mass
was used and calculating the lattice spacing
requires the light dynamical quark limit, they were only able to
produce a rough estimate for the lattice spacing. We use the DBW2
action at $\beta = 0.72$ on a $16^3\times 32\times 8$ lattice as our
starting point.

The RBC~\cite{Wu:1999cd} and CP-PACS~\cite{AliKhan:1999zn,AliKhan:2000iv}
collaborations noted that the residual chiral symmetry breaking for
domain wall fermions is reduced for the Iwasaki action compared to
the Wilson gauge action ($c_1 = 0$) for quenched simulations.  This
was studied further in~\cite{Aoki:2002vt}, where it was found that
the DBW2 action markedly reduces residual chiral symmetry breaking
on quenched lattices with $a^{-1} \approx 2$ GeV.  For 2-flavor
dynamical simulations with domain wall fermions, it was found in~\cite{Wu:1999cd}
that the Iwasaki gauge action was not
much better than the Wilson gauge action for very coarse lattices
with $a^{-1} \approx 700$ MeV.  The recent 2-flavor simulations of
QCD with domain wall fermions with $a^{-1} \approx 1.7$ GeV
\cite{Aoki:2004ht} used the DBW2 gauge action, but there was not
sufficient computer power available then to test this choice.  One of
our goals in this paper is to pursue this question further to see how
the gauge action choice affects residual chiral symmetry breaking, the
tunnelling of topological charge and algorithmic performance.

Thus, we explore three classes of gauge actions with different relative
admixtures of the rectangle term in Eq.~(\ref{eq:gauge_action}):
Iwasaki, DBW2 and actions with even larger rectangle coefficients. All
the ensembles were generated with a lattice size of $16^3 \times 32
\times 8$. The Iwasaki and DBW2 ensembles were generated with the RHMC
algorithm and are described in Table~\ref{tab:rhmc_ensembles}.  All
these ensembles have 2+1 flavors with the strange quark mass held
fixed at approximately its physical value, and two or three values for
the equal $u$ and $d$ quark masses, allowing for rudimentary extrapolations
to the chiral limit.
In addition to the quenched result that the
DBW2 gauge action has excellent chiral properties~\cite{Aoki:2002vt},
exploratory work~\cite{Levkova:2004xw} showed that a
negative rectangle coefficient ($c_1$ in Eq.~(\ref{eq:gauge_action}))
has the effect of making the gauge fields smoother, thus rendering the
residual mass smaller. We study this effect further, by increasing the
magnitude of $c_1$, choosing $\beta$ to keep the lattice scale fixed
to approximately 1.6 GeV. These ensembles were generated with
3 degenerate flavors  using the inexact $R$ algorithm, with the quark mass
approximately that of the physical strange quark, and are
described in Table~\ref{tab:PR_datasets}. The three data sets at the
bottom are at finer lattice spacings, which we generated to study the
effect of this on the residual mass and the rate of change of topology.

The RBC collaboration found that the
combination of $\beta$ and $c_1$ that gives roughly the same lattice
scale falls on the curve
\begin{equation}
 \frac{\beta_R}{\beta_P} = -0.125 + A(a) \beta + B(a) {\beta}^2,
 \label{eq:plaq_rect_fit}
\end{equation}
where $\beta_R = c_1$ and $\beta_P = 1 - 8c_1$. This is shown in
Figure~\ref{fig:plaq_rect_plane}, where early simulations with $0$ and
$2$ flavors of dynamical fermions and the simulations presented in
this paper are included. The solid curve is a fit to the quenched
simulations at an inverse lattice spacing of 2.0 GeV with different values of
$\beta_R/\beta_P$. The ability to predict the lattice spacings in this
manner allows us to do simulations at the desired coupling without the
need for extensive searching in the parameter space.

While the QCDOC computers provide some of the most powerful resources
currently available for lattice QCD, each parameter choice requires
substantial computing resources.  For example a 1,500 trajectory
simulation takes a few weeks to generate on 1,024 nodes of QCDOC,
depending on the quark masses and algorithms used.  In our study, the
choice of parameters often depended on ensembles running at that time, so it
was important to generate ensembles as quickly as possible.  We
exploited naive parallelism and the availability of several smaller
machines by spawning short Markov chains from the original chain.
{\it i.e.} starting from a configuration in the original chain, a
second distribution of random numbers (different from those in the
original evolution) was generated and these were used to
evolve a new branch. These branches were `farmed' out to
several machines in parallel. This had the advantage of increasing statistics to an
acceptable level while reducing the `wall clock' time.

We used the decomposition in Eq.~(\ref{eq:DeterminantDecomp}) for the
(D, 0.72, 0.01/0.04) and (D, 0.72, 0.04/0.04) ensembles, whereas 
for all the other ensembles, including the (D, 0.72, 0.02/0.04) ensemble, 
we combined the two 1/4 power determinants in the
denominator of Eq.~(\ref{eq:DeterminantDecomp}) into a single 1/2 power
determinant and used one less pseudofermion field.

The parameters that enter into the RHMC algorithm control the accuracy
of the rational approximation and the range of eigenvalues for which it is valid.
Table~\ref{tab:RHMC_parameters} gives the value for the
parameters we used.  The maximum and minimum eigenvalues,
$\lambda_{\rm max} $ and $\lambda_{\rm min}$, of ${\cal D}(m_i) $ are
used to determine the eigenvalue range for which the rational
approximation has to be valid. The degree of the rational polynomial
determines the accuracy of the approximation over this range.  We use
a more accurate rational approximation for the accept/reject step than
for the molecular dynamics integration, since the accept/reject step
removes any errors in the approximation arising during the
integration.

Table \ref{tab:RHMC_0_72_evol_stat} gives values for $\langle \delta H
\rangle$, $\langle e^{-\delta H} \rangle$ and the acceptance for the
ensembles generated with the RHMC algorithm.  $\langle e^{-\delta H}
\rangle$ is equal to $1$ within errors, indicating that the algorithm is working
correctly.  Also shown in Table \ref{tab:RHMC_0_72_evol_stat} are the
ensemble averages for the plaquette, where typically the first 1000
trajectories of the ensemble were excluded from the averages.

\subsection{The static quark potential}
\label{sec:static}

The static quark potential depends relatively weakly on the sea quark masses.
Consquequently, the chiral limit can be taken with reasonable confidence and provides an
estimate of the lattice spacing that is relatively free of systematic uncertainty, compared to
hadron masses.

The static potential $V(\vec{r})$ between a quark and anti-quark pair
at relative spatial displacement $\vec{r}$ is calculated from the
Wilson loop expectation value $\langle W(\vec{r},t) \rangle$, where
\be
\langle W(\vec{r},t) \rangle = C(\vec{r}) e^{-V(\vec{r})t}
+{\rm excited} \ {\rm states}.  
\ee
Here $\langle W(\vec r,t) \rangle$ is the average
of the standard Wilson loop with spatial side of length $\vec r$ and
temporal extent $t$.  The static quark potential is then given by the
ratio
\begin{equation}
V(\vec r,t) = \ln \Bigl\{\frac{W(\vec r,t)}{W(\vec r,t+1)}\Bigr\}.
\label{eq:v_r_t}
\end{equation}
The time dependence in $V(\vec r,t)$ should disappear for sufficiently 
large $t$, so an important requirement is that $V(\vec r)$ be determined
from a plateau seen in the quantity $V(\vec{r},t)$ as $t$ increases for fixed 
$\vec{r}$.

We compute lattice values for the Wilson loops $\langle W(\vec{r},t) \rangle$
using the method of Bolder {\it et al.}~\cite{Bolder:2000un}.
This approach evaluates all separations $\vec r$, including those
which do not lie along a lattice axis.  For such off-diagonal separations
the Bresenham algorithm is used to determine the sequence of spatial
gauge links that make up the spatial lines joining the two time 
segments of the Wilson loop.  For both the on- and off-axis cases, these 
spatial links are constructed by APE smearing~\cite{Albanese:1987ds} 
to improve the ratio of signal to 
noise. To be precise, at each step of APE smearing we replace each spatial link by
\newcommand{\csmear}{c_\mathrm{smear}}
\begin{eqnarray}
\nonumber
U^\prime_\mu(x) \equiv \mathbb{P}\left[ 
U_\mu(x) + 
c_{\rm smear,3d} \sum_{\nu=1,2,3 \mu\not=\nu} \right. && \left. \left( 
U_\nu(x) U_\mu(x+\hat{\nu}) U^\dagger_\nu(x+\hat{\mu}) + \right. \right. 
\\ &&  \left.
\!\!\!\!\!\!\!\!\!\!\!\!\!\!\!\!\!\!\!\!U^{\dagger}_{\nu}(x-\hat{\nu})U_{\mu}(x-\hat{\nu})U_{\nu}(x+\hat{\mu}-\hat
{\nu})
\right) \Bigg] \ \mu=\{1,2,3\} ,
\label{eq:APESmear}
\end{eqnarray}
where we include all spatial directions and $\mathbb{P}$
denotes a projection back onto $SU(3)$.  
This process is then repeated $N_{\rm smear,3d}$ times.  The smearing 
coefficient and number of smearing steps are tuned to be 
$(c_{\rm smear,3d},N_{\rm smear,3d})=(0.50, 20\sim35)$, to maximise the overlap 
with the ground state of the Wilson loop, $C(\vec{r})$.

This approach was also used by the RBC
collaboration~\cite{Aoki:2004ht,Hashimoto:2004rs} and an earlier
description of some of the results presented here can be found in
Ref.~\cite{Hashimoto:2005re}. The results for $V(\vec{r})$ obtained by
this procedure were checked by an independent calculation using only
on-axis loops and Chroma code~\cite{Edwards:2004sx}.  While the errors
were much larger for times $t=5$ and 6, the results are consistent.

Physical parameters are obtained by fitting the lattice value of 
$V(\vec{r})$ to the function
\begin{equation}
V(\vec{r}) = V_{0}-{\alpha\over|\vec{r}|}+\sigma |\vec{r}|.
\label{eq:pot_fit}
\end{equation}
Finally, the parameters $\alpha$ and $\sigma$ can be used to determine
the Sommer scale~\cite{Sommer:1993ce,Guagnelli:1998ud}
\be
 r_0(m_l) = \sqrt{(1.65-\alpha) \over \sigma}
\label{eq:r0def}
\ee
for each gauge action and gauge coupling $\beta$, in lattice units. The value of $r_0$ 
in the chiral limit can then be used to define the lattice spacing.
To express this in physical units we take $a=0.5~{\rm fm}/r_{0}$.

\subsection{Topological Charge}
\label{sec:TopologicalCharge}

Continuum Yang-Mills gauge fields can be divided into classes
characterized by an integer-valued winding number known as the
Pontryagin index or topological charge:
\begin{equation}
Q = \frac{1}{32 \pi^2} \int d^4x \; \epsilon_{\mu \nu \sigma \tau}
\tr F_{\mu \nu}(x) F_{\sigma \tau}(x) \in \mathbb{Z}
\label{eqn_top_gluonic}
\end{equation}
which is stable under smooth deformations of the gauge field.
The Atiyah-Singer index theorem \cite{Atiyah:1963_cd,Atiyah:1968_cd}
predicts that the Dirac operator on a gauge background with 
topological charge $Q$ will have at least $Q$ exact zero
modes in the zero quark-mass limit,
\be
  Q = n_+-n_-,
\label{eqn:indextheorem}
\ee
where $n_+(n_-$) is the number of positive (negative) chirality zero modes.

On the lattice the situation is much more complicated; gauge fields are far
from smooth and the low-energy form of the Dirac operator can be distorted by
the explicit breaking of the continuum symmetries.
Nonetheless, when using domain wall fermions, zero modes of the Dirac operator
are apparent and their numbers have been found to correspond well with the value
of the topological charge calculated from the gauge field (by methods
described below)\cite{Blum:2000kn,Blum:2001qg,Aoki:2002vt}.  
Given the numerical cost in solving for the low-lying eigenvalues of the domain
wall Dirac operator, we apply two gluonic definitions of topological charge:
\begin{enumerate}
\item calculate the topological charge via Eq.~(\ref{eqn_top_gluonic})
using a classically $O(a^2)$-improved definition of the field strength
tensor built from plaquette and rectangle clover-leaf terms (precise details
  can be found in \cite{Aoki:2004ht});
\item the `5Li' definition of \cite{deForcrand:1995qq,deForcrand:1997sq}, which
combines the $1\times 1$, $2\times 2$, $1\times 2$, $1\times 3$ and $3\times 3$
clover-leaf terms to give a classically $O(a^4)$-improved definition of the
field strength tensor and, therefore, of the topological charge using 
Eq.~(\ref{eqn_top_gluonic}).
\end{enumerate}

Since both these methods expand about the classical continuum limit,
they cannot be directly applied to lattice gauge field configurations.
Rather than deal with the subtle issue of constructing renormalised
operators, we smooth the gauge field configurations so that expanding
about the continuum limit is sensible. When using the first definition
we follow \cite{DeGrand:1997ss} in using APE smearing
\begin{eqnarray}
\nonumber
U^\prime_\mu(x) \equiv \mathbb{P}\left[ 
(1 - \csmear) U_\mu(x) + 
\frac{\csmear}{6} \sum_{\nu
\not = \mu} \right. && \left. \left( 
U_\nu(x) U_\mu(x+\hat{\nu}) U^\dagger_\nu(x+\hat{\mu}) + \right. \right. 
\\ && \left. \left.
U^\dagger_\nu(x-\hat{\nu}) U_\mu(x-\hat{\nu}) 
U_\nu(x+\hat{\mu}-\hat{\nu})
\right) \right] \; ,
\label{eq:APESmear2}
\end{eqnarray}
with $\csmear=0.45$ in which the temporal link is also smeared
in contrast to Eq.~(\ref{eq:APESmear}). We quote the value of $Q$
after 20 steps, although we calculate up to 30 steps to check the stability of
the extracted value. 
When using the second definition of topological charge, we follow
\cite{deForcrand:1995qq,deForcrand:1997sq} and cool the configurations
using the 5Li action. This takes the same combination of loops as used
for the 5Li definition of the topological charge (although not
clover-leaf symmetrized) to construct the action. Again, this
definition is $O(a^4)$-improved about the classical continuum
limit. It is also chosen so that the size of the instantons is
invariant under cooling for instantons of size a few lattice
spacings. We have used up to 50 cooling steps. The results given in
this paper are determined after 30 steps, beyond which the answers are
stable.

A third method we have used to monitor the topological charge is to
calculate the fermionic operator $\langle \bar{q} \gamma_5 q \rangle$
using a stochastic estimator. We only used a single stochastic source
per configuration and, therefore, this measurement has large
fluctuations when $Q \not = 0$, since the overlap of the noisy
estimator with the small number of topological zero modes has a large
variance. These fluctuations can be reduced by averaging measurements
from nearby lattices in the Markov chain. At
molecular dynamics time $\tau = i$, the smoothed value of $\langle
\bar{q} \gamma_5 q \rangle$ is found by averaging over measurements in
the range $i-n/2 \le \tau < i+n/2$. We refer to this as a smoothing window of size $n$. 
Smoothing suppresses topological fluctuations
that exist for only a short time in the evolution and should give a
good estimator of the topological charge if the topology only changes
significantly on a molecular dynamics timescale that is larger than
$n$. The three methods give measurements of $Q$ that agree extremely well.

\subsection{Hadron masses}
\label{sec:hadrons}

Hadronic correlation functions are constructed from
quark propagators which take the following form
\be
S_{AB}(x,y)=\langle q_A(x) \overline{q}_B(y) \rangle,  
\label{tech:prop}
\ee
where the indices $A$ and $B$ represent different smearings of the
quark field, $q(x)$. All the smeared quark fields make use of the
same construction of the four-dimensional quark field from the five-dimensional 
domain wall fermion field $\Psi(x,s)$. In
the case of the local ($L$) quark field, $q_L(x)$, 
this takes the form
\be
q_{L}(x) = P_L \Psi(x,s=0) + P_R \Psi(x,s=(L_s-1)) \,
\ee
where $P_{R/L} = \frac{1}{2}\left(1 \pm \gamma_5\right)$ are the chiral
projectors. 
To construct the Coulomb gauge-fixed wall sources ($W$), we
replace the quark field in Eq.~(\ref{tech:prop}) with the non-local
field
\be
q_{W}(t)_c =  \sum_{r,c'} V\left(r,t\right)_{c,c'} q_L(r,t)_{c'} 
\ee
where $V(r,t)$ is the color matrix which transforms the spatial
links at $x=(r,t)$ into Coulomb gauge and  $c$ and $c'$ are color 
indices. 
The third type of smearing ($S$) uses a hydrogenic wavefunction~\cite{Boyle:1999gx}.
The quark field is convolved with a spatial smearing function
after fixing the gauge fields to the Coulomb gauge as follows,
\begin{eqnarray}
q_{S}(t)_c &=& \sum_{r,r',c'} S\left(r,r'\right)V\left(r',t\right)_{c,c'} 
q_L(r',t)_{c'} \, , \\
&&S(r,r') = e^{(-\Delta r/R)} \delta_{r,r_0} \, ,
\label{tech:smearing_rad}
\end{eqnarray}
where $\Delta r$ is the modulus of the minimum distance between 
the center of the source, $r_0$, and $r'$ (taking into account
the periodic boundary conditions), and $R$ is a tunable parameter. 
We give detailed comparisons of results for hadron masses obtained with different
smearings in Section \ref{sec:Results}.

Correlation functions for mesonic operators are constructed as 
follows
\be
  C_{ij}(t,\vec{p})=\sum_{\vec{x}}e^{i\vec{p}\cdot\vec{x}}
	\langle \Omega_i(\vec{x},t) \Omega^{\dag}_j(\vec{0},0)
        \rangle
\label{eq:mesonCorr}
\ee
where
\begin{equation}
  \Omega_i(\vec{x},t)=q(\vec{x},t)\Gamma_i\bar{q}(\vec{x},t)
\end{equation}
and, for instance, $\Gamma_i=\gamma_5$ for the pseudoscalar meson and
$\Gamma_i=\gamma_k$ for the vector meson.
The spectrum can then be extracted by fitting these correlation functions at
zero momentum to the form
\be
  C(t)=\sum_n A_n\left( e^{-m_nt}+e^{-m_n(T-t)}\right)
\ee
where $T$ is the size of the lattice in the time direction.
For sufficiently large $t$, the ground state will dominate
the correlation function.

The standard baryon interpolating operator is composed of 
a local diquark operator and a spectator-like quark field:
\be
  \Omega_{ijk,B}(x)=\epsilon_{abc}\left[q^{T}_{a,i}(x) C \Gamma q_{b,j}(x)\right]q_{c,k}(x),
\ee
where $\Gamma$ stands for one of the possible 16 Dirac matrices and $C$
is the charge conjugation matrix.  The superscript $T$ denotes
transpose and the indices $abc$ and $ijk$ label color and flavor,
respectively.  For the $(I,J)=(\frac{1}{2}, \frac{1}{2})$ baryons (the
nucleon), an iso-scalar diquark is chosen, {\it i.e} $\{\Gamma=1,
\gamma_5, \gamma_5\gamma_\mu\}$. However, only two of the three
operators are independent, as they are related to each other through
a Fierz transformation. In this work, the two nucleon operators are 
chosen to be
\begin{eqnarray}
\Omega_{B_1}& = & \epsilon_{abc}\left[u^{T}_{a}C\gamma_5d_{b}\right]u_{c}, \\
\Omega_{B_2}& = & \epsilon_{abc}\left[u^{T}_{a}Cd_{b}\right] u_{c} .
\end{eqnarray}
The intrinsic parity of these operators is defined by the parity
transformation of the internal quark fields. The $\Omega_{B_1}$
($\Omega_{B_2}$) operator transforms as a positive (negative) parity
operator. However, this parity assignment is easily flipped by
multiplication of the local baryon interpolating operator by
$\gamma_5$. Therefore, a two-point correlation function composed of
either $\Omega_{B_1}$ or $\Omega_{B_2}$ operators possesses both the
positive- and negative-parity nucleon contributions. For details, see
Ref.~\cite{{Sasaki:2001nf},{Sasaki:2005ug}}. The $\Omega_{B_1}$
operator is conventionally used in lattice QCD for the $J^P=1/2^+$
nucleon ($N$). However, it is also of interest to examine the
$J^P=1/2^-$ nucleon ($N^*$), so both $\Omega_{B_1}$ and $\Omega_{B_2}$
operators are utilized for $(I,J)=(\frac{1}{2}, \frac{1}{2})$ baryon
spectroscopy.  

Taking the trace of the baryon two-point correlator with the relevant
projection operator, $P_+=(1+\gamma_4)/2$, the two-point correlator in
a finite box with (anti-)~periodic boundary conditions
is given by
\be
  C_B(t) = A_\eta e^{-m_\eta t} - A_{-\eta}e^{-m_{-\eta}(T-t)},
  \label{ch6:eqn:baryonCorr}
\ee
where the parity of the forward propagating state $\eta=+$ ($-$) 
is the same as the intrinsic parity of the interpolating operator, $\Omega_{B_1}$ ($\Omega_{B_2}$), 
whilst the backward propagating state has the opposite parity. 

According to an intensive study of the nucleon excited states in
Ref.~\cite{Sasaki:2001nf}, the $\Omega_{B_2}$ operator has a poor
overlap with the nucleon ground state. Therefore, in this study, the
masses of the positive and negative parity states were determined by a
simultaneous fit to the following forms
\ba
 C_{B_{1}}(t) &=& A_+e^{-m_+t} - A_-e^{-m_-(T-t)},\\\nn
 C_{B_{2}}(t) &=& A^{\;'}_{-}e^{-m_-t}.\nn
\ea

\subsection{Residual mass}
\label{subsec:mres_def}

While at short distances the domain wall fermion formulation is a
five-dimensional lattice theory, at long distances and for large $L_s$
it is expected to appear identical to continuum QCD, with chiral
symmetry broken only by the explicit mass term $m_f$, introduced in
Eq.~(\ref{eq:D_perp}).  The deviations from this ideal behaviour can be
easily described by a continuum, Symanzik effective Lagrangian.
Because of the finite size of the fifth dimension, this effective
Lagrangian will contain explicit chiral symmetry breaking terms.
While the coefficients of these terms are suppressed as $L_s$
increases, they are important because of their chiral properties and
because they are of lower dimension (three and five) than the
dimension-six, chirality conserving, $O(a^2)$ terms:
\begin{equation}
{\cal L}_{\rm eff} = L_{\rm QCD}(m_f=0) + (m_f + \mres)\overline{q}q
         + c_5 \overline q \sigma^{\mu\nu} F^{\mu\nu} q.
\label{eq:syman_eff}
\end{equation}

The leading term is simply an additional mass term whose coefficient
is called the residual mass, labelled $\mres$.  As is conventional, the
normalization of $\mres$ is fixed by requiring the input parameter
$m_f$ and $\mres$ to multiply the same $\overline{q}q$ mass operator
as in Eq.~(\ref{eq:syman_eff}).  In this paper we will ignore the
effects of the Sheikholeslami-Wohlert, $c_5$ term, which is suppressed
by both large $L_s$ and one power of the lattice spacing.  As will be
seen below, methods for determining $\mres$ typically determine a
quantity which depends on the input quark mass $m_f$.  This $m_f$
dependence is a lattice artefact and represents the
$O(a)$ or $O(a^2)$ ambiguity in our definition of $\mres$.  A more
careful treatment, which requires the analysis of the $c_5$ term in
Eq.~(\ref{eq:syman_eff}) and similar lattice artefacts in
the quantities being used to compute $\mres$ is beyond the scope of
this paper.

The residual mass can be computed from the additional
contribution to the partially conserved axial current, $J^a_{5q}$, at the
mid-point of the fifth
dimension~\cite{Furman:1994ky,Blum:1998ud,Blum:2000kn}.  Assuming that
this mid-point contribution can also be described using the Symanzik
effective theory of Eq.~(\ref{eq:syman_eff}), then to lowest order in
the quark mass and lattice spacing $J^a_{5q} = \mres J^a_5 = \mres
\overline{q}\gamma^5 t^a q$, where $t^a$ is a generator of the flavor
symmetry.  We can then compute $\mres$ by averaging over time the
quantity $R(t)$, which is defined by
\begin{equation}
  R(t) = \frac{\langle \sum_{\vec{x}} J^a_{5q} (\vec{x}, t) \pi^a(0)
  \rangle } {\langle \sum_{\vec{x}} J^a_5(\vec{x}, t) \pi^a(0) \rangle
  } .  
\label{eq:mres_ratio} 
\end{equation} 
Here $\pi^a(0)$ is a (possibly smeared) pseudoscalar interpolating
field at $t=0$.  The minimum time used in the average need only be
large enough to remove any contribution to the correlators in $R(t)$
from unphysical states, since $\mres$ should affect all physical
states equally.  In the following we will refer to this average of
$R(t)$ as $\mres^\prime(m_f)$ to explicitly display its dependence on
$m_f$ and to allow a clear discussion of how we deal with this small,
non-zero lattice spacing ambiguity in determining the constant $\mres$
which appears in the effective Lagrangian of Eq.~(\ref{eq:syman_eff}).

\subsection{Determining the pseudoscalar decay constant}

The pseudoscalar decay constant is defined by
\begin{equation}
  Z_A \langle 0| \, \bar{q}(x) \gamma^\mu \gamma_5 q(x) \, | \pi, \vec{p} \rangle
  = -i f_{P} p^\mu e^{-ip \cdot x}
  \label{eq:axial_current_me}
\end{equation}
where $Z_A$, the renormalization factor for the local axial vector current,
can easily be determined for domain wall fermions since $Z_A A^a_\mu
= {\cal A}^a_\mu$ and ${\cal A}^a_\mu$ is the (partially) conserved axial
current.  In particular, we follow \cite{Blum:2000kn} and define
\begin{eqnarray}
C(t+1/2) &=& \sum_{\vec x} \langle {\cal A}^a_0(\vec x ,t)
         \; \pi^a(\vec{0}, 0) \rangle \nonumber \\ 
L(t)     &=& \sum_{\vec x} \langle A^a_0 (\vec x,t) 
         \; \pi^a(\vec{0}, 0) \rangle.
\label{eq:eq:Ct_Lt}
\end{eqnarray}
We then calculate $Z_A$ from these correlators, using a combination
that is free of ${\cal O}(a)$ errors and minimizes ${\cal O}(a^2)$
errors.  This gives us the following explicit form for $Z_A$.
\begin{equation}
Z_A = \frac{1}{1+t_{\rm max} - t_{\rm min}}
  \sum_{t=t_{\rm min}}^{t_{\rm max}}
        \frac{1}{2}
        \left\{ \frac{C(t+1/2)+C(t-1/2)}{2\ L(t)} +
        \frac{2\ C(t+1/2)}{L(t)+L(t+1)} \right\}.
\label{eq:ZA_def}
\end{equation}

We have calculated $f_P$ in three ways.  The simplest
uses the local-local axial current in Eq.~(\ref{eq:mesonCorr}), giving
\begin{equation}
C^{LL}_{A_0, A_0}(t) = \sum _{\vec x} \langle 0 | A_0(\vec{x},t) \;
A_0(0,0) | 0 \rangle \to A^{LL}_{A_0, A_0}( e^{-mt} +
 e^{-m(T-t)}).
\label{eq:fpv1}
\end{equation}
The amplitude $A^{LL}_{A_0, A_0}$ is determined from fitting the correlator to
Eq.~(\ref{eq:fpv1}), which in turn using Eq.~(\ref{eq:axial_current_me}) yields $f_P$ from
\begin{equation}
  f_P = Z_A \sqrt{\frac{2 A^{LL}_{A_0, A_0}}{m_P}}.
  \label{eq:fpMeth1}
\end{equation}
In the second method, we use the axial Ward-Takahashi identity which,
including the midpoint contribution as $J^a_{5q} = \mres J^a_5$,
gives
\begin{equation}
  m_P Z_A \langle 0 | A_0 | \pi \rangle = 2(m_f + \mres) 
  \langle 0 | P | \pi \rangle
\end{equation}
where $P$ is the pseudoscalar density.  From the local-local
pseudoscalar correlator, we have
\begin{equation}
C^{LL}_{P,P}(t) = \sum _{\vec x} \langle 0 | P(\vec{x},t) \;
P(0,0) | 0 \rangle \to A^{LL}_{P,P}( e^{-m_Pt} +
 e^{-m_P(T-t)})
\end{equation}
which, combined with Eq.~(\ref{eq:axial_current_me}), gives
\begin{equation}
  f_P = \frac{2 (m_f + \mres)}{m_P} \;
   \sqrt{\frac{2 A^{LL}_{P,P}}{m_P}}.
  \label{eq:fpMeth2}
\end{equation}

The third method uses the local axial current pseudoscalar
correlator
\begin{equation}
C^{LL}_{A_0,P}(t) = \sum_{\vec{x}}\langle 0 | A_0(\vec{x},t)
 \; P(\vec{0},0)| 0 \rangle \to
\frac{\langle 0| A_0| \pi \rangle \langle \pi|P|0 \rangle}{2m_{P}}
  ( e^{-m_Pt} + e^{-m_P(T-t)}).
\end{equation}
The ratio of this correlator and the pseudoscalar density correlator is then fitted to the 
following form,
\begin{equation}
\frac{C_{A,P}(t)}{C_{P,P}(t)} = \frac{\langle 0|A_{0}| \pi 
\rangle \langle \pi|P|0 \rangle}{\langle 0|P|\pi \rangle \langle \pi|P|0 \rangle} 
\approx \frac{\langle 0|A_{0}|\pi \rangle}{\langle 0|P|\pi \rangle}
\tanh\left[m_{P}(\frac{T}{2}-t)\right],
\end{equation}
to yield
\be
A_{A_0,P} = \frac{\langle 0|A_{0}|\pi \rangle}{\langle 0|P|\pi \rangle} .
\ee
Using the amplitude from the fit to the pseudoscalar density
correlator, together with the value of $Z_{A}$, we obtain an
expression for the pseudoscalar decay constant as
\begin{equation}
  f_P=Z_{A} \sqrt{\frac{2A_{P,P}}{m_P}} A_{A_0,P}.
  \label{eq:fpMeth3}
\end{equation}
Comparing these different extractions allows us to probe
the systematic error due to both mass and amplitude extraction,
and the degree to which the chiral Ward identity is satisfied
after shifting $m_f \rightarrow m_f + m_{\rm res}$.

\subsection{Autocorrelation length}
The integrated autocorrelation
time is defined as~\cite{Allton:1998gi,Autocorr}
\begin{equation}\label{framework:eq:tint}
\tau^{{\rm int}}_{A}=\frac{1}{2}\sum^{\infty}_{-\infty} \rho_{A}(t) = \frac{1}{2} + \sum^{\infty}_{t=1}\rho_{A}(t)
\end{equation}
where the autocovariance function, $\rho_A(t)$ defines the exponential autocorrelation time $t_{exp}$
\begin{equation}
\rho_{A}(t)=\frac{\Gamma_{A}(t)}{\Gamma_{A}(0)}, \quad \rho_{A}(t)\stackrel{t \rightarrow \infty}{\rightarrow} e^{-t/\tau_{{\rm exp}}}
\end{equation}
and the autocovariance of an observable $A$ is
\begin{equation}\label{framework:eq:autocov}
\Gamma_{A}(t) = \langle(A_{s}-\langle A\rangle)(A_{s+t}-\langle A\rangle)\rangle.
\end{equation}
The subscripts $t$ and $s$ label the Monte Carlo time and the outer
average is over all pairs separated by $t$. 
It is standard to consider configurations separated by $2\tau_{A}^{{\rm int}}$
to be statistically independent.

In practice, we truncate the sum in Eq.~(\ref{framework:eq:tint}) at
some finite value, $t_{{\rm max}}$, and define the cumulative
autocorrelation time to be
\begin{equation}
\tau_{A}^{{\rm cum}} = \frac{1}{2} + \sum_{t=1}^{t_{\rm max}}\rho_A(t).
\end{equation}
In the limit of sufficiently large $t_{\rm max}$, this will be a good approximation
to $\tau_A^{\rm int}$. In order to obtain reliable
estimates for $\tau_{{\rm exp}}$ and $\tau^{{\rm cum}}_{A}$, autocorrelations
should ideally be measured using ensembles containing many more
configurations than the value of $\tau^{{\rm int}}_{A}$. Conversely, it must
be noted that the measurement of the autocorrelation time is not
sensitive to correlations over ranges of trajectories that are an
appreciable fraction of the ensemble length, or greater.


\section{Results}
\label{sec:Results}
 
Figures~\ref{framework:plot:tint} and \ref{framework:plot:tint2} shows
example plots for the autocorrelation function and the integrated
autocorrelation length for the plaquette and the pseudoscalar meson
correlator at timeslice 12.  The statistical errors plotted for
$\rho_A(t)$ and $\tau_{{\rm cum}}$ were estimated using a jackknife
procedure. In order to take into account the effects of
autocorrelations in the error estimates for $\rho(t)$ and $\tau_{{\rm
cum}}$ themselves, the original data for $\rho(t)$ and $\tau_{{\rm
cum}}$ were grouped in bins of size $b$. Bin size was increased until
the size of the jackknife errors stabilised to give the error bands
shown on the plots.
The final results can be found in
Table~\ref{framework:tab:autocorrs}, for both the DBW2 $\beta=0.72$
and Iwasaki $\beta=2.13$ RHMC datasets (the other ensembles are much
shorter, and so the results are not given). 

\subsection{Static quark potential}

The results for the values of the parameters $V_0$, $\alpha$, $\sigma$
and $r_0$ in Eq.~(\ref{eq:pot_fit}) and (\ref{eq:r0def}) are given in
Table~\ref{tab:sq_pot_data}\footnote{The results for (D,0.764,
0.02/0.04) and (D, 0.764, 0.04/0.04) are obtained using Chroma code
with same ($c^3_{\rm smear}, N^3_{\rm smear}$) and fit ranges but t=4.}.  These are
obtained by using the fitting range $r \in [\sqrt{3},8]$ and the
choice $t=5$ in Eq.~(\ref{eq:v_r_t}).  The first error in $r_0$ given
in the table is statistical and the second is an estimate of the
systematic error in the fitting procedure.  This estimate of
systematic error is determined from the shift in the central value
when the limits of the fitting range in $r$ are swept through $r_{\rm
min} \in [\sqrt{2},\sqrt{6}]$ and $r_{\rm max} \in [7,9]$ and $t$
is changed from 5 to 6.  In Figure~\ref{fig:plateaux} we show some
example plateaux found for $r_0$ as $t$ is varied between 3 and 6,
suggesting a reasonably stable result and good suppression of excited
states by $t=5$.

In Table~\ref{tab:sq_pot_chiral} we list the result for $r_0$ that we obtain
from a simple linear extrapolation to vanishing light sea quark mass for the 
four ensembles where we have two or more values for the light sea quark mass.
Figure~\ref{fig:0.72_chiral} compares
two extrapolations for the DBW2 $\beta=0.72$ ensembles where we use either the
two lightest mass values, $m_l = 0.01$ and 0.02 or all three 
$m_l = 0.01$, 0.02, and 0.04. 

\subsection{Topological Charge}

The time histories of the plaquette $\langle \bar{q} q \rangle$,
$\langle \bar{q} \gamma_5 q \rangle$ and the topological charge, are
plotted in Figure~\ref{fig:dbw2_0.72_01_04_evol} for the (D, 0.72,
0.01/0.04) ensemble.  Measured values are plotted every 10
trajectories (5 molecular dynamics time units).  This ensemble has the
smallest values for $m_l$, and these time histories make our choice of
500 time units for thermalization appear reasonable.

The topological charge in the bottom panel of Figure~\ref{fig:dbw2_0.72_01_04_evol}
was measured using gluonic method 1, as described in Section \ref{sec:TopologicalCharge}.
We can compare this with a fermion based definition of topology using $\langle \bar{q}
\gamma_5 q \rangle $.  To make this comparison, Figure
\ref{fig:dbw2_0.72_01_04_pbg5p_evol} gives the evolution of a smoothed
version of $\langle \bar{q} \gamma_5 q \rangle $ for the (D, 0.72,
0.01/0.04) ensemble, with smoothing windows of size 25, 50, 100 and
200 time units. Comparing these evolutions with the unsmoothed $ \langle \bar{q}
\gamma_5 q \rangle $ evolution in the third panel of 
Figure~\ref{fig:dbw2_0.72_01_04_evol} reveals noise is substantially
reduced by smoothing.  The relevance of the resulting signal can be
seen by comparing with the evolution of the topology as measured from
the gauge field.
Figure~\ref{fig:dbw2_0.72_01_04_tcharge_pbg5p_compare} directly
compares this estimate of the evolution of topological charge with the
evolution of $\langle \bar{q} \gamma_5 q \rangle $, which has been
smoothed with a window size of 50.  The data are strongly correlated
and show good agreement between topology as determined from smeared
gauge links and from the Dirac operator.
Since smoothing the data changes its normalization, we have rescaled
the smoothed data such that its largest value is equal to that of the
largest value of topological charge from the gauge field.

A comparison of 5Li and classically improved methods has been
performed on the (I, 2.13, 0.04/0.04) ensemble.
Figure~\ref{fig:topo_method_history} shows the topological charge
history from the two methods As can be seen they track
each other quite well, although they disagree in places by as much as
$\sim 3$ units of topological charge. It can also be noted that the
5Li method give results which are much better aligned with integers
than the classically improved method.

As a whole, while the topological charge is certainly not completely
decorrelated from configuration to configuration, this set of
ensembles are sampling the topological sectors quite well. 
This can also be seen in the histograms shown in Figure~\ref{fig:dbw20.72_tens}.
 
Figure~\ref{fig:plaq_rect_top_charge_evol} shows the gluonic measure
of the topological charge for the large rectangle simulations at fixed
lattice spacing. Although the topological charge is sampled well
between $-15$ and 15 for the (D, 0.72, 0.04/0.04, R) ensemble (top
panel), as we make $c_1$ more negative, the range of the topological
charge we are able to sample tends to narrow. For the (C7.47, 0.16,
0.04/0.04, R) ensemble (bottom panel), the topological charge
fluctuates only in the range of $\pm 5$, and it takes a large number
of trajectories to move between different topology sectors. The
situation is worse at weaker couplings, where the topological charge
evolution is much slower. This can be seen in
Figure~\ref{fig:plaq_rect_top_charge_weak_coupling}, where the top three panels have a
rectangle coefficient of $c_1 = -3.57$ and the bottom one has $c_1 = -2.3$.  In the
absence of a dramatic improvement in the chiral properties of domain wall fermions,
this is a compelling reason to avoid larger rectangle coefficients. As
should be expected, these trends with rectangle coefficient and
coupling also apply to the DBW2 and Iwasaki actions. Figure~\ref{fig:top_action_comp}
shows -- from top to bottom -- representative topological charge
histories for the DBW2 action with $\beta=0.764$ and $\beta=0.78$, and
the Iwasaki action with $\beta=2.13$ and $\beta=2.2$.

\subsection{Hadron masses}
\label{sec:Hmass}

Hadronic correlation functions were constructed from quark propagators
which were smeared at the source and local at the sink, as described
in Section~\ref{sec:hadrons}. We use the following notation to
describe the combinations of smearing: ${\rm S}_1 {\rm F}_1 - {\rm S}_2 {\rm F}_2$, where ${\rm S}_i$
denotes source and ${\rm F}_i$ sink for the quark propagators $i=\{1,2\}$ in
the meson correlator. In particular, four combinations were computed:
LL-LL, SL-LL, SL-SL, and WL-WL.  The first combination has the
attractive property that, as both meson interpolating operators are the
same, the ground state and all the excited state contributions to the
correlation function are positive. The effective mass,
\be
  M^{\rm eff}(t) = \log{\left[\frac{C(t)}{C(t+1)}\right] },
\ee
which is used to determine when the ground state is dominating the
correlation function, then approaches a plateau from above and can be
used to determine unambiguously a fitting range for a single
exponential function. Smeared and wall sources have the advantage that
the plateau, and hence the fitting range that can be used, starts earlier.

Figure \ref{fig:b0.72_rhmc_01_04_mpi_src_compare_pp} shows the
pseudoscalar effective mass for four different source/sink
combinations as determined from the pseudoscalar-pseudoscalar
correlator for the (D, 0.72, 0.01/0.04) ensemble, while Figure
\ref{fig:b0.72_rhmc_01_04_mrho_src_compare} shows a similar plot for
the vector meson effective mass. In both cases, the mesons are constructed
from degenerate mass quarks with $m_{\rm val}=0.01$.  These plots use
a radius of 3.0 for the smearing function ($R$ in
Eq.(~\ref{tech:smearing_rad})), which was found to be
approximately optimal for both the pseudoscalar meson and the vector
meson. This can be seen from Figure~\ref{ch6:plot:smearing}, which
shows a comparison of the vector meson effective mass for various
different smearing radii on the (I, 2.2, 0.02/0.04) ensemble with a valence
quark mass of 0.04.
When multiple sources are available, we have also extracted the mass
by simultaneously fitting a pair of correlators with different sources to
the ground and first excited state. The advantage of this approach is
that the systematic uncertainty in the ground state mass arising from the choice of
fit range is reduced. In summary, with careful choices of fitting ranges
all the sources used in this work give consistent results.

The extracted pseudoscalar meson masses for the cases where valence and sea quark masses
are equal, {\it i.e.} unitary data, are given in Table~\ref{ch6:tab:ps} for all of the
RHMC ensembles. For the DBW2 gauge action at $\beta=0.72$ we also have data for which the valence quarks are 
not degenerate with the sea quarks. These results are given in Table~\ref{dbw2:llll_ps}. 
The corresponding results for the vector meson masses are given in Tables \ref{ch6:tab:rho} and
\ref{dbw2:llll_vec} respectively.

As with the construction of the meson operators, smeared quark fields
are used to construct improved nucleon interpolating fields. In
particular, all the data presented were extracted using correlators
in which one interpolating operator is constructed from local quark
fields, while the other interpolating operator is either constructed
from smeared quark fields (denoted SL-SL-SL), or gauge-fixed
wall sources (denoted WL-WL-WL).  Figure~\ref{ch6:fig:nuc} shows a
typical effective mass plot for the nucleon. The backwards
propagating, negative parity state in the standard nucleon correlator
has been reflected about the middle of the time axis.  Both states
are shown together in Figure~\ref{ch6:fig:nuc}, where solid symbols
represent data obtained from the WL-WL-WL correlators while open
symbols correspond to the SL-SL-SL correlators.  Fits to either type
of smearing produce the same nucleon mass within the measured
uncertainties.

The mass of the negative-parity nucleon is less well determined due to
the poor signal and has some dependence on the type of smearing
chosen. Moreover, the physical volume of these lattices is likely to
be too small to extract accurately orbitally excited baryon states
such as the $N^{\star}$. However, the data presented can be used in
combination with larger volume runs to estimate the size of
the finite volume effects on the remaining spectrum. The extracted
values of the nucleon mass are collected in Table~\ref{ch6:tab:nucs}.
The three fit ranges given are for the forwards propagating state
$(N_{+})$, the backwards propagating state $(N_{-})$, and the negative
parity correlator $(N'_{-})$. 

\subsection{The residual mass}

The ratio $R(t)$ given in Eq.~(\ref{eq:mres_ratio}) is shown in
Figure~\ref{fig:b0.72_rhmc_04_rt_wl} for the ensemble (D, 0.72 0.01/0.04).
Since the correlator is time symmetric, the data shown is the average
of the two halves of the lattice. This ratio should be constant as
long as the separation in time is large enough that the effects
of non-physical states/non-locality are eliminated. For the LL-LL
correlator this is from $t \ge 7$ in the figure, but it is earlier for the
WL-WL correlator as the overlap of the smeared operator with the
physical states is larger in this case.  As discussed above in Section
\ref{subsec:mres_def}, we refer to this average of $R(t)$ as
$\mres^\prime(m_f)$, to distinguish it from the constant $\mres$ which
appears in the Lagrangian of Eq.~(\ref{eq:syman_eff}).  The values of
$\mres^\prime(m_f)$ from the unitary data are given in
Table~\ref{tab:mres} and from the non-degenerate data in
Table~\ref{tab:RHMC_0_72_rhmc_rt}.

Table~\ref{tab:plaq_rect_rho_mres} contains the results for
$\mres^\prime(m_f)$ and the vector meson mass as the magnitude of the rectangle
coefficient $c_1$ is increased. In this case $\mres^\prime(m_f)$ has
been estimated by averaging $R(t)$ from $t = 4$ to $t=16$. The
decrease in $\mres$ as $|c_1|$ increases is expected, since the
non-locality introduced by the large rectangle term suppresses the
fluctuations of the gauge fields on the scale of two lattice spacings, 
and the number of low-lying localized modes is moderately
reduced~\cite{Antonio:2005wm}. If a small residual mass were the only
consideration for our parameter choices, our goal could be
accomplished by increasing the magnitude of the rectangle coefficient.
However, we would forfeit our ability to tunnel between
different topological sectors, as shown in the previous section. It is
for this reason that we concentrated on the Iwasaki and DBW2 actions
in our more extensive action studies.

\subsection{$Z_A$ and the pseudoscalar decay constant}
\label{sec:ZAfP}

The value of $Z_A$ is determined from Eq.~(\ref{eq:ZA_def}), using
either LL-LL or SL-SL correlators. In the case when smearing is
used, this corresponds to using a smeared-smeared pseudoscalar 
interpolating field in Eq.~(\ref{eq:eq:Ct_Lt}). Since the axial
current operator is unchanged, this still corresponds to an extraction
of the renormalization factor for the local axial current, $A_\mu^a$. 
Figure~\ref{ch6:fig:ZA} shows a comparison of two such extractions
for the (D, 0.72, 0.02/0.04) ensemble, with good agreement being found.
The values of $Z_A$ for the unitary data are given in Table~\ref{ch6:tab:ZA}.
For the (D, 0.72, 0.01/0.04) ensemble we also have some partially quenched
data, extracted using LL-LL correlators. The values are given in
Table~\ref{dbw2:llll_za}. 

The value of the pseudoscalar decay constant has been determined in
three ways according to Eq.~(\ref{eq:fpMeth1}), (\ref{eq:fpMeth2}) and
(\ref{eq:fpMeth3}). These results are presented for both the unitary and non-unitary data in 
Tables~\ref{ch6:tab:fpi2}-\ref{ch6:tab:fpi3}. Figure~\ref{ch6:fig:fpi3} shows
an example of the third method (Eq.~(\ref{eq:fpMeth3})) for the DBW2 $\beta=0.72$
ensembles.

\subsection{Chiral extrapolations and results for phenomenological quantities}

To estimate values of phenomenologically relevant quantities requires
extrapolating our data to the physical $u$ and $d$ quark masses. For the DBW2 $\beta=0.72$
ensembles we have three sea-quark mass values, as well as non-degenerate valence quark
masses, enabling us to explore the extent to which our data agrees with the predictions of
leading order (LO) or next-to-leading order (NLO) (partially quenched) chiral perturbation theory (PQ$\chi$PT).
Preliminary work on this data can be found in~\cite{Lin:2005gh}.
For the remaining DBW2 and Iwasaki ensembles, we only have two sea-quark mass values and so
only linear extrapolations are possible. For the larger rectangle ensembles, only one
sea-quark mass was generated, so valence only extrapolations are performed.

For each of these extrapolations we must include the effects of the
residual quark mass which should be treated as an addition to the
explicit quark mass $m_f$.  Thus, the chiral limit should be defined
as that value of $m_f$ for which $m_f +\mres=0$.  Our values for the residual
mass, $\mres^\prime(m_f)$, determined from the midpoint term in the
axial current divergence, themselves depend on $m_f$, a lattice
artefact.  We treat this $m_f$ dependence in two different ways where
we have three sea-quark masses. Firstly we extrapolate linearly in
$m_f$ to define $\mres = \mres^\prime(m_f=0)$ for the unitary data,
giving $\mres=0.0106(1)$. This is shown in
Figure~\ref{DBW2:mres}. Secondly we construct the quark mass 
\be 
  m_q = m_f + \mres^\prime(m_f) 
\ee 
and extrapolate hadronic quantities versus
$m_q$ to $m_q=0$.  For data where we have two sea-quark masses we
follow the second procedure.  These two procedures adopt slightly
different choices for the residual mass: $\mres= \mres^\prime(m_f=0)$
and $\mres= \mres^\prime(m_f=-\mres)$.  The difference between these
two approaches is at most a few percent and reflects a non-zero lattice
spacing error that we do not attempt to control in this paper.

Concentrating on the DBW2 $\beta=0.72$ ensembles,
Figure~\ref{DBW2:psmass_sqr} plots the square of the pseudoscalar
meson mass from the unitary data versus $m_f$. Lowest order chiral perturbation theory would suggest that
this should be linear in the mass, and extrapolate to zero in the
$m_f=-m_{\rm res}$ limit. This is not exactly obeyed, $m_P^2$
extrapolates to $-0.0144(24)$ at $m_f=-m_{\rm res}$. This is due to
the relatively small size of the fifth dimension. Conversely,
vanishing pseudoscalar meson mass occurs at a value of
$m_f=-0.0078(5)$. Whilst this represents a relatively large
discrepancy with the definitions of zero quark mass used above, and is
much bigger than the discrepancy in the definitions of $\mres$,
it is itself a small effect when compared to the statistical
uncertainties of other hadron masses and matrix elements extrapolated
to the chiral limit. Ultimately, the uncertainty in the value of
$\mres$ is not the dominant source of uncertainty in the value of any
results presented here.

At NLO in PQ$\chi$PT the quark mass dependence of the pseudoscalar
meson mass and decay constant has both a linear and non-linear
component. For the lightest two DBW2 $\beta=0.72$ ensembles we have
generated non-degenerate data for these two quantities, detailed in
sections~\ref{sec:Hmass} and~\ref{sec:ZAfP}. Consequently, we can
attempt fits to NLO partially quenched chiral formula to this data. A
preliminary study of these fits was first reported
in~\cite{Lin:2005gh} where $m_P$ and $f_P$ were fitted to the chiral
formula independently. Here we update the results by performing
combined fits to both $m_P$ and $f_P$. While we observed consistency
between PQ$\chi$PT and our data with valence quark masses as heavy as
0.03 in the independent fits of ~\cite{Lin:2005gh}, this is not the
case for the combined fits, shown in Figure~\ref{fig:simultNLO_mpi}
and \ref{fig:simultNLO_fpi}. The fitted curves miss the data points
badly at large quark masses. This is likely due to the fact that the
quark masses are so heavy that the next-to-next-to-leading order
contribution becomes important.
Given the limited statistics, large residual mass, relatively small
volume and coarse lattice spacing, our study on the chiral fits is far
from conclusive. Further investigations on these lattice artifacts are
in progress~\cite{Lin:2006cf}.

For all the RHMC datasets, we extrapolate linearly in 
quark mass ($m_q$), the unitary data for the quantity $m_X$ as follows
\be
  m_X(m_q) = A_X + B_X m_q \ ,
\ee
where $m_X=\{ m_P^2, m_V, m_N, m_{N^\star}, f_P\}$.  Examples of these
extrapolations for various datasets are shown in 
figures~\ref{fig:IWmPchiral}-\ref{fig:DBW2fPchiral}.

To predict physical quantities our strategy is as follows: The $u$ and
$d$ quark masses are set to zero, and the lattice spacing is
determined, either from $r_0$ or by setting the vector meson mass in this
limit to be equal to that of the physical rho meson.  The $s$ quark
mass is then set from the physical kaon mass in lattice units and the
pseudoscalar meson mass made from two quarks each with a mass of half
the strange quark mass. Predictions for the values of the
following quantities can then be made: 
$\{m_{K^\star}, m_N, m_{N^\star}, f_\pi, f_K \}$. Preliminary results
were first presented in~\cite{Antonio:2005jm,Antonio:2005yh}.

The results for the value of the vector meson in the chiral limit are
listed in Table~\ref{tab:chirho}. The lattice spacing can then be
determined from the physical rho meson mass. Whilst not ideal, as the rho
meson can decay in nature and has a large width, and because we should
extrapolate our data to the $u$ and $d$ quark masses, not the chiral
limit, this method should be accurate enough for our purposes. The
values of the lattice spacing determined in this way are also
displayed in Table~\ref{tab:chirho}. One can compare these values for
the lattice spacings with those obtained from the potential by
setting $r_0=0.5$~fm, in Table~\ref{tab:sq_pot_chiral}. There is crude
agreement, but the lattice spacing from the potential is
systematically finer.  Alternatively, we can estimate the value of
$r_0$ from the potential and the lattice spacing determined from vector meson mass.
This is also displayed in
Table~\ref{tab:chirho}, where the errors quoted have been obtained by
adding the errors for $r_0/a$ and $a^{-1}$ from the vector in
quadrature. A crude estimate for the value of $r_0$ in the continuum
can be made by an average, weighted by the square of the errors:
\be
  r_0=0.554(21)\ {\rm fm}.
\ee
This is equivalent to a constant continuum extrapolation and
ignores the different systematic uncertainties for each ensemble,
as well as the issues described above for extrapolation
of the vector meson mass, and so should be only considered as a
qualitative, rather than quantitative result. The determinations
of the lattice spacing from the potential and the vector meson
mass are consistent, given the uncertainty
in the physical value of $r_0$. In general, despite its problematic nature,
we used the vector meson mass to set the scale, 
where physical units were required and $r_0$ for scaling
analyses as the value of $r_0/a$ is properly controlled 
for each ensemble.

For the larger rectangle data, where  only valence chiral
extrapolations are possible, we set the lattice spacing from
the vector meson mass. The results are displayed in 
Table~\ref{tab:plaq_rect_chiral_limit}.  Extrapolating $m_V$ with respect to the
dynamical mass instead of the valence mass will decrease the value of
$am_\rho$ by 10\% to 15\%, and hence increase the value of $a^{-1}$ by the
same percentage. However, this is likely to effect all four
datasets by a similar amount, so the valence chiral $m_V$
is good enough to compare the relative lattice spacings. We
see that they are approximately matched, making the comparison
of the residual mass meaningful.

The approximate value for the bare $s$ quark mass is extracted by
requiring that a pseudoscalar meson made up of two quarks of mass
$m_s/2$ has the experimental kaon mass. In doing this, we are
neglecting both next-to-leading order chiral perturbation theory
effects and the masses of the $u$ and $d$ quarks.
Table~\ref{ch6:tab:chips} collects together the intercepts for
the linear extrapolations, plus the extracted strange quark masses.
These chiral extrapolations were performed on the multiple correlator fits tabulated in
Table~\ref{ch6:tab:ps}. As can be seen, the canonical heavy dynamical quark
mass we have been using (0.04) is very close to the value of the
$s$ quark mass extracted from this method.

The values for $K^\star$, the nucleon and its negative
parity partner are shown in Table~\ref{tab:mkstarNucleons}.
The determination of $m_{K^\star}$ can be seen for the DBW2
ensembles in Figure~\ref{fig:DBW2mVchiral}. We examined the 
scaling behaviour of dimensionless ratios versus $(a/r_0)^2$.
For the vector masses, this ratio was $m_{K^\star}/m_\rho$ and
is shown in Figure~\ref{fig:r0kstar}. Given the change in
lattice volumes, the relatively large values of residual mass, 
and only two sea-quark masses, the scaling behaviour appears
reasonable. 

The $J$ parameter~\cite{Lacock:1995tq} is defined as 
\be
J = m_{V} \frac{dm_{V}}{dm_{P}^{2}}
\ee
and is determined at the experimental ratio
\be
\frac{M_{K^{*}}}{M_{K}} = 1.8.
\ee
The linear fitting forms for the chiral extrapolations
of the vector ($m_V$) and pseudoscalar ($m_P$) masses 
implying that a plot of $m_{V}$ versus $m_{P}^{2}$ will be a straight
line for varying quark masses. Here we are working along the `unitary
trajectory' where $m_{\rm{sea}}=m_{\rm{val}}$. Figure~\ref{fig:J}
shows a plot of $m_{V}$ versus $m_{P}^{2}$ for the DBW2
$\beta$=0.72 and Iwasaki $\beta$=2.13 cases and, indeed, for the
$\beta$ =0.72 case, where there are three points, approximate linear
behaviour is seen. The intersection of this line with
$m_{V}=1.8m_{S}$ (starred points) determines the reference value
$m_{V}$ which is to be multiplied by the slope to yield
$J$. Figure~\ref{fig:J-2} shows the value of the $J$
parameter on all the datasets. Within the large errors, agreement with 
the experimental value is observed. 

Figure~\ref{fig:scalenuc}
shows the dependence of the baryon spectrum at the chiral limit, in
dimensionless units, on the lattice spacing. As the size of lattice is
fixed to be $16^3\times 32$, the physical extent of the box in fm is
smaller for the smaller $(a/r_0)^2$ in
Figure~\ref{fig:scalenuc}. The scaling behaviour of the spectrum
appears promising despite the change in volume. A continuum
extrapolation is not attempted. The $N^\star$ mass at the finest
lattice spacings is approximately two sigma smaller than that at the
coarsest lattice spacing, which is consistent with the theoretical
expectation that the $N^\star$ becomes degenerate with the $N$ in a
small enough box \cite{Sasaki:2001nf}. This would also suggest that
finite size effects may be beginning to affect the $N$ for the finest
ensembles. These finite size effects would
tend to increase the mass of the $N$ \cite{Fukugita:1992jj}. With the
current statistical resolution it is not possible to judge whether
this is really the case.

The Edinburgh plot~\cite{Bowler:1985hr} is shown in Figure~\ref{fig:edinburgh}.
This is a useful way of
comparing results for different actions without the need for any
extrapolations of the data. Shown on the graph are the experimental
ratios and the values obtained in the static quark limit, where the
hadron mass is equal to the sum of the valence quark masses. The curve
obtained from the phenomenological model for the hadron masses
described in~\cite{Ono:1977ss} is shown as a guide for the eye. It is
remarkable that, even at relatively coarse lattice spacing, with a
small fifth dimension and moderate chiral symmetry
breaking, the data follows the phenomenological curve, albeit with
relatively large statistical error. This very promising result
suggests that a reliable chiral extrapolation could be
performed with more data.

The values of the pseudoscalar meson decay constant in the chiral limit for
all datasets are displayed in Table~\ref{tab:chifpi}.  Setting this
equal to the physical value of $f_\pi$ provides another estimate of
the lattice spacing. This is also tabulated in
Table~\ref{tab:chifpi}. The different methods for extracting the decay
constant give consistent answers. For the Iwasaki $\beta=2.13$ data
set, the $AA$ correlator gives slightly higher values than the other
methods. However, at weaker coupling all the definitions agree, perhaps
suggesting that this effect is a lattice artefact. 
Figure~\ref{fig:r0fKfpi} shows the scaling behaviour of the dimensionless
ratio $f_K/f_\pi$ from the $PP$ method. This shows excellent scaling,
albeit with large errors and, moreover, is in agreement with the
experimental value.


\section{Conclusions}
\label{sec:Conclusions}
Our criteria to be able to simulate 2+1-flavor QCD on QCDOC using
domain wall fermions are as follows. The topological charge
autocorrelation length should not be significantly greater than one
hundred HMC trajectories. Too infrequent changes in topological
charge, particularly on the small lattices we are using, would signal
poor ergodicity. On the other hand, the greater the presence of
dislocations which drive topological change, the larger the density of
localised near-zero modes which contribute to $m_{\rm res}$. To reach
the chiral regime under theoretical control $\mres$ has to be small
compared to the explicit $u$ and $d$ quark masses. Thus, a
balance has to be struck between increasing the rate of local
topological fluctuations and decreasing $m_{\rm res}$, {\it i.e.} the
level of chiral symmetry breaking. Finally, to control both lattice
artefacts and finite volume effects for light hadron physics, we need
to simulate at lattice spacings in the range 0.09-0.13~fm and spatial
extent of at least 2.5-3~fm.

We find the Iwasaki action satifies all these criteria. While the DBW2
gauge action produces a smaller $\mres$ at a given lattice spacing, we find the
sampling of topological charge, especially at the finer lattice
spacing to be too slow. Increasing the coefficient of the rectangle
further produces only a modest decrease in $\mres$, while further
supressing topology change.

We have presented results for the the static interquark potential,
light meson and baryon masses, and light pseudoscalar decay
contants. We find that, even on our small volumes and small extent in
the 5th dimension, with rather heavy sea-quark masses and crude
approach to chiral extrapolation, both the Iwasaki and DBW2 actions,
reproduce experimental values, albeit with large statistical
uncertainties. The scaling behavior appears promising over the range
of lattice spacings from 0.09-0.13~fm, except for the baryons, where
it appears to be spoiled by finite size effects for the smallest
volume.

We conclude that, using the Iwasaki gauge action, there is a range
of values for the quark masses and lattice spacings such that the chiral regime 
for 2+1 flavor QCD with domain wall fermions on QCDOC is accessible, while maintaining 
control of lattice artefacts and finite volume effects for a range of light 
hadron physics.


\section*{Acknowledgements}

We thank Dong Chen, Calin Cristian, Zhihua Dong, Alan Gara, Andrew
Jackson, Changhoan Kim, Ludmila Levkova, Xiaodong Liao, Guofeng
Liu, Konstantin Petrov and Tilo Wettig for developing with us the QCDOC
machine and its software. This development and the resulting computer
equipment used in this calculation were funded by the U.S.\ DOE grant
DE-FG02-92ER40699, PPARC JIF grant PPA/J/S/1998/00756 and by RIKEN. This work
was supported by DOE grants DE-FG02-92ER40699 and DE-AC02-98CH10886 and PPARC grants
PPA/G/O/2002/00465, PP/D000238/1 and PP/C504386/1. AH is supported by the UK Royal Society.
AS and CJW were supported in part by DOE grant DE-AC02-98CH10886.
We thank BNL, EPCC, RIKEN, and the U.S.\ DOE for
supporting the computing facilities essential for the completion of this work.
We also thank the RIKEN Super Combined Cluster at RIKEN, for the
computer resources used for the static quark potential calculation.
KH thanks RIKEN-BNL Research Center for its hospitality where this work
was performed


\pagebreak
\bibliographystyle{apsrev}
\bibliography{paper}

\pagebreak


\input{tab.tex}

\clearpage
\pagebreak


\input{fig.tex}

\end{document}

%% file: user_def.tex
\usepackage{color}
\usepackage{ulem}
\definecolor{DeepBlue}{rgb}{0,0.0,0.5}
\definecolor{DeepGreen}{rgb}{0,0.5,0}
\definecolor{DeepRed}{rgb}{0.5,0,0}

\newcommand\old[1]{}

\newcommand{\ba}{\begin{eqnarray}}
\newcommand{\ea}{\end{eqnarray}}
\newcommand{\bas}{\begin{eqnarray*}}
\newcommand{\eas}{\end{eqnarray*}}
\newcommand{\be}{\begin{equation}}
\newcommand{\ee}{\end{equation}}
\newcommand{\bes}{\begin{equation*}}
\newcommand{\ees}{\end{equation*}}
\newcommand{\bi}{\begin{itemize}}
\newcommand{\ei}{\end{itemize}}

\newcommand{\nn}{\nonumber}

\newcommand{\bcentre}{\begin{center}}
\newcommand{\ecentre}{\end{center}}
 
\font\tenmsb=msbm10 scaled\magstep1
\font\sevenmsb=msbm7 scaled\magstep1
\font\fivemsb=msbm5 scaled\magstep1
\newfam\msbfam
\textfont\msbfam=\tenmsb
\scriptfont\msbfam=\sevenmsb
\scriptscriptfont\msbfam=\fivemsb

\newcommand{\mres}{m_{\rm res}}

\reversemarginpar   
\makeatletter

\let\adkendequation=\endequation%
\def\endequation{\adkendequation\adklabel\global\@ignoretrue}
\let\adkendeqnarray=\endeqnarray%
\def\endeqnarray{\adkendeqnarray\adklabel\global\@ignoretrue}
\newbox\marglabbox
\def\adklabel{\ifvoid\marglabbox\else\marginpar{\unhbox\marglabbox}\fi}

\def\rmsub#1#2{#1_{\mbox{\tiny #2}}}	    

\def\ln{\mathop{\rm ln}}		    
\def\tr{\mathop{\rm tr}}		    
\def\det{\mathop{\rm det}}		    

\def\su3{SU(3)}

\def\tD{\mbox{D}\kern-0.65em\raise0.15ex\hbox{/}\kern0.15em} 
\def\sD{\mbox{\scriptsize D}\kern-0.5em\raise0.15ex\hbox{\scriptsize/}}
\def\ssD{\mbox{\tiny D}\kern-0.42em\raise0.15ex\hbox{\tiny/}}

\def\dslash{\hbox{\(\partial\)}\kern-0.5em\raise0.15ex\hbox{/}} 

\def\mres{\rmsub{m}{res}}



%% file: tab.tex
\clearpage

\begin{table}
\caption{2+1-flavor ensembles and the mnemonics used to describe them.
A $\star$ denotes that the ensemble was obtained by farming. 
A $\dag$ denotes that the ensemble was farmed from a thermalised $R$ algorithm
dataset. We generate $N_{\rm therm}$ trajectories (of length $\tau=1/2$) and a total of
$N_{\rm traj}$ trajectories for each ensemble.\label{tab:rhmc_ensembles}} 
\begin{center}
\begin{tabular}{lr@{.}lccccccr}
\hline
Action & \multicolumn{2}{c}{$\beta$} & $\frac{m_{l}}{m_{s}}$ & Alg. & $\delta \tau$ &
  Mnemonic & $N_{\rm traj}$ & $N_{\rm therm}$ &\\
\hline
DBW2                 & 0&72 & 0.01/0.04 & RHMC & $\frac{1}{50}$ & (D, 0.72, 0.01/0.04) & 6000 & 1000 \\
DBW2                 & 0&72 & 0.02/0.04 & RHMC & $\frac{1}{54}$ & (D, 0.72, 0.02/0.04) & 6000 & 1000 \\
DBW2                 & 0&72 & 0.04/0.04 & RHMC & $\frac{1}{50}$ & (D, 0.72, 0.04/0.04) & 3395 & 1600 \\
DBW2${}^\star$       & 0&764 & 0.02/0.04 & RHMC & $\frac{1}{70}$ & (D, 0.764, 0.02/0.04) & 2940 & 800 \\
DBW2${}^{\star\dag}$ & 0&764 & 0.04/0.04 & RHMC & $\frac{1}{70}$ & (D 0.764, 0.04/0.04) & 5320 & 100 \\
DBW2                 & 0&78 & 0.02/0.04 & RHMC & $\frac{1}{70}$ & (D, 0.78 , 0.02/0.04) & 1505 & 800 \\
DBW2                 & 0&78 & 0.04/0.04 & RHMC & $\frac{1}{70}$ &  (D, 0.78 , 0.04/0.04) & 1620 & 800 \\
\hline
Iwasaki              & 2&13 & 0.02/0.04 & RHMC & $\frac{1}{50}$ & (I, 2.13 , 0.02/0.04) & 3595 & 1000 \\
Iwasaki              & 2&13 & 0.04/0.04 & RHMC & $\frac{1}{50}$ & (I, 2.13 , 0.04/0.04) & 3595 & 1000 \\
Iwasaki              & 2&2 & 0.02/0.04 & RHMC & $\frac{1}{50}$ & (I, 2.2 , 0.02/0.04) & 5900 & 800 \\
Iwasaki${}^\star$    & 2&2 & 0.04/0.04 & RHMC & $\frac{1}{50}$ & (I, 2.2 , 0.04/0.04) & 5800 & 800 \\
\hline
\end{tabular}
\end{center}
\end{table}

\begin{table}
\caption{2+1-flavor datasets used in our exploration of gauge
actions with different plaquette and rectangle contributions
and the ensemble mnemonics we use
to describe them. The top three datasets have the same lattice spacing, 
and the others are finer. Here MD time refers to Molecular Dynamics evolution time.\label{tab:PR_datasets}}
\begin{center}
\begin{tabular}{r@{.}lr@{.}lcccccc}
\hline
\multicolumn{2}{c}{$c_1$} & \multicolumn{2}{c}{$\beta$} & $\frac{m_l}{m_{s}}$ & Alg. & $\delta \tau$ &
  Mnemonic & MD time & $N_{\rm therm}$ \\
\hline
-2&3  & 0&48 & 0.04/0.04 & R & 0.01 & (C2.3, 0.48, 0.04/0.04, R) &
  800 & 300 \\

-3&57 & 0&32 & 0.04/0.04 & R & 0.01 & (C3.57, 0.32, 0.04/0.04, R) & 800
  & 300 \\

-7&47 & 0&16 & 0.04/0.04 & R & 0.01 & (C7.47, 0.16, 0.04/0.04, R) & 760
  & 300 \\
\hline
-3&57 & 0&333 & 0.04/0.04 & R & 0.01 & (C3.57, 0.33, 0.04/0.04, R) & 760
  & 300 \\
-3&57 & 0&36 & 0.04/0.04 & R & 0.01 & (C3.57, 0.36, 0.04/0.04, R) & 690
  & 300 \\
-2&3  & 0&53 & 0.04/0.04 & R & 0.01 & (C2.3, 0.53, 0.04/0.04, R) &
  700 & 300 \\
\hline
\end{tabular}
\end{center}
\end{table}


\begin{table}
\caption{Parameter values for the 2+1 flavour RHMC algorithm
simulations.
$\lambda_{\rm min}$ and $ \lambda_{\rm max}$ are the maximum
and minimum eigenvalues of ${\cal D}(m_i)$, which are needed for
the rational approximation.  $n_{\rm MD}$ is the degree of the rational
approximation used in the molecular dynamics evolution and $n_{\rm MC}$
is the rational approximation degree used in the
Monte Carlo accept/reject step.  The conjugate gradient stopping
condition for the evolution was $1.0 \times 10^{-6}$ and for
the accept/reject step was $1.0 \times 10^{-10}$.\label{tab:RHMC_parameters}}
\begin{center}
\begin{tabular}{ccccccccccc}
\hline
Ensemble & & \multicolumn{3}{c}{$m_l$} & \multicolumn{3}{c}{$m_s$} & \multicolumn{3}{c}{$m_{PV}$} \\
         & $\lambda_{\rm max}$ & $\lambda_{\rm min}$ & $n_{\rm MD}$ & $n_{\rm MC}$
       			       & $\lambda_{\rm min}$ & $n_{\rm MD}$ & $n_{\rm MC}$
       			       & $\lambda_{\rm min}$ & $n_{\rm MD}$ & $n_{\rm MC}$ \\ \hline
(D, 0.72, 0.01/0.04) &2.4 & $6\times 10^{-5}$ & 11 & 18 & $6\times 10^{-4}$ & 9 & 14 &
                            $4\times 10^{-2}$ & 6 & 9 \\ 
(D, 0.72, 0.02/0.04) &2.4 & $2\times 10^{-4}$ & 10 & 15 & $4\times 10^{-4}$ & 9 & 14 &
                            $3\times 10^{-2}$ & 6 & 9 \\
(D, 0.72, 0.04/0.04) &2.4 & $6\times 10^{-4}$ & 9 & 14 & $6\times 10^{-4}$ & 9 & 14 &
                            $3\times 10^{-2}$ & 6 & 9 \\  \hline
(D, 0.764, 0.02/0.04) &2.42 & $1\times 10^{-4}$ & 10 & 15 & $3\times 10^{-4}$ & 9 & 14 &
                            $2\times 10^{-2}$ & 5 & 8 \\
(D, 0.764, 0.04/0.04) &2.42 & $1\times 10^{-4}$ & 10 & 15 & $3\times 10^{-4}$ & 9 & 14 &
                            $2\times 10^{-2}$ & 5 & 8 \\  \hline
(D, 0.78, 0.02/0.04) &2.42 & $1\times 10^{-4}$ & 10 & 15 & $3\times 10^{-4}$ & 9 & 14 &
                            $2\times 10^{-2}$ & 5 & 8 \\
(D, 0.78, 0.04/0.04) &2.42 & $1\times 10^{-4}$ & 10 & 15 & $3\times 10^{-4}$ & 9 & 14 &
                            $2\times 10^{-2}$ & 5 & 8 \\  \hline
(I, 2.13, 0.02/0.04) &2.4 & $1\times 10^{-4}$ & 10 & 15 & $3\times 10^{-4}$ & 9 & 14 &
                            $3\times 10^{-2}$ & 5 & 8 \\
(I, 2.13, 0.04/0.04) &2.4 & $1\times 10^{-4}$ & 10 & 15 & $3\times 10^{-4}$ & 9 & 14 &
                            $3\times 10^{-2}$ & 5 & 8 \\  \hline
(I, 2.2, 0.02/0.04) &2.4 & $2\times 10^{-4}$ & 10 & 15 & $4\times 10^{-4}$ & 9 & 14 &
                            $3\times 10^{-2}$ & 5 & 8 \\
(I, 2.2, 0.04/0.04) &2.4 & $2\times 10^{-4}$ & 10 & 15 & $4\times 10^{-4}$ & 9 & 14 &
                            $3\times 10^{-2}$ & 5 & 8 \\  \hline
\end{tabular}
\end{center}
\end{table}


\begin{table}
\caption{Statistics for 2+1 flavour DBW2 ensembles with $\beta = 0.72$
generated with the RHMC algorithm.\label{tab:RHMC_0_72_evol_stat}}
\begin{center}
\begin{tabular}{cccccr@{(}lr@{.}l}
\hline
Ensemble & $< \delta H > $ & $< e^{-\delta H}>$ & Accept 
& $\overline{P_{\mu\nu}}$ & \multicolumn{2}{c}{$\langle \bar{q} q \rangle$} & 
\multicolumn{2}{c}{$\langle \bar{q} \gamma_5 q  \rangle \times 10^{-5}$ }  \\
\hline
(D, 0.72, 0.01/0.04) & 0.388(14) & 0.991(14) & 0.658(6) &
 0.608201(21) & 0.00315&1) &   4&7(145) \\
(D, 0.72, 0.02/0.04) & 0.311(13) & 1.000(15) & 0.693(5) &
0.608094(18) & 0.00422&9) &  9&6(7.9)  \\
(D, 0.72, 0.04/0.04) & 0.398(16) & 0.992(15) & 0.651(9)  &
0.607788(10) & 0.00612&11) &  -0&6(8.4)  \\
\hline
\end{tabular}
\end{center}
\end{table}

\begin{table}
\begin{center}
\caption{Estimates of autocorrelation times for the average plaquette and the pseudoscalar meson on timeslice 12. 
Note plaquettes are separated by 1 trajectory while pseudoscalar meson correlators are measured
on $N_{\rm meas}$ configurations each separated by 5 trajectories. }
\label{framework:tab:autocorrs}
\begin{tabular}{ccccccc}
\hline
Dataset&\multicolumn{2}{c}{Ensemble size}&
\multicolumn{2}{c}{$\langle P_{\mu,\nu}\rangle$} & \multicolumn{2}{c}{$C_P(t=12)$}\\
\hline
($\beta : \frac{m_l}{m_s} $)& $N_{\rm traj}$ & $N_{\rm meas}$ & $\tau^{\rm cum}$ & $\tau^{\rm exp}$&$\tau^{\rm cum}$ & $\tau^{\rm exp}$\\
\hline
(D, 0.72 , 0.01/0.04) & 6000 & 1000  & $20(10)$ & $>4$ & $13(10)$ & $>8$\\ 
(D, 0.72 , 0.02/0.04) & 6000 & 1000 & $10(5)$ & $>5$ & $15(10)$ & $>12$\\ 
(I, 2.13 , 0.02/0.04) & 3595 & 520  & $6(3)$ & $>6$ & $5(3)$ & $>5$\\ 
(I, 2.13 , 0.04/0.04) & 3595 & 520  & $6(4)$ & $>4$ & $7(3)$ & $>9$\\ 
\hline
\end{tabular}
\end{center}
\end{table}

\begin{table}
\begin{center}
\caption{Results for the coefficients appearing in the fit to the static quark potential 
given in Eq.~(\ref{eq:pot_fit}) and the implied values of $r_0$ and $1/a$.  In computing the 
latter, the value $r_0=0.5$fm is used.  The configurations used to obtain the results in 
this table are separated by 5 Monte Carlo steps.\label{tab:sq_pot_data}}
\begin{tabular}{ccr@{.}lr@{.}lr@{.}lr@{.}lr@{.}l}
\hline
  Ensemble Mnemonic          & $N_{\rm meas}$   & \multicolumn{2}{c}{$V_0$} & \multicolumn{2}{c}{$\alpha$} &
 \multicolumn{2}{c}{$\sigma$} & \multicolumn{2}{c}{$r_0$}&  \multicolumn{2}{c}{$1/a$ GeV} \\ \hline
(D, 0.72, 0.01/0.04)          &900 &0&880(10) &0&466(13) &0&0754(19) &3&962(30)(16) 
&1&564(12)(6)  \\
(D, 0.72, 0.02/0.04)         &800 &0&853(11) &0&434(13) &0&0817(21) &3&858(32)(75)  
&1&523(13)(30) \\
(D, 0.72, 0.04/0.04)         &280 &0&783(17) &0&365(22) &0&1018(32) &3&554(31)(67)  
&1&403(12)(27) \\
\hline
(D, 0.764, 0.02/0.04)        & 294 & 0&816(20) & 0&417(27) & 0&0571(33)&  4&646(84)(21)&  1&833(33)(08) \\
(D, 0.764, 0.04/0.04)        & 300 & 0&793(19) &0&398(26)  &0&0648(31)  &4&397(60)(18)&  1&735(24)(07) \\
\hline
(D, 0.78 , 0.02/0.04)        & 180 &0&791(9)  &0&412(11) &0&0524(16) &4&863(54)(142) 
&1&929(21)(56) \\
(D, 0.78 , 0.04/0.04)        & 165 &0&761(10) &0&378(13) &0&0601(18) &4&600(49)(16)  
&1&815(19)(6)  \\
\hline
(I, 2.13 , 0.02/0.04)        &420 &0&834(10) &0&406(13) &0&0668(18) &4&315(41)(97)  
&1&703(16)(38) \\
(I, 2.13 , 0.04/0.04)        & 240 &0&792(14) &0&350(18) &0&0770(25) &4&110(41)(21)  
&1&622(16)(8)  \\
(I, 2.2 , 0.02/0.04)         &280 &0&804(7)  &0&393(9)  &0&0506(12) &4&982(45)(30)  
&1&966(18)(23) \\
(I, 2.2 , 0.04/0.04)         &320 &0&781(6)  &0&365(9)  &0&0560(11) &4&788(35)(30)  
&1&890(14)(12) \\
(I, 2.3 , 0.04/0.04)         &165 &0&741(5)  &0&348(8)  &0&0397(9)  &5&729(55)(137) 
&2&261(22)(54) \\
\hline
(C2.3, 0.48 , 0.04/0.04, R)  & 180 &0&814(16) &0&413(19) &0&0773(32) &4&002(57)(60)  
&1&579(22)(24) \\
(C2.3, 0.53 , 0.04/0.04, R)  & 160 &0&739(6)  &0&384(8)  &0&0492(12) &5&074(48)(48)  
&2&002(19)(19) \\
(C3.57, 0.32 , 0.04/0.04, R) & 180 &0&828(15) &0&433(18) &0&0773(28) &3&968(49)(43)  
&1&566(19)(17) \\
(C3.57, 0.33 , 0.04/0.04, R) & 145 &0&792(12) &0&416(15) &0&0657(21) &4&335(47)(102) 
&1&711(19)(40) \\
(C3.57, 0.36 , 0.04/0.04, R) & 135 &0&719(9)  &0&371(12) &0&0500(17) &5&060(66)(33)  
&1&997(26)(13) \\
(C7.47, 0.16 , 0.04/0.04, R) & 165 &0&832(15) &0&448(19) &0&0733(30) &4&050(57)(430)
&1&598(22)(17) \\
\hline
\end{tabular}
\end{center}
\end{table}

\begin{table}
\begin{center}
\caption{Results for the lattice spacing extrapolated to the chiral limit
for the two light quarks, $m_f = - m_{\rm res}$, while the mass of the third 
(strange) quark is held fixed at $m_f=0.04$.  The masses column lists those
sea quark masses used in the extrapolation.  \label{tab:sq_pot_chiral}} 
\begin{tabular}{lr@{.}lccc}
\hline
  Action & \multicolumn{2}{c}{$\beta$} & masses           & $r_0$ &  $1/a$ [GeV] \\
\hline
DBW2     & 0&72    & 0.01, 0.02, 0.04 & 4.260(52)(12)   & 1.681(20)(5)   \\
DBW2     & 0&72    & 0.01, 0.02       & 4.177(115)(105) & 1.648(41)(45)  \\
DBW2     & 0&764   & 0.02, 0.04       & 5.030(230)(040) & 1.983(92)(16)  \\
DBW2     & 0&78    & 0.02, 0.04       & 5.184(134)(335) & 2.046(53)(132) \\
Iwasaki  & 2&13    & 0.02, 0.04       & 4.628(121)(239) & 1.826(48)(94)  \\
Iwasaki  & 2&20    & 0.02, 0.04       & 5.239(114)(111) & 2.068(44)(45)  \\
\hline
\end{tabular}
\end{center}
\end{table}


\begin{table}
\caption{Results for pseudoscalar meson masses from fits to unitary RHMC data.
 \label{ch6:tab:ps}}
\begin{center}
\begin{tabular}{lccccr}
\hline
 Ensemble Mnemonic & Fit range &\multicolumn{4}{c}{Results}\\
\hline
& $t_{\rm min}-t_{\rm max}$ &$m_{\rm{val}}$ & $m_P$
& $\chi^{2}/dof$ & correlators \\ 
\hline
(D, 0.72 , 0.1/0.04)   & 4-14  & 0.01 &$0.303(3)$  & $66/16$ & SL-SL, SS-SS\\
(D, 0.72 , 0.02/0.04)  & 5-16  & 0.02 &$0.3742(9)$ & $34/18$& LL-LL, WL-WL\\
(D, 0.72 , 0.04/0.04)  & 5-16  & 0.04 &$0.4916(10)$& $50/18$ & LL-LL, WL-WL\\ 
(D, 0.764 , 0.02/0.04) & 7-15  & 0.02 &$0.311(1)$  & $20/12$ & LL-LL, WL-WL\\
(D, 0.764 , 0.04/0.04) & 6-16  & 0.04 &$0.4203(7)$ & $23/16$ & LL-LL, WL-WL\\
(D, 0.78 , 0.02/0.04)  & 11-16 & 0.02 &$0.288(6)$  & $5/4$   & WL-WL \\
(D, 0.78 , 0.04/0.04)  & 11-16 & 0.04 &$0.400(7)$  & $8/4$   & LL-LL \\
\hline
(I, 2.13 , 0.02/0.04)  & 10-16 & 0.02 &$0.362(2)$  & $10/8$  & LL-LL, SL-SL\\
(I, 2.13 , 0.04/0.04)  & 9-16  & 0.04 &$0.4665(9)$ & $14/10$ & SL-SL, SS-SS\\
(I, 2.2 , 0.02/0.04)   & 5-16  & 0.02 &$0.315(2)$  & $29/18$ & LL-LL, WL-WL\\
(I, 2.2 , 0.04/0.04)   & 6-16  & 0.04 &$0.425(1)$  & $21/16$ & LL-LL, WL-WL\\
\hline
\end{tabular}
\end{center}
\end{table}

\begin{table}
\caption{Results for pseudoscalar meson masses from fits to non-degenerate LL-LL
   DBW2 $\beta=0.72$ RHMC data.  \label{dbw2:llll_ps}}
\begin{center}
\begin{tabular}{lcccc}
\hline
$m_{\rm val}$ & Fit range &\multicolumn{3}{c}{$m_l$} \\ \hline
              &  $t_{\rm min}-t_{\rm max}$ & 0.01 & 0.02 & 0.04 \\ \hline

0.005 & 8-16  & $0.2619(26)$ & $0.2769(30)$ &    \\
0.01 & 8-16  &  & $0.3125(27)$ &    \\
0.015 & 8-16  & $0.3321(23)$ & $0.3450(25)$ &    \\
0.02 & 8-16  & $0.3627(22)$ &  & $0.3882(17)$ \\
0.025 & 8-16  & $0.3914(21)$ & $0.4031(22)$ &    \\
0.03 & 8-16  & $0.4186(21)$ & $0.4298(21)$ & $0.4419(16)$ \\
0.035 & 8-16  & $0.4445(20)$ & $0.4552(20)$ &    \\
0.04 & 8-16  & $0.4693(20)$ & $0.4795(19)$ &  \\
\hline
\end{tabular}
\end{center}
\end{table}

\begin{table}
\caption{Results for the vector meson masses from fits to unitary RHMC data.
\label{ch6:tab:rho}}
\begin{center}
\begin{tabular}{lcccr}
\hline
Ensemble Mnemonic & Fit range &\multicolumn{3}{r}{Results}\\
\hline
& $t_{\rm min}-t_{\rm max}$ &  $m_V$ & $\chi^{2}/dof$& correlators\\ 
\hline
(D, 0.72 , 0.01/0.04)  & 6-14 & $0.580(10)$  & $13/12$ &LL-LL,SL-SL\\
(D, 0.72 , 0.02/0.04)  & 6-16 & $0.635(4)$ & $27/16$ &LL-LL,WL-WL\\
(D, 0.72 , 0.04/0.04)  & 6-13 & $0.703(5)$ & $11/10$ &LL-LL,WL-WL\\
(D, 0.764 , 0.02/0.04) & 8-14 & $0.543(5)$ & $13/8$ &LL-LL,WL-WL\\
(D, 0.764 , 0.04/0.04) & 7-15 & $0.607(3)$ & $21/12$ &LL-LL,WL-WL\\
(D, 0.78 , 0.02/0.04)  & 10-15& $0.48(3)$  & $2/4$   & WL-WL\\
(D, 0.78 , 0.04/0.04)  & 8-15 & $0.575(5)$ & $24/13$ & LL-LL,WL-WL\\
\hline
(I, 2.13 , 0.02/0.04)  & 5-14 & $0.581(4)$ & $14/8$ & SS-SS\\
(I, 2.13 , 0.04/0.04)  & 6-14 & $0.661(3)$ & $21/12$ & SS-SS,SL-SL\\
(I, 2.2 , 0.02/0.04)   & 7-16 & $0.493(6)$ & $11/14$ & LL-LL,WL-WL\\
(I, 2.2 , 0.04/0.04)   & 7-15 & $0.586(5)$ & $19/12$ & LL-LL,WL-WL\\
\hline
\end{tabular}
\end{center}
\end{table}

\begin{table}
\caption{Results for the vector meson masses from fits to non-degenerate LL-LL 
   DBW2 $\beta=0.72$ RHMC data. 
\label{dbw2:llll_vec}}
\begin{center}
\begin{tabular}{lcccc}
\hline
$m_{\rm val}$ & Fit range &\multicolumn{3}{c}{$m_l$} \\ \hline
              &  $t_{\rm min}-t_{\rm max}$ & 0.01 & 0.02 & 0.04 \\ \hline

0.005 & 8-16  & $0.574(22)$ & $0.586(18)$ &    \\
0.01 & 8-16  &  & $0.596(15)$ &    \\
0.015 & 8-16  & $0.595(13)$ & $0.610(13)$ &    \\
0.02 & 8-16  & $0.610(11)$ &  & $0.648(12)$ \\
0.025 & 8-16  & $0.6258(90)$ & $0.6405(97)$ &    \\
0.03 & 8-16  & $0.6419(78)$ & $0.6564(87)$ & $0.6744(94)$ \\
0.035 & 8-16  & $0.6579(69)$ & $0.6723(78)$ &    \\
0.04 & 8-16  & $0.6740(62)$ & $0.6880(71)$ &  \\
\hline
\end{tabular}
\end{center}
\end{table}

\begin{table}
\caption{Results for the nucleon and negative parity partner from fits to the unitary RHMC data. \label{ch6:tab:nucs}}
\begin{center}
\begin{tabular}{lcccccc}
\hline
Dataset & \multicolumn{3}{c}{$t_{\rm min}-t_{\rm max}$} &\multicolumn{3}{c}{Results}\\
\hline
Ensemble & $N$ from $\Omega_{B_1}$ & 
$N^\star$ from $\Omega_{B_1}$ & 
$N^\star$ from $\Omega_{B_2}$ & 
$m_{N}$ & $m_{N^\star}$ & $\chi^{2}/dof$ \\ 
\hline
\hline
(D, 0.72 , 0.02/0.04 )  & 10-16 & 20-27 & 6-11 &$0.904(8)$ 
& $1.18(2)$ & $26/16$ \\
(D, 0.72 , 0.04/0.04 )  &  8-14 & 23-25 & 4-8 &$1.021(4)$ 
& $1.28(2)$ & $13/10$ \\
(D, 0.764 , 0.02/0.04 ) & 11-15 & 20-23 & 6-10 & $0.76(2)$ 
& $0.96(2)$ & $16/9$ \\
(D, 0.764 , 0.04/0.04 ) & 10-16 & 20-23 & 7-12 &$0.888(3)$ 
& $1.17(1)$ & $23/12$ \\
\hline
( I, 2.13 , 0.02/0.04 ) & 9-14 & 21-25 & 7-9 &$0.82(1)$    
& $1.14(2)$ & $11/9$ \\
( I, 2.13 , 0.04/0.04 ) & 8-12 & 24-25 & 6-11 &$0.984(5)$ 
& $1.28(2)$ & $8/8$  \\
( I, 2.2 , 0.02/0.04 ) & 8-16 & 21-25 & 6-10 &$0.729(3)$ 
& $0.90(1)$ & $15/14$ \\
( I, 2.2 , 0.04/0.04 ) & 10-15 & 21-26 & 6-11 &$0.860(4)$ 
& $1.051(5)$ & $19/13$ \\
\hline
\end{tabular}
\end{center}
\end{table}
\clearpage

\begin{table}
\caption{Results for the residual mass for unitary data on the RHMC data.}
\label{tab:mres}
\begin{center}
\begin{tabular}{lccc}
\hline
 Ensemble Mnemonic & Fit range &\multicolumn{2}{c}{Results}\\
\hline
& $t_{\rm min}-t_{\rm max}$  &
$m^\prime_{\rm res}(m_f)$ & $\chi^{2}/dof$ \\ 
\hline
(D, 0.72 , 0.01/0.04)  & 5-16 & $0.01089(4)$ & $51/11$  \\  
(D, 0.72 , 0.02/0.04)  & 9-16 & $0.01092(7)$ &  $18/15$ \\
(D, 0.72 , 0.04/0.04)  & 8-16 & $0.01146(8)$ &  $31/15$ \\
(D, 0.764 , 0.02/0.04) & 9-15 & $0.00535(2)$ & $18/13$ \\
(D, 0.764 , 0.04/0.04) & 10-14 & $0.00540(1)$ & $12/9$ \\
(D, 0.78 , 0.02/0.04)  & 11-15 & $0.00428(2)$ & $6/4$ \\ 
(D, 0.78 , 0.04/0.04)  & 4-15 & $0.00427(2)$ & $9/11$ \\
\hline
(I, 2.13 , 0.02/0.04)  & 10-15 & $0.01127(3)$ & $12/11$ \\
(I, 2.13 , 0.04/0.04)  & 9-15  & $0.01175(5)$ & $20/13$ \\
(I, 2.2 , 0.02/0.04)   & 10-15 & $0.00688(2)$ & $19/11$ \\
(I, 2.2 , 0.04/0.04)   & 12-16 & $0.00711(2)$ & $17/9$ \\ 
\hline
\end{tabular}
\end{center}
\end{table}

\begin{table}
\caption{Results for the residual mass, $m^\prime_{\rm res}(m_f)$ for the LL non-degenerate DBW2 $\beta=0.72$ RHMC data.
\label{tab:RHMC_0_72_rhmc_rt}}
\begin{center}
\begin{tabular}{lrrr}
\hline
$m_l^{\rm val}$ & $m_l^{\rm dyn} = 0.01 $ & $m_l^{\rm dyn} = 0.02 $
  & $m_l^{\rm dyn} = 0.04 $   \\\hline
0.005 &  0.01101(6) & 0.01120(7) & \\
0.010 &   & 0.01109(7) & \\
0.015 &  0.01078(5) & 0.01097(6) & \\
0.020 &  0.01068(5) & & 0.01191(10) \\
0.025 &  0.01059(5) & 0.01077(6) & \\
0.030 &  0.01050(5) & 0.01068(6) & 0.01167(9)\\
0.035 &  0.01042(5) & 0.01060(6) & \\
0.040 &  0.01035(5) & 0.01053(6) & \\
\hline
\end{tabular}
\end{center}
\end{table}

 \begin{table}
 \centering
 \caption{The vector meson masses and the residual masses computed for the four data sets: (D, 0.72, 0.04/0.04, R), (C2.3, 0.48, 0.04/0.04, R), (C3.57, 0.32, 0.04/0.04, R), and (C7.47, 0.16, 0.04/0.04, R). }\label{tab:plaq_rect_rho_mres}

\begin{tabular}{lccc}
\hline

Ensemble & $m_{\rm val}$  &  $m_V$  &$m^{\prime}_{\rm res}(m_f)$\\

\hline
(D, 0.72, 0.04/0.04, R) &$0.02$ & $0.626(10)$ & $0.01141(6)$\\
&$0.03$ & $0.655(8)$ &  $0.01119(6)$\\
&$0.04$ & $0.685(7)$ &  $0.01100(5)$\\
&$0.05$ & $0.715(6)$ & $0.01084(5)$\\

\hline
(C2.3, 0.48, 0.04/0.04, R) &$0.02$ & $0.612(10)$ & $0.00781(6)$\\
&$0.03$ & $0.636(10)$& $0.00768(5)$\\
&$0.04$ & $0.661( 8)$ & $0.00756(5)$\\
&$0.05$ & $0.689( 6)$ & $0.00747(4)$\\

\hline
(C3.57, 0.32, 0.04/0.04, R) &$0.02$ & $0.598(11)$ & $0.00754(5)$\\
&$0.03$ & $0.628( 9)$ &  $0.00742(5)$\\
&$0.04$ & $0.657( 8)$ &  $0.00732(5)$\\
&$0.05$ & $0.686( 7)$ &  $0.00723(4)$\\

\hline
(C7.47, 0.16, 0.04/0.04, R) &$0.02$ & $0.599(10)$ &  $0.00654(5)$\\
&$0.03$ & $0.623( 8)$ &  $0.00642(5)$\\
&$0.04$ & $0.650( 7)$ & $0.00634(4)$\\
&$0.05$ & $0.678( 6)$ & $0.00626(4)$\\
\hline

\end{tabular}
\end{table}

\begin{table}
\caption{Results for $Z_{A}$ from fits to unitary RHMC data.}
\label{ch6:tab:ZA}
\begin{center}
\begin{tabular}{lcccc}
\hline
Ensemble Mnemonic & Fit range &\multicolumn{3}{c}{Results}\\
\hline
($\beta , m_l/m_s$)& $t_{\rm min}-t_{\rm max}$ & $Z_{A}$ & $\chi^{2}/dof$ & correlators\\ 
\hline
( D, 0.72 , 0.01/0.04 ) & 5-11 & $0.7335(2)$  & $18/13$ & SL-SL\\
( D, 0.72 , 0.02/0.04 ) & 5-11 & $0.7347(2)$  & $19/13$ & SL-SL\\
( D, 0.72 , 0.04/0.04 ) & 8-13 & $0.7373(1)$  & $11/11$ & SL-SL\\
( D, 0.764 , 0.02/0.04 ) & 6-12 & $0.75521(5)$  & $25/13$& WL-WL \\
( D, 0.764 , 0.04/0.04 ) & 6-14 & $0.75722(7)$  & $21/17$& WL-WL \\
( D, 0.78 , 0.02/0.04 ) & 8-14 & $0.7625(3)$  & $14/13$ & WL-WL\\
( D, 0.78 , 0.04/0.04 ) & 8-14 & $0.7662(2)$  & $17/13$ & WL-WL\\
\hline
( I, 2.13 , 0.02/0.04 ) & 7-14 & $0.73376(10)$  & $28/15$& SL-SL \\
( I, 2.13 , 0.04/0.04 ) & 6-11 & $0.7357(1)$  & $22/11$ & SL-SL \\
( I, 2.2 , 0.02/0.04 ) & 10-15 & $0.74563(9)$  & $13/11$ & WL-WL\\
( I, 2.2 , 0.04/0.04 ) & 10-15 & $0.74820(7)$  & $23/11$ & WL-WL\\
\hline
\end{tabular}
\end{center}
\end{table}

\begin{table}
\caption{Results for $Z_{A}$ from fits to non-degenerate LL-LL DBW2 $\beta=0.72$
RHMC data. \label{dbw2:llll_za}}
\begin{center}
\begin{tabular}{lcccc}
\hline
$m_{\rm val}$ & Fit range &\multicolumn{3}{c}{$m_l$} \\
              &  $t_{\rm min}-t_{\rm max}$ & 0.01 & 0.02 & 0.04 \\ \hline

0.005 & 8-16  & $0.73207(32)$ & $0.73382(29)$ &    \\
0.01 & 8-16  &  & $0.73413(26)$ &    \\
0.015 & 8-16  & $0.73338(22)$ & $0.73461(24)$ &    \\
0.02 & 8-16  & $0.73416(20)$ &  & $0.73518(29)$ \\
0.025 & 8-16  & $0.73498(19)$ & $0.73592(20)$ &    \\
0.03 & 8-16  & $0.73583(19)$ & $0.73668(20)$ & $0.73651(22)$ \\
0.035 & 8-16  & $0.73670(19)$ & $0.73745(19)$ &    \\
0.04 & 8-16  & $0.73759(19)$ & $0.73825(18)$ & \\
\hline
\end{tabular}
\end{center}
\end{table}

%
%

\begin{table}
\caption{Fitted values for the pseudoscalar meson decay constant using the value 
of $Z_{A}$ and the axial-axial correlator for 
unitary RHMC data, Eq.~(\ref{eq:fpMeth1}).\label{ch6:tab:fpi2}}
\begin{center}
\begin{tabular}{lcccc}
\hline
Dataset & Fit range &\multicolumn{3}{c}{Results}\\
\hline
& $t_{\rm min}-t_{\rm max}$ & $A_{A_{0}A_{0}}$ & $f_{P}$ & $\chi^{2}/dof$ \\ 
\hline
( D, 0.72 , 0.01/0.04 ) & 9-15 & $0.028(2)$  & $0.099(3)$  & $9./6$ \\
( D, 0.72 , 0.02/0.04 ) & 9-15 & $0.044(1)$  & $0.111(2)$  & $11/6$ \\
( D, 0.72 , 0.04/0.04 ) & 9-15 & $0.075(3)$  & $0.128(3)$  & $11/6$ \\
( D, 0.764 , 0.02/0.04 ) & 8-16 & $0.0219(5)$ &$0.0890(8)$  & $14/7$ \\
( D, 0.764 , 0.04/0.04 ) & 8-15 & $0.043(1)$  & $0.107(1)$  & $8/6$ \\
( D, 0.78 , 0.02/0.04 ) & 11-16 & $0.016(2)$  & $0.083(3)$  & $9/4$ \\
( D, 0.78 , 0.04/0.04 ) & 12-16 & $0.026(2)$  & $0.086(2)$  & $5/4$ \\
\hline
( I, 2.13 , 0.02/0.04 ) & 5-16 & $0.0307(7)$  & $0.096(1)$  & $13/11$ \\
( I, 2.13 , 0.04/0.04 ) & 10-16 & $0.057(2)$  & $0.114(2)$  & $3/6$ \\
( I, 2.2 , 0.02/0.04 ) & 9-16 & $0.0195(5)$  & $0.0825(10)$  & $10/6$ \\
( I, 2.2 , 0.04/0.04 ) & 10-16 & $0.0407(7)$  & $0.1024(9)$  & $14/6$ \\
\hline
\end{tabular}
\end{center}

\end{table}

\begin{table}
\caption{Fitted values for the pseudoscalar meson decay constant using the value of $Z_A$ and axial-axial 
correlator for non-degenerate DBW2 $\beta=0.72$ RHMC data, Eq.~(\ref{eq:fpMeth1}). \label{tab:RHMC_0_72_rhmc_fpi_AA_RBC}}
\begin{center}
\begin{tabular}{lrrr}
\hline
$m_l^{\rm val}$ & $m_l^{\rm dyn} = 0.01 $ & $m_l^{\rm dyn} = 0.02 $
  & $m_l^{\rm dyn} = 0.04 $ 
  \\
\hline
0.005 & 0.0950(24) &  0.1017(23) & \\
0.010 &  &  0.1043(22) & \\
0.015 & 0.1009(22) &  0.1070(22) & \\
0.020 & 0.1038(21) &  & 0.1213(29)\\
0.025 & 0.1066(21) &  0.1125(21) & \\
0.030 & 0.1094(21) &  0.1152(21) & 0.1260(28)\\
0.035 & 0.1121(20) &  0.1180(21) & \\
0.040 & 0.1147(20) &  0.1206(21) & \\
\hline
\end{tabular}
\end{center}
\end{table}

\begin{table}
\caption{Fitted values for the pseudoscalar meson decay constant using the residual mass and the pseudoscalar density correlator, Eq.~(\ref{eq:fpMeth2}) for the unitary RHMC data.\label{ch6:tab:fpi1}}
\begin{center}
\begin{tabular}{lcccc}
\hline
Dataset & Fit range &\multicolumn{3}{c}{Results}\\
\hline
& $t_{\rm min}-t_{\rm max}$ & $A_{PP}$ & $f_{P}$ & $\chi^{2}/dof$ \\ 
\hline
(D, 0.72 , 0.01/0.04 ) & 11-16 & $0.76(5)$  & $0.097(3)$  & $18/5$ \\
(D, 0.72 , 0.02/0.04 ) & 8-16 & $0.93(3)$  & $0.115(2)$  & $7/8$ \\
(D, 0.72 , 0.04/0.04 ) & 10-16 & $0.96(4)$  & $0.129(3)$  & $9/5$ \\
(D, 0.764 , 0.02/0.04 ) & 10-16 & $0.49(1)$  & $0.0884(9)$  & $12/5$ \\
(D, 0.764 , 0.04/0.04 ) & 11-16 & $0.55(1)$  & $0.108(1)$  & $7/4$ \\
(D, 0.78 , 0.02/0.04 ) & 11-16 & $0.36(3)$  & $0.080(3)$  & $2/4$ \\
(D, 0.78 0.04/0.04)& 11-16 & $0.35(2)$ &$0.089(3)$ & $6/4$ \\
\hline
(I, 2.13 , 0.02/0.04 ) & 10-16 & $0.53(2)$  & $0.092(1)$  & $8/5$ \\
(I, 2.13 , 0.04/0.04 ) & 9-16 & $0.60(2)$  & $0.112(2)$  & $7/7$ \\
(I, 2.2 , 0.02/0.04 ) & 9-16 & $0.39(1)$  & $0.0833(8)$  & $10/7$ \\
(I, 2.2 , 0.04/0.04 ) & 11-16 & $0.48(1)$  & $0.1033(9)$  & $8/4$ \\
\hline
\end{tabular}
\end{center}
\end{table}

\begin{table}
\caption{Fitted values for the pseudoscalar meson decay constant using
  $Z_{A}$ and the local axial correlator, Eq.~(\ref{eq:fpMeth3}).
\label{ch6:tab:fpi3}}
\begin{center}
\begin{tabular}{lccc}
\hline
Dataset & Fit range &\multicolumn{2}{c}{Results}\\
\hline
& $t_{\rm min}-t_{\rm max}$ & $f_{P}$ & $\chi^{2}/dof$  \\ 
\hline
( D, 0.72 , 0.01/0.04 ) & 10-16 & $0.11(2)$  & $1/6$  \\
( D, 0.72 , 0.02/0.04 ) & 10-16 & $0.113(2)$  & $7/6$ \\
( D, 0.72 , 0.04/0.04 ) & 12-16 & $0.127(3)$  & $5/4$  \\
( D, 0.764 , 0.02/0.04 ) & 11-16 & $0.089(1)$  & $4/4$  \\
( D, 0.764 , 0.04/0.04 ) & 11-16 & $0.107(1)$  & $6/4$\\
( D, 0.78 , 0.02/0.04 ) & 12-16 & $0.076(3)$  & $8/4$  \\
( D, 0.78 , 0.04/0.04 ) & 10-16 & $0.088(3)$  & $0.3/5$  \\
\hline
( I, 2.13 , 0.02/0.04 ) & 11-16 & $0.092(1)$  & $4/5$ \\
( I, 2.13 , 0.04/0.04 ) & 10-16 & $0.111(2)$  & $3/5$  \\
( I, 2.2 , 0.02/0.04 ) & 11-16 & $0.0830(8)$  & $7/5$ \\
( I, 2.2 , 0.04/0.04 ) & 10-16 & $0.102(1)$  & $10/5$ \\
\hline
\end{tabular}
\end{center}
\end{table}
\clearpage


\begin{table}
\caption{Vector meson masses in the chiral limit and
 $r_0$}
\label{tab:chirho}
\begin{center}
\begin{tabular}{lcccc}
\hline
Dataset &\multicolumn{4}{c}{Results}\\
\hline
dataset & $m_V$ & $a^{-1}$ (GeV) & $\chi^{2}/dof$ & $r_0$(fm)\\ 
\hline
(D,0.72)   &  $0.52^{+1}_{-1}$ & $1.49^{+3}_{-3}$ & $1/1$ & 0.564(13) \\
(D,0.764)  &  $0.45^{+2}_{-1}$ & $1.71^{+5}_{-6}$ &-&  0.580(20)\\
(D,0.78)   &  $0.38^{+3}_{-4}$ & $2.1^{+2}_{-2}$ & - & 0.487(57)\\
\hline
(I,2.13)   &  $0.46^{+1}_{-1}$ & $1.68^{+4}_{-4}$ & - & 0.544(34)\\
(I,2.2)    &  $0.37^{+1}_{-2}$ & $2.08^{+9}_{-6}$ & - & 0.497(28)\\
\hline
\end{tabular}
\end{center}
\end{table}


\begin{table}[ht]
\centering
\caption{Results for the residual mass and the vector meson masses with different 
rectangle contributions. The values shown are extrapolations to the
valence chiral limit: $m_{val} = 0$ for the residual masses and
$m_{val} = -m_{\rm res}$ for vector meson masses. The inverse lattice
spacings determined from the vector meson masses are shown in the last
column.\label{tab:plaq_rect_chiral_limit} }
\begin{tabular}{lccc}
\hline
Ensemble & $m^\prime_{\rm res}$ & $m_V$ & $1/a$ (GeV) \\
\hline
(D, 0.72, 0.04/0.04, R) & 0.01176(7) & 0.530(15) & 1.450(41)\\
(C2.3, 0.48, 0.04/0.04, R) & 0.00802(6) & 0.539(16) & 1.430(43) \\
(C3.57, 0.32, 0.04/0.04, R) & 0.00773(6) & 0.517(16) & 1.489(45) \\
(C7.47, 0.16, 0.04/0.04, R) & 0.00670(6) & 0.525(16) & 1.467(44) \\
\hline
(C3.57, 0.333, 0.04/0.04, R) & 0.00480(9)& 0.509(26) & 1.51(8) \\
(C3.57, 0.36, 0.04/0.04, R) & 0.002306(21) & 0.400(12)& 1.92(7)\\
(C2.3, 0.53, 0.04/0.04, R) & 0.002683(30) & 0.405(17) & 1.90(8)\\
\hline
\end{tabular}
\end{table}

\begin{table}
\caption{Pseudoscalar meson mass squared in the chiral limit and 
 the bare $s$ quark mass. \label{ch6:tab:chips}}
\begin{center}
\begin{tabular}{lccc}
\hline
Ensembles  & \multicolumn{3}{c}{Results}\\ \hline
           & $m_{P}(m_{q}\rightarrow0)$ & $\chi^{2}/dof$ & $m_{s}(m_{\rho})$\\ 
\hline
(D,0.72)   & $-0.014(2)$  & $0.006/1$ & $0.049(2)$ \\
(D,0.764)  & $-0.004(2)$  & -         & $0.044(3)$ \\
(D,0.78)   & $-0.011(10)$ & -         & $0.036(5)$ \\
\hline
(I,2.13)   & $-0.002(4)$  & -         & $0.042(2)$ \\
(I,2.2 )   & $-0.008(3)$  & -         & $0.032(2)$ \\
\hline
\end{tabular}
\end{center}
\end{table}

\begin{table}
\caption{Results for $m_{K^\star}$, the nucleon and its negative partity partner. \label{tab:mkstarNucleons}}
\begin{center}
\begin{tabular}{lccc}
Dataset & $m_{K^\star}$ & $m_N$ & $m_{N^\star}$  \\
\hline
(D,0.72) & $0.588(7)$ & $0.73(2)$  & $1.02(5)$ \\ 
(D,0.764) & $0.51(1)$ & $0.59(4)$  & $0.80(3)$ \\ 
(D,0.78) & $0.46(2)$  & - & - \\
\hline
(D,2.13) & $0.529(10)$ & $0.58(3)$  & $0.92(5)$ \\
(D,2.2)  & $0.443(9)$ & $0.554(9)$ & $0.71(3)$ \\ 
\hline
\end{tabular}
\end{center}
\end{table}

\clearpage
\begin{table}
\begin{center}
\caption{Pseudoscalar meson decay constant in the 
chiral limit and corresponding lattice spacing from the physical 
value of $f_{\pi}$.\label{tab:chifpi}}
\begin{tabular}{ccccccc}
\hline
Dataset &\multicolumn{6}{c}{Results}\\
\hline
$\beta$ & $f_{P}^{PP}$ & $a^{-1}$ (GeV) & $f_{P}^{AA}$ & $a^{-1}$ (GeV) & $f_{P}^{AP}$ & $a^{-1}$ (GeV) \\ 
\hline
0.72   & $0.080(5)$ & $1.6(1)$  & $0.082(4)$ & $1.59(8)$ & $0.082(4)$ & $1.59(8)$\\
0.764  & $0.064(2)$ & $2.04(6)$ & $0.067(2)$ & $1.94(7)$ & $0.065(3)$ & $1.99(9)$ \\
0.78   & $0.068(8)$ & $1.9(3)$  & $0.07(1)$  & $1.8(3)$   & $0.062(8)$ & $2.1(3)$ \\
\hline
2.13   & $0.062(3)$ & $2.10(9)$ & $0.067(5)$ & $1.9(1)$   & $0.061(3)$ & $2.1(1)$ \\
2.2    & $0.057(2)$ & $2.29(8)$ & $0.056(3)$ & $2.31(10)$ & $0.057(3)$ & $2.3(1)$ \\
\hline
\end{tabular}
\end{center}
\end{table}


\clearpage

%% file: fig.tex
\clearpage

%
%

\begin{figure}
\begin{center}
\epsfig{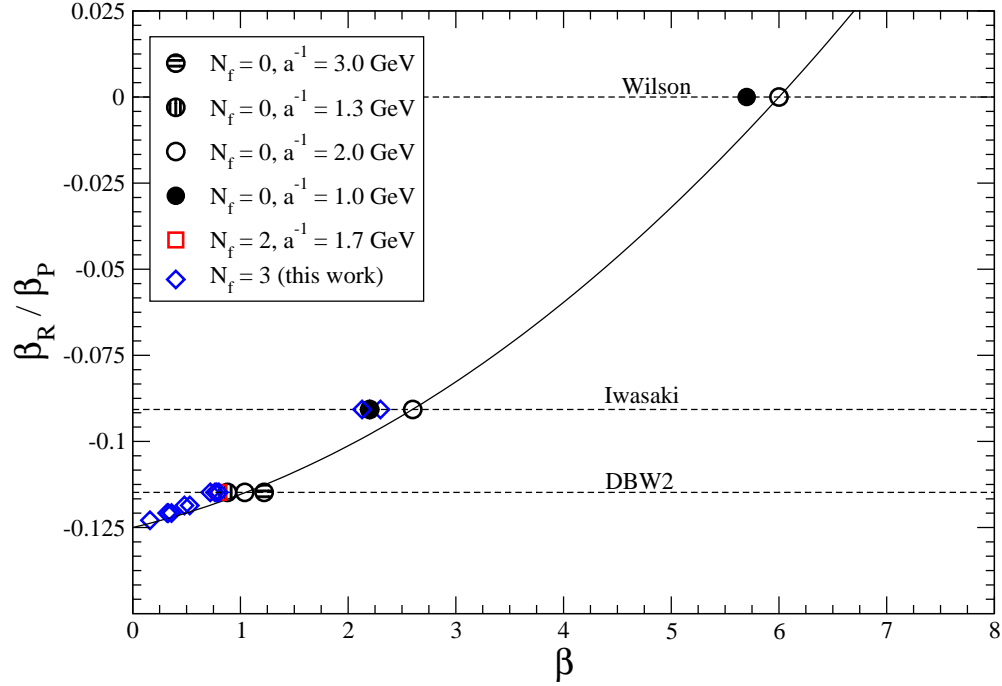}
\caption{Parameters of $\beta_P$ and $\beta_R$ for quenched (circles)
  and $2$ flavor (squares) simulations with same lattice spacings and
  the choices for the parameters used for the simulations reported in
  this paper (diamonds).}
\label{fig:plaq_rect_plane}
\end{center}
\end{figure}

\begin{figure}
\begin{center}
\epsfig{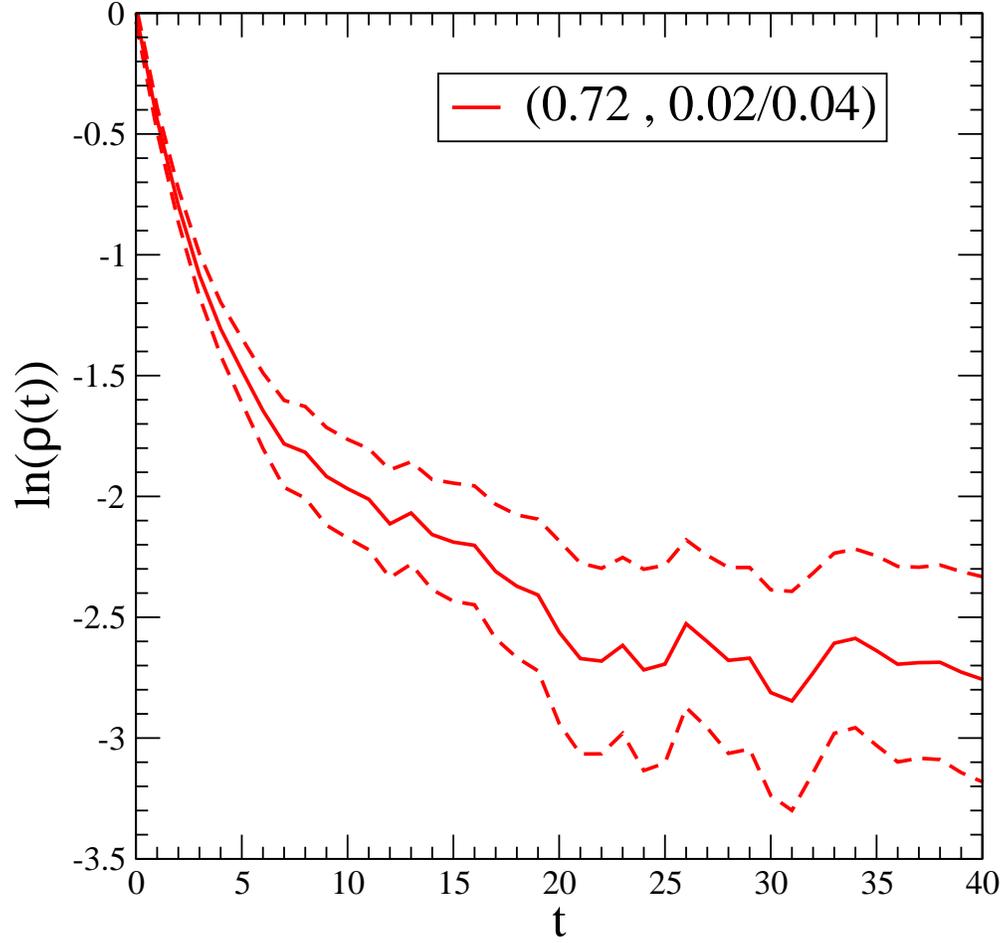}
\caption{Logarithm of the normalised autocorrelation
  function for the plaquette on the (D, 0.72 , 0.02/0.04) dataset.
  $\tau_{exp}$ is found from the slope at early $t$. }
\label{framework:plot:tint}
\end{center}
\end{figure}

\begin{figure}
\begin{center}
\epsfig{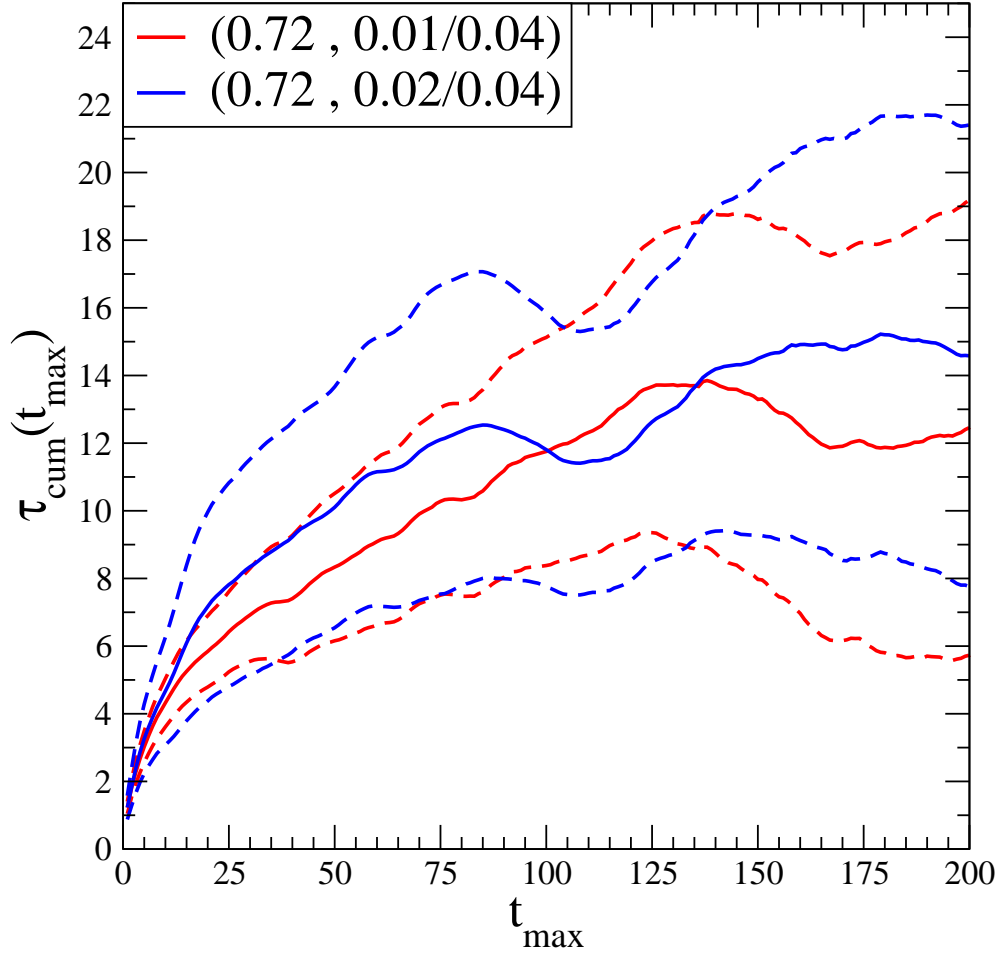}
\caption{Integrated autocorrelation time for the pseudoscalar meson on the
  DBW2 $\beta$=0.72 datasets with the longest single Monte Carlo
  chains. The separation for decorrelated configurations should be
  $2\tau_{\rm cum}$ and the measurement is made every 5 trajectories.}
\label{framework:plot:tint2}
\end{center}
\end{figure}

\begin{figure}
\begin{center}
\epsfig{file=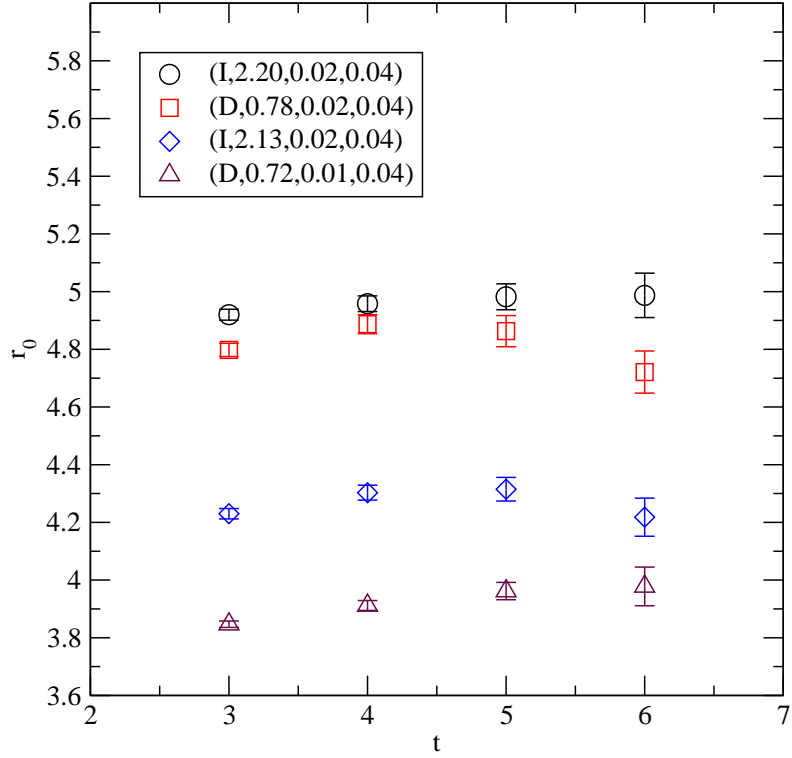, width=0.8\textwidth}
\caption{Values for $r_0$ found for four choices of the time variable $t$ in 
Eq.~(\ref{eq:v_r_t}).  These show that a reasonable plateau has been reached 
by $t=5$, the value we adopt to determine $r_0$.}
\label{fig:plateaux}
\end{center}
\end{figure}

\begin{figure}
\begin{center}
\epsfig{file=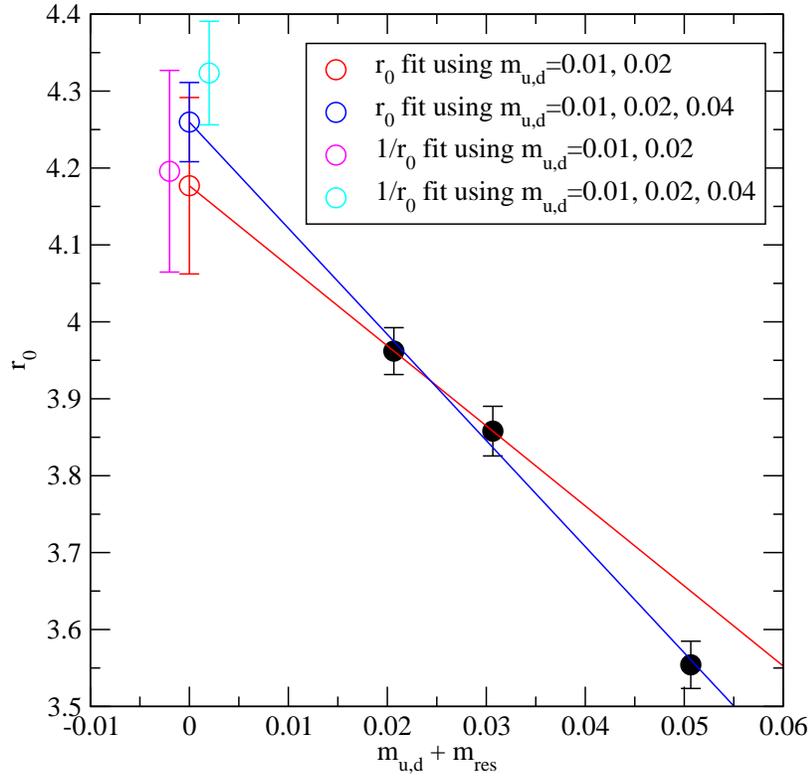, width=0.8\textwidth}
\caption{Chiral extrapolation of the Sommer parameter $r_0$ to the limit
$m_{\rm u,d} = m_l + m_{\rm res} \rightarrow 0$ using results from the
three DBW2, $\beta=0.72$ ensembles with $m_l = 0.01, 0.02$ and 0.04.  
Extrapolations are shown using both the two lightest masses and all three.}
\label{fig:0.72_chiral}
\end{center}
\end{figure}

%
%
\begin{figure}
\begin{center}
\epsfig{file=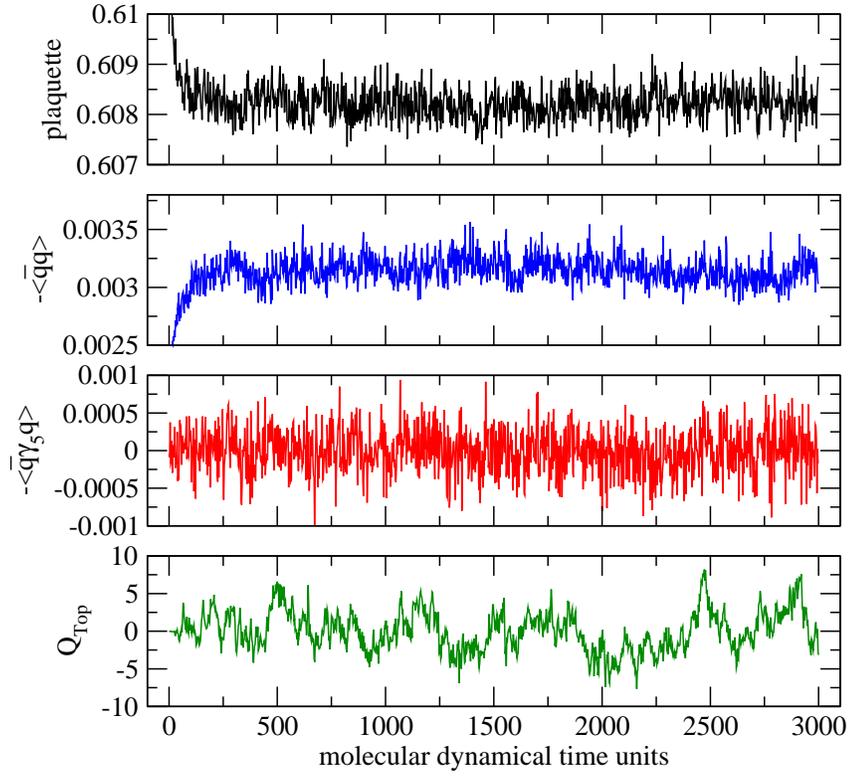, width=0.8\textwidth}
\caption{From top to bottom, the panels give the time history of
the plaquette, $\langle \bar{q} q \rangle$ and $\langle \bar{q}
\gamma_5 q \rangle$ and the toplogical charge for the (D, 0.72,
0.01/0.04) dataset.  The values plotted are measured every 5 time
units.}
\label{fig:dbw2_0.72_01_04_evol}
\end{center}
\end{figure}
%

%
%
\begin{figure}
\begin{center}
\epsfig{file=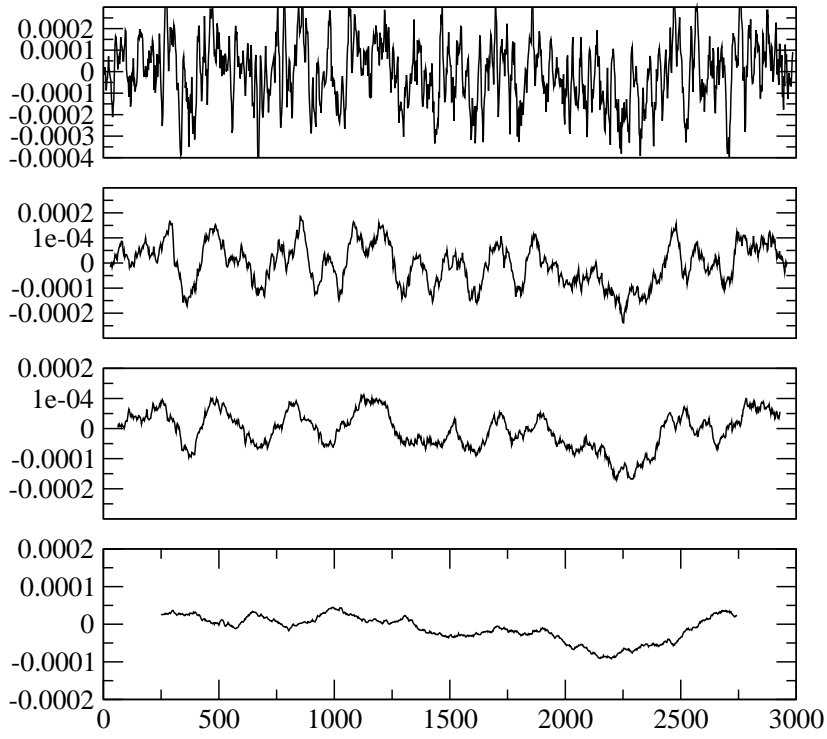, width=0.8\textwidth}
\caption{The time history of $\langle \bar{q} \gamma_5 q \rangle$
is shown for the (D, 0.72, 0.01/0.04) dataset, with different sizes
for the smoothing window.  If a smoothing window of size $s$ is
used, the data plotted at time unit $n$ is an average of data from
$n - s/2$ to $n + s/2 - 1$.  From top to bottom, the panels have a
smoothing window of size 25, 50, 100 and 200 time units. }
\label{fig:dbw2_0.72_01_04_pbg5p_evol}
\end{center}
\end{figure}

%
%
\begin{figure}
\begin{center}
\epsfig{file=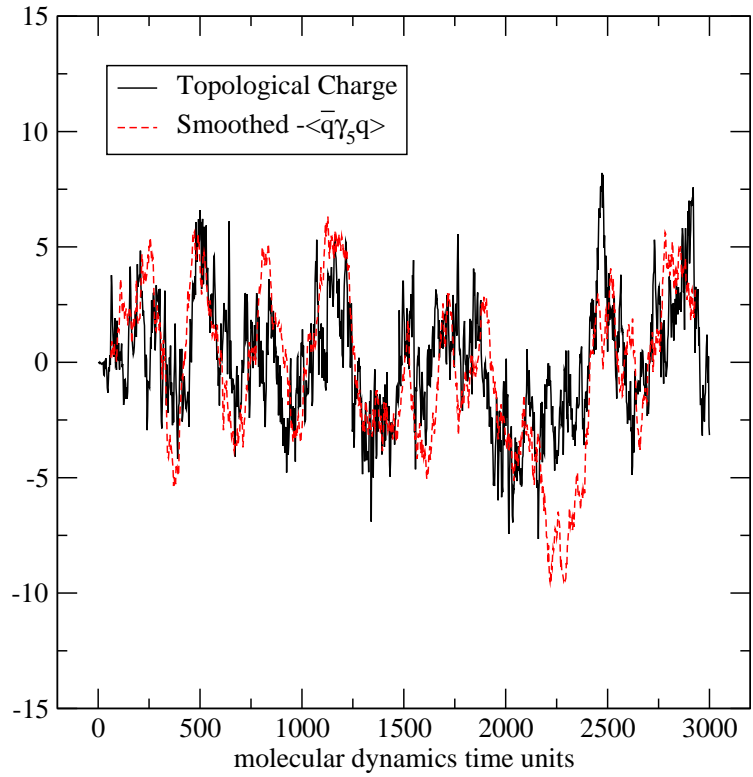, width=0.8\textwidth}
\caption{The time history of the topological charge and
$\langle \bar{q} \gamma_5 q \rangle$, with a smoothing window of
50, is shown for the (D, 0.72, 0.01/0.04) dataset.
The evolutions are very similar.}
\label{fig:dbw2_0.72_01_04_tcharge_pbg5p_compare}
\end{center}
\end{figure}

\begin{figure}
\begin{center}
\epsfig{file=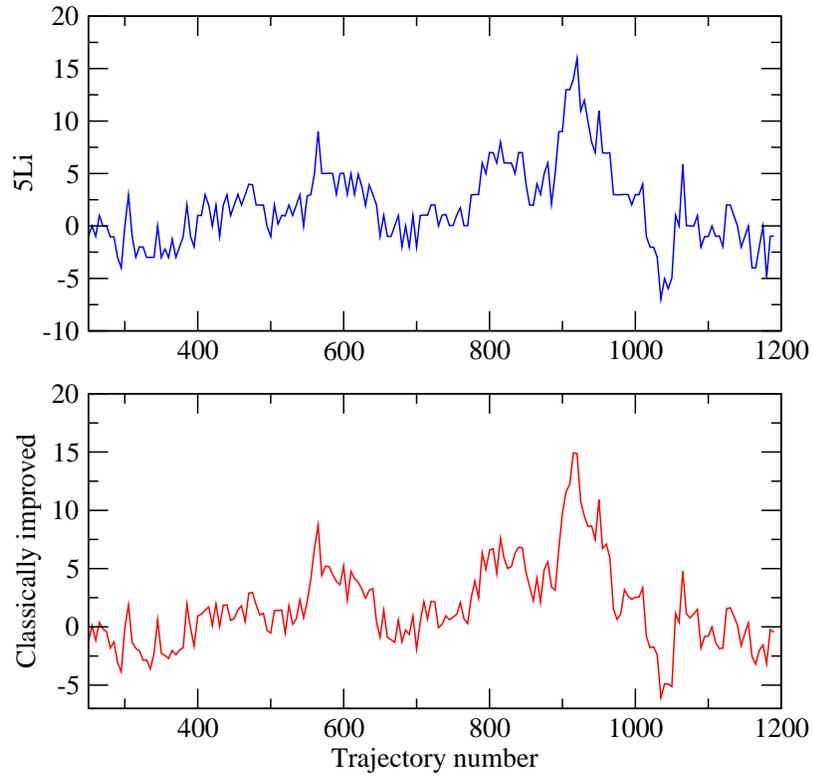, width=0.8\textwidth}
\caption{Comparison of the 5Li and classically improved methods
of calculating the topological charge for $\sim 1000$ HMC trajectory
lengths on the (I, 2.13, 0.04/0.04) ensemble.\label{fig:topo_method_history}}
\end{center}
\end{figure}

\begin{figure}
\begin{center}
\epsfig{file=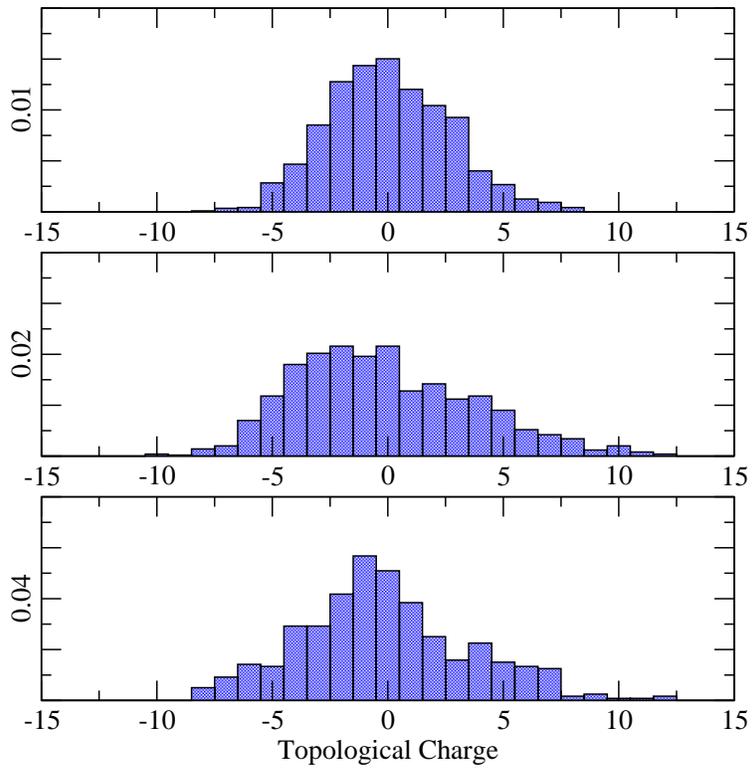, width=0.8\textwidth}
\caption{The distribution of the topological charge for (from top to bottom) the
(D,0.72,0.01/0.04), (D,0.72,0.02/0.04) and (D,0.72,0.04/0.04)
ensembles, taken from the classically improved method of 
calculating the topological charge.}
\label{fig:dbw20.72_tens}
\end{center}
\end{figure}
%

%
%
\begin{figure}
\begin{center}
  \epsfig{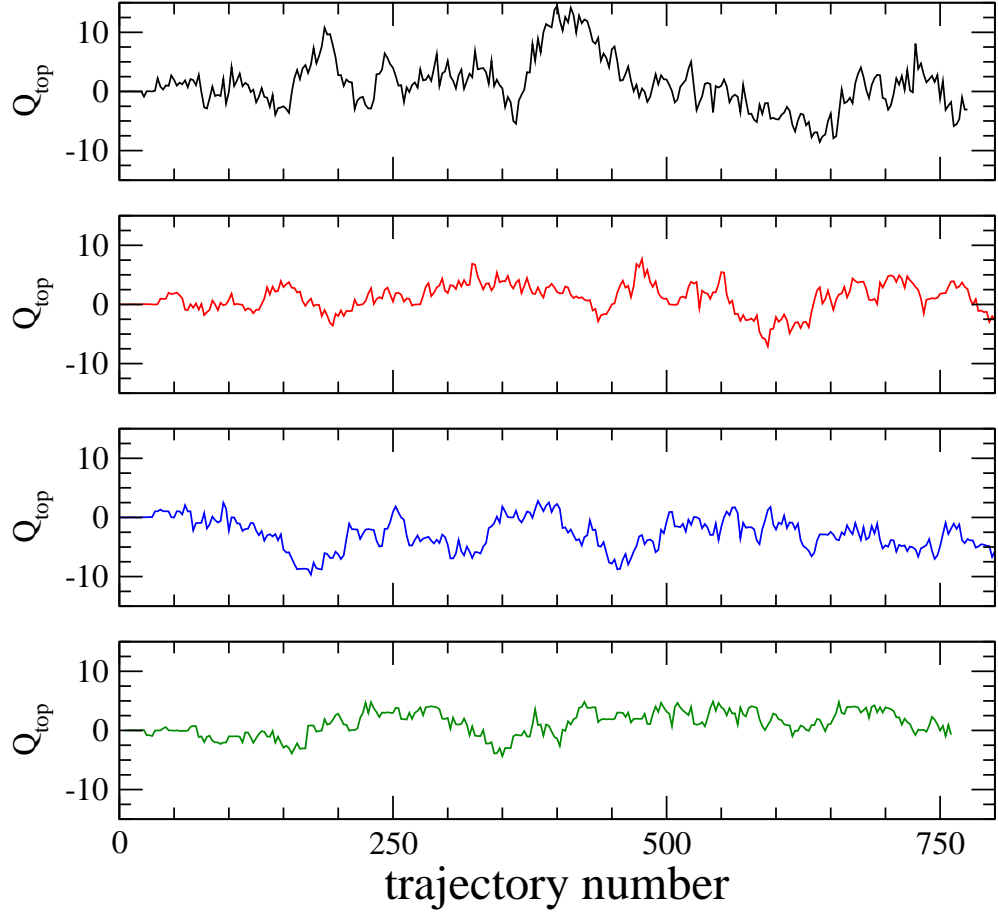}
  \caption{The evolution of topological charge for
   the four simulations (D, 0.72, 0.04/0.04, R) (top panel),
  (C2.3, 0.48, 0.04/0.04, R), (C3.57, 0.32, 0.04/0.04, R), and (C7.47,
  0.16, 0.04/0.04, R) (bottom panel).}
  \label{fig:plaq_rect_top_charge_evol}
\end{center}
\end{figure}

%
%
\begin{figure}
\begin{center}
\epsfig{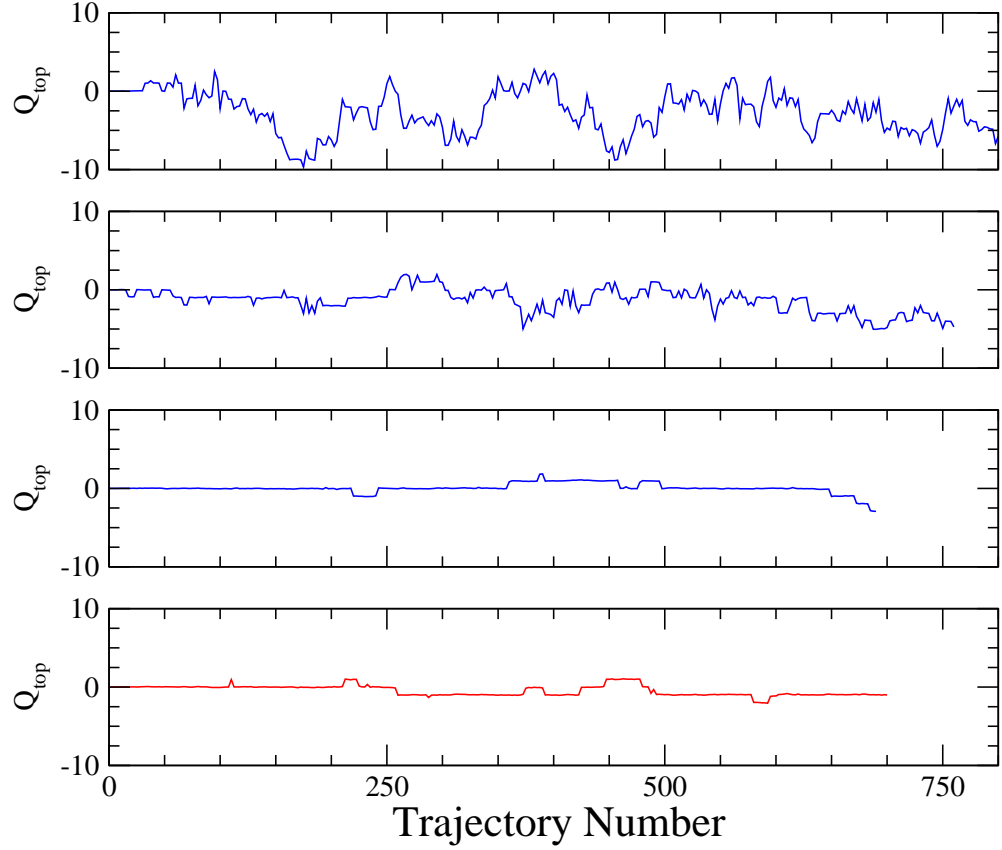}
\caption{The evolution of topological charge for
   the four simulations (C3.57, 0.32, 0.04/0.04, R)(top panel),
  (C3.57, 0.333, 0.04/0.04, R), (C3.57, 0.36, 0.04/0.04, R) and (C2.3,
  0.53, 0.04/0.04, R) (bottom panel). Note that the topology stops evolving as we go to weaker couplings.}
  \label{fig:plaq_rect_top_charge_weak_coupling}
\end{center}
\end{figure}

\begin{figure}
\begin{center}
\epsfig{file=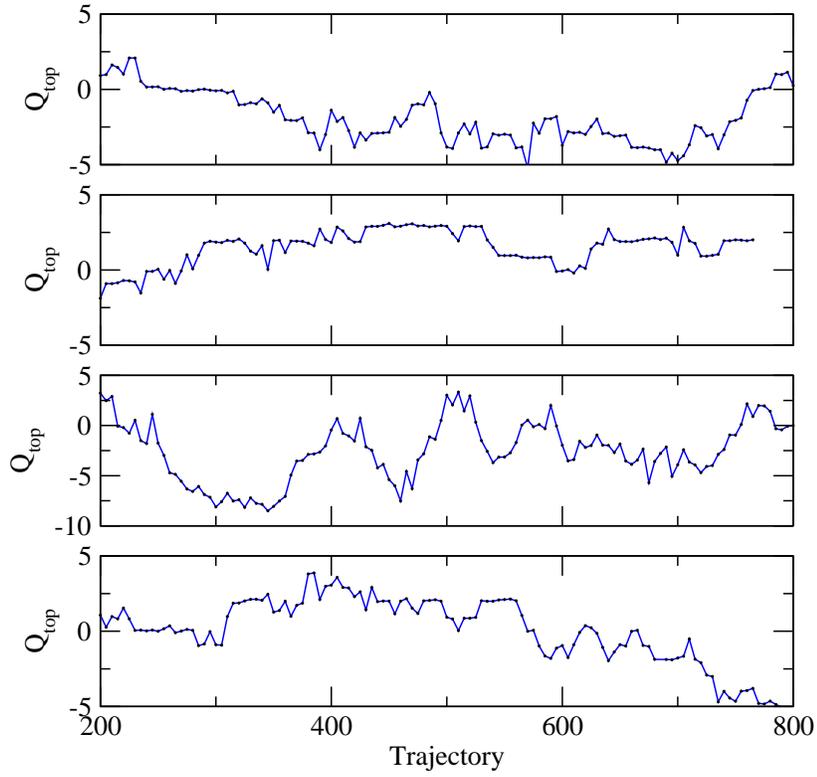,width=.8\textwidth}
\caption{Representative topological charge histories for the (D,0.764), (D,0.78), (I,2.13) 
and (I,2.2) actions (top to bottom). As can be seen, the rate of topological charge tunnelling 
decreases both when moving between the Iwasaki and DBW2 actions, and when moving to smaller
lattice spacings.}
\label{fig:top_action_comp}
\end{center}
\end{figure}

%
%
\begin{figure}
\begin{center}
\epsfig{file=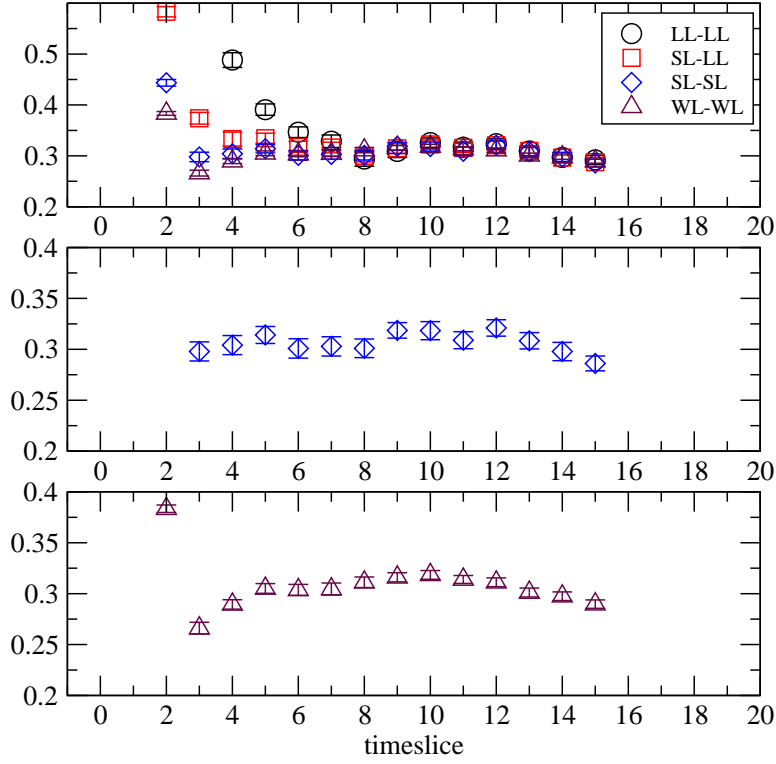, width=0.8\textwidth}
\caption{The effective mass from the
pseudoscalar meson correlator for the (D, 0.72, 0.01/0.04)
ensemble with $m_l^{\rm val} = 0.01$ for different sources.  The top
panel shows four different source/sink combinations, the middle panel
is for a pseudoscalar meson correlator made of two SL quark propagators and the
bottom panel is the WL-WL case. 
\label{fig:b0.72_rhmc_01_04_mpi_src_compare_pp}}
\end{center}
\end{figure}

%
%
\begin{figure}
\begin{center}
\epsfig{file=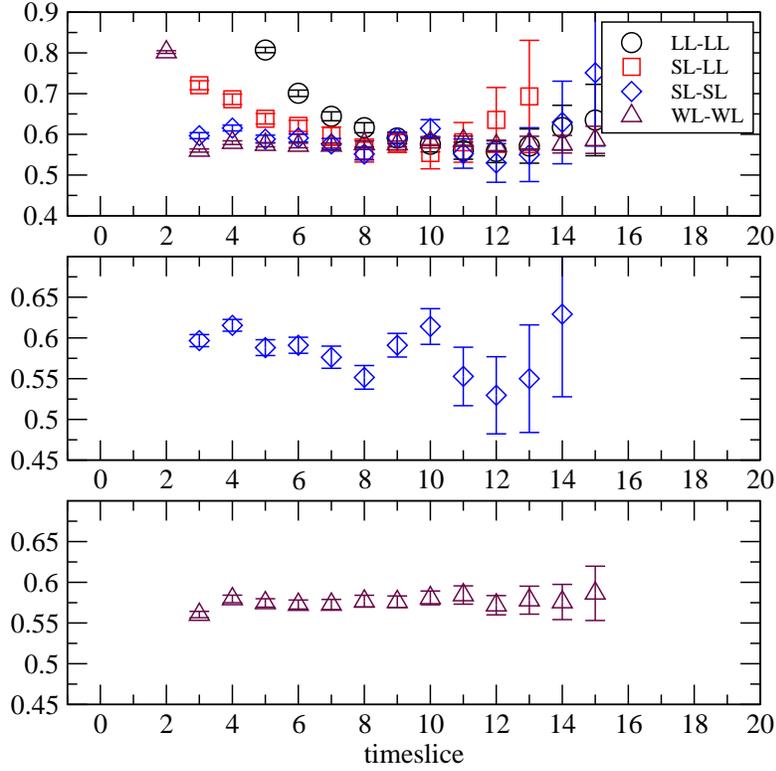, width=0.8\textwidth}
\caption{The vector meson effective mass for the (D, 0.72, 0.01/0.04) ensemble
with $m_l^{\rm val} = 0.01$ for different sources. The top
panel shows four different source/sink combinations, the middle panel
is for a vector meson correlator made of two SL quark propagators and the
bottom panel is the WL-WL case.}
\label{fig:b0.72_rhmc_01_04_mrho_src_compare}
\end{center}
\end{figure}

%
%
\begin{figure}
\begin{center}
\epsfig{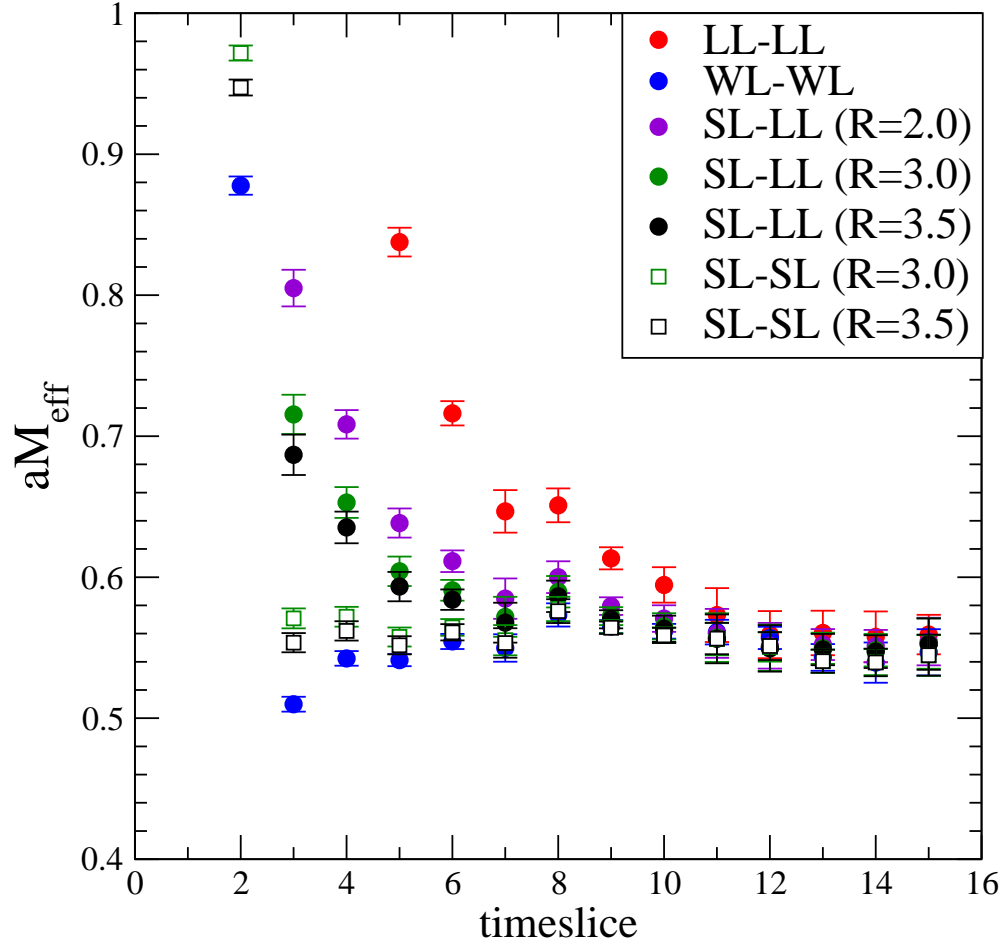}
\caption{Comparison of smearing functions for a vector meson constructed from 
  valence quarks with mass $m=0.04$ using 
  72 Iwasaki $\beta=2.2$ configurations with
  $m_{l}=0.02$ and $m_{s}=0.04$. 10 configurations were averaged into
  each bin and then a full correlated analysis performed on the binned
  data.
\label{ch6:plot:smearing}}
\end{center}
\end{figure}

\begin{figure}
\begin{center}
\epsfig{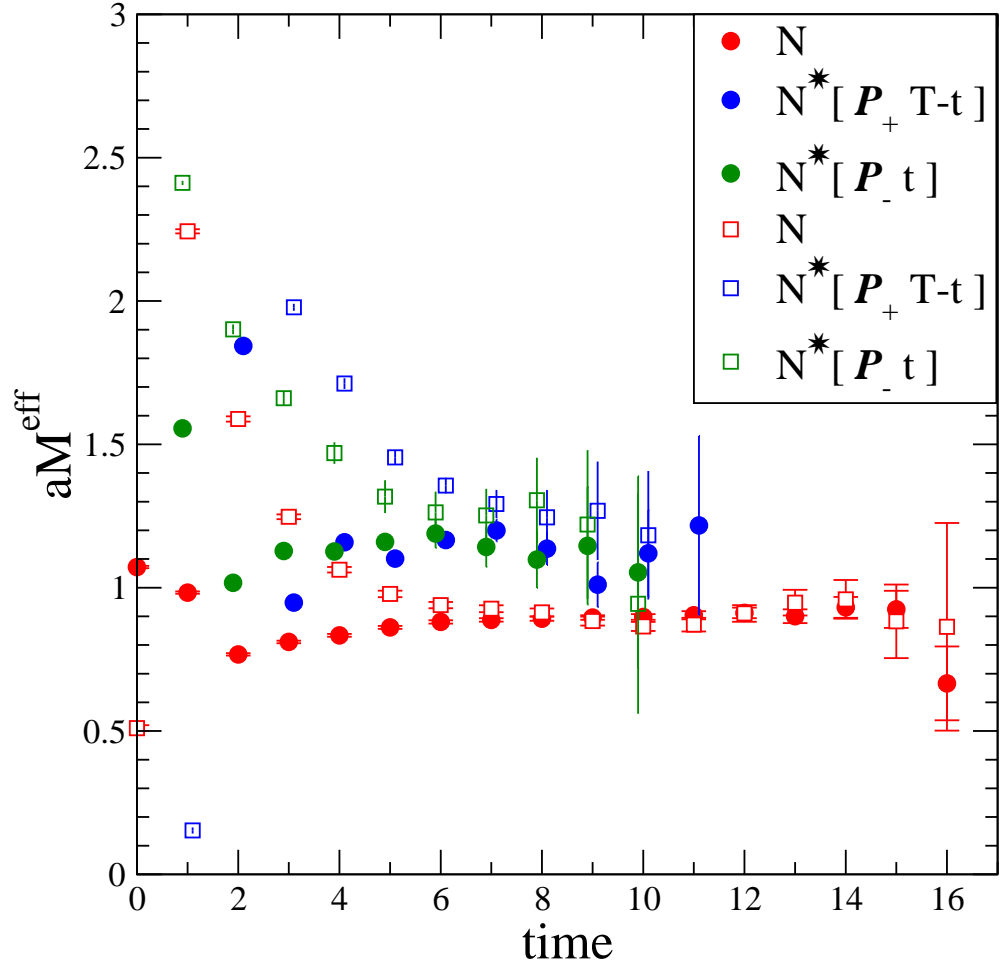}
\caption{Nucleon effective mass
  plot for the (D, 0.72, 0.02/0.04) dataset. Circles correspond to
  the WL-WL-WL calculations, the squares for the SL-SL-SL calculations.}
\label{ch6:fig:nuc}
\end{center}
\end{figure}
\clearpage

%
%
\begin{figure}
\begin{center}
\epsfig{file=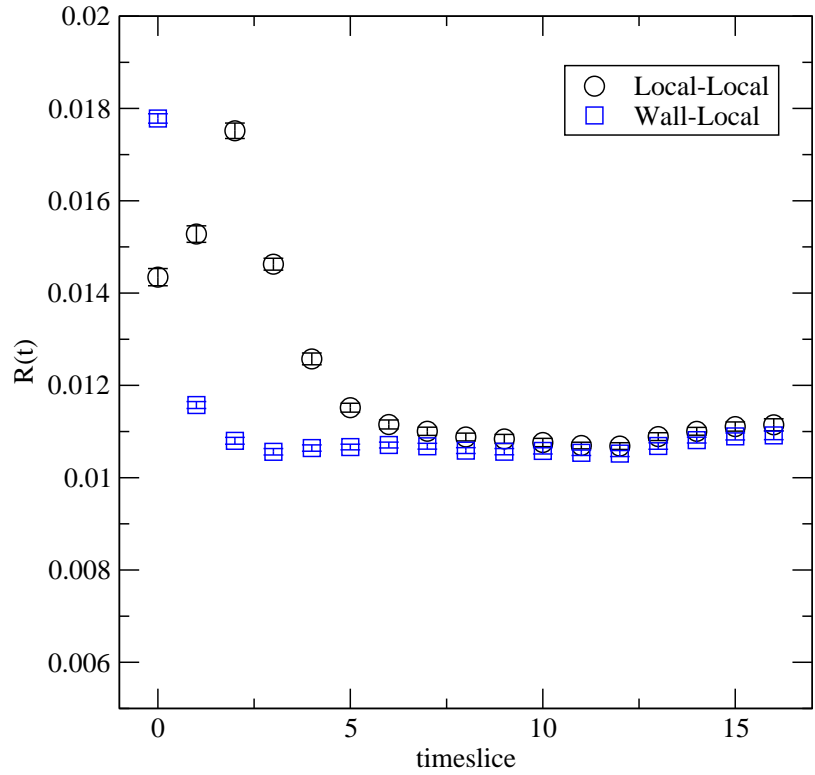, width=0.8\textwidth}
\caption{The ratio $R(t)$ that determines $\mres$ for the
(D, 0.72, 0.01/0.04) ensemble, for pseudoscalars made from LL
quark propagators and WL quark propagators.}
\label{fig:b0.72_rhmc_04_rt_wl}
\end{center}
\end{figure}

\clearpage	
\begin{figure}
\begin{center}
	\epsfig{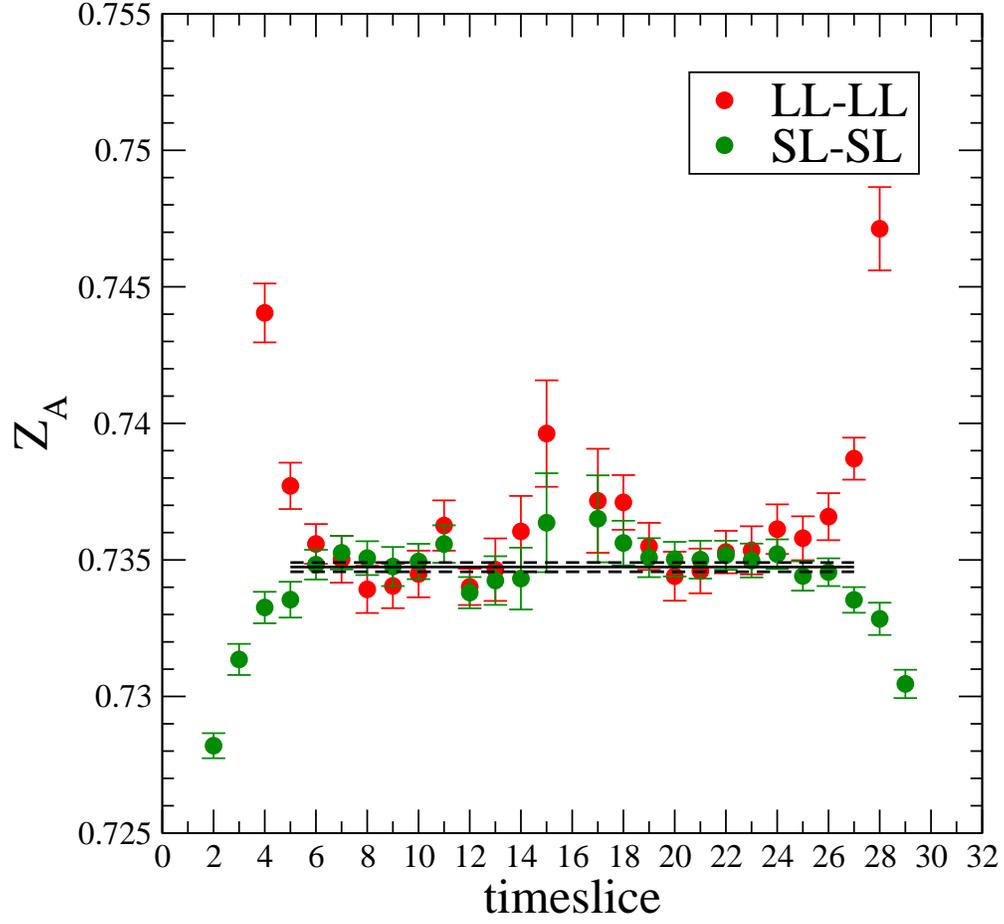}
        \caption{ $Z_{A}$ for the (D, 0.72, 0.02/0.04) dataset.
          Different colors correspond to the different smearings. The
          lines shown are a fit to the SL-SL plateau.
\label{ch6:fig:ZA}}
\end{center}
\end{figure}

\begin{figure}
\begin{center}
	\epsfig{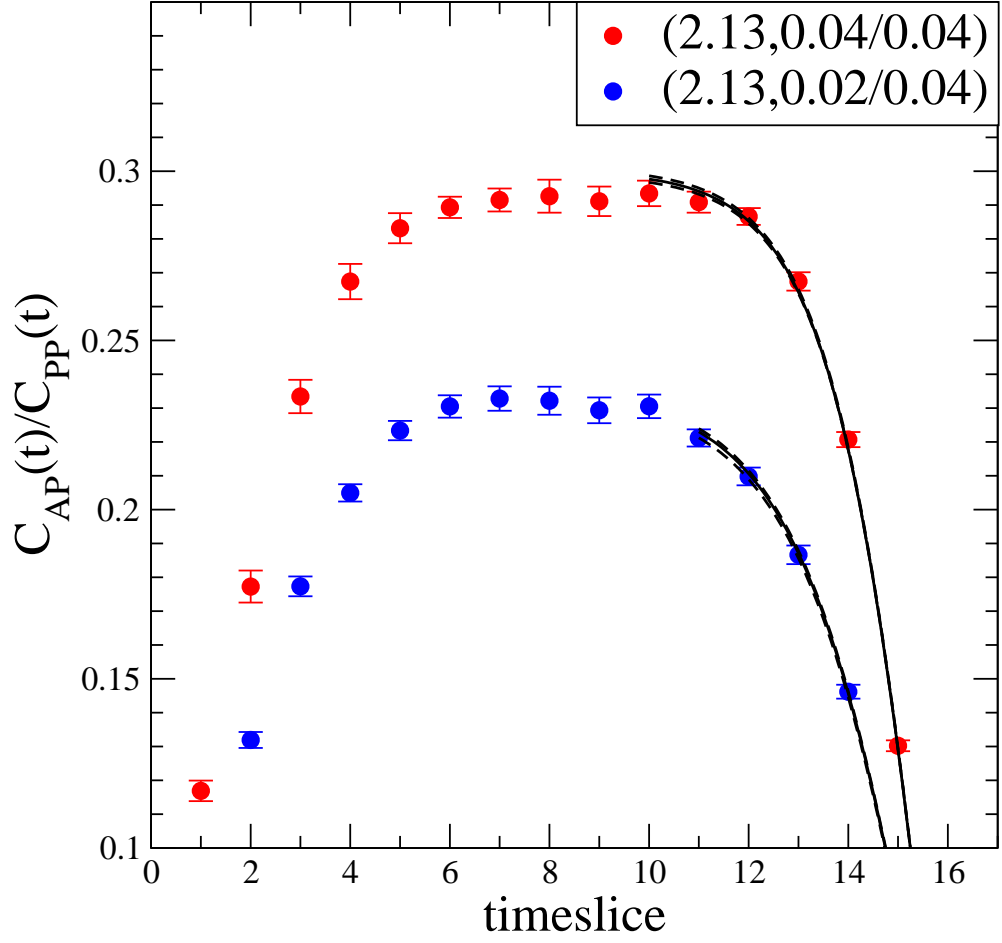}
\caption{The ratio $C_{AP}(t)/C_{PP}(t)$ versus time for
  the (I, 2.13, 0.02/0.04) and  (I, 2.13, 0.04/0.04)
  datasets. The lines shown
  are the tanh fit to the LL-LL correlators. }
\label{ch6:fig:fpi3}
\end{center}
\end{figure}

\clearpage	

\begin{figure}
\begin{center}
\epsfig{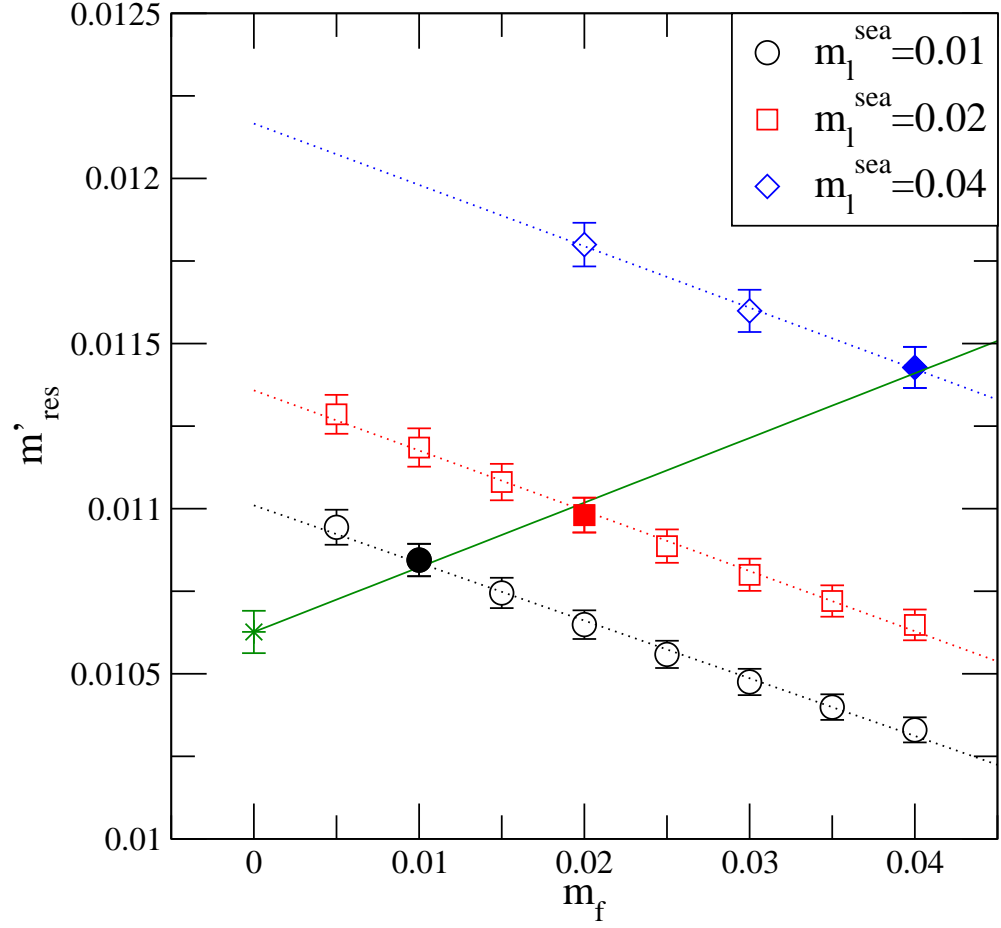}
\caption{The $m^{\prime}_{\rm res}$ dependence on $m_f$
for the DBW2 $\beta=0.72$ dataset. The solid symbols show
the unitary points used in linear extrapolation.}
\label{DBW2:mres}
\end{center}
\end{figure}

\begin{figure}
\begin{center}
\epsfig{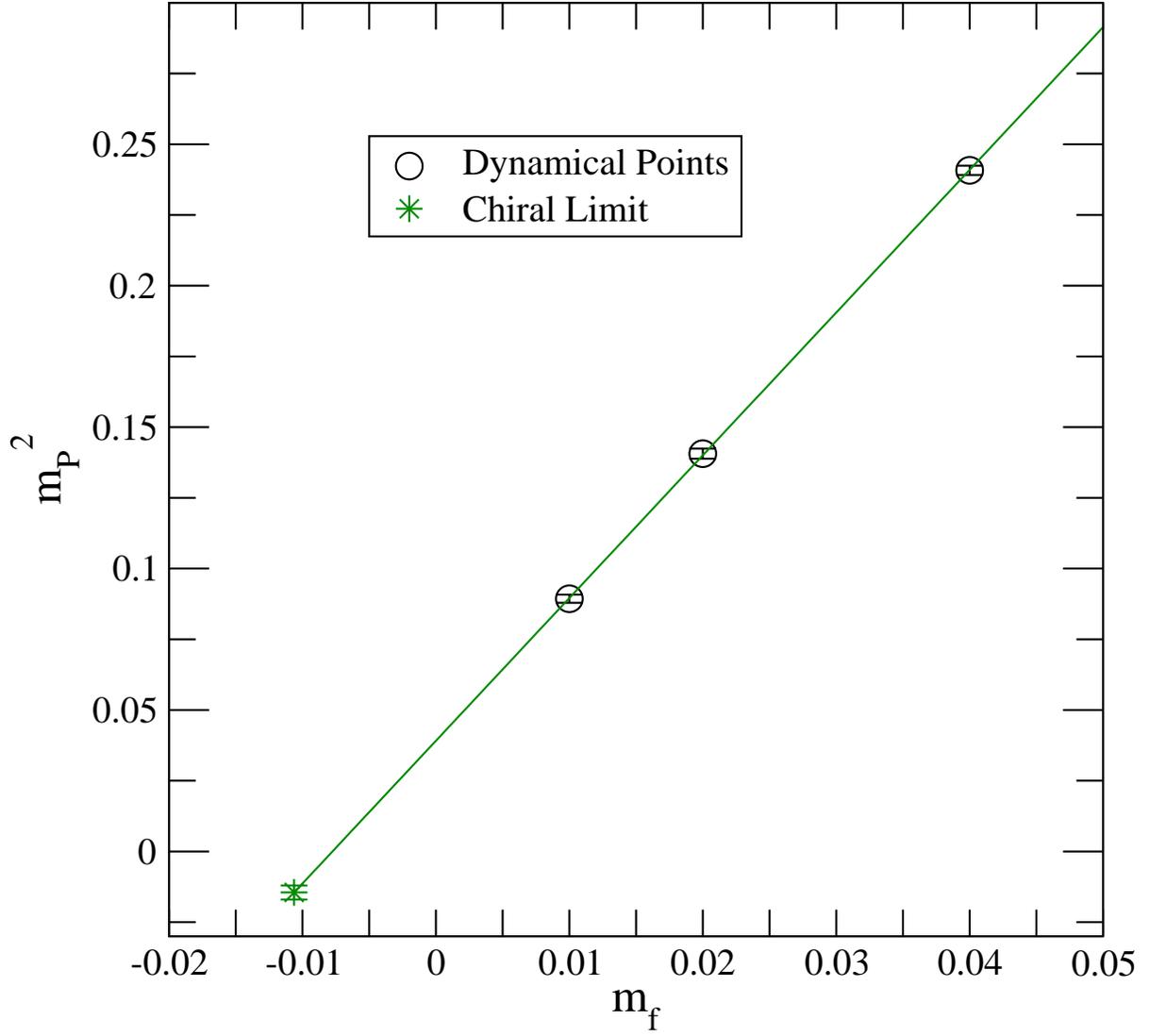}
\caption{The linear extrapolation of $m_{P}^2$ for the unitary
($m_{\rm val} = m_{l}$) points.}
\label{DBW2:psmass_sqr}
\end{center}
\end{figure}

\clearpage
\begin{figure}
\begin{center}
	\epsfig{file=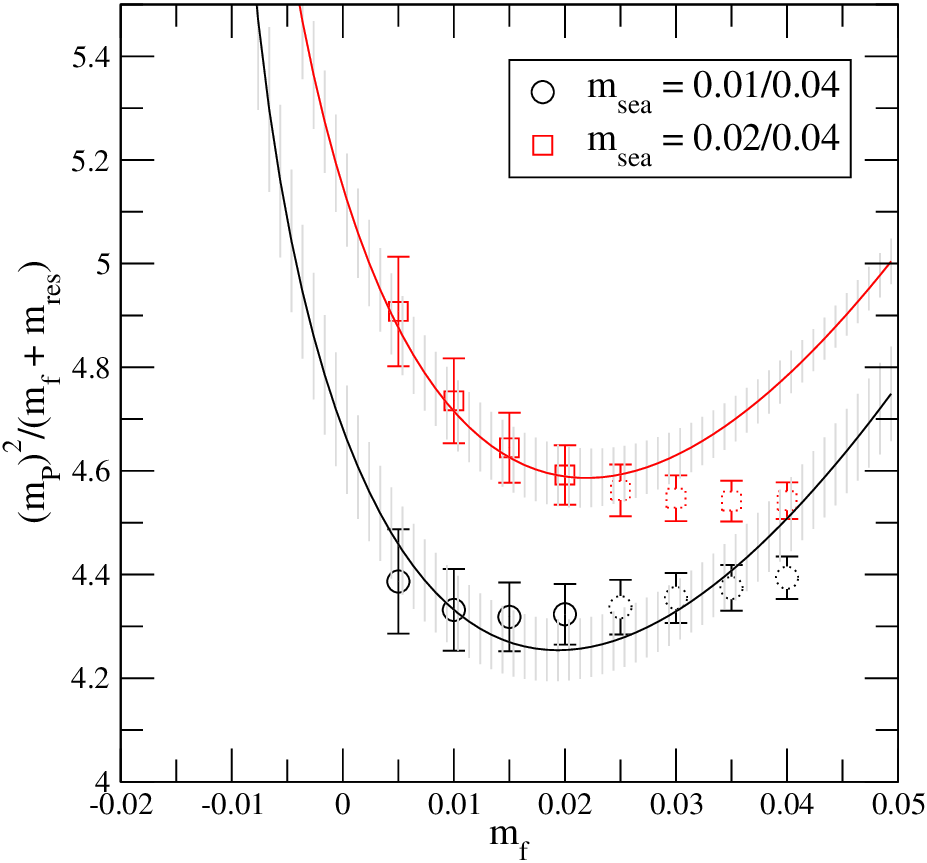,width =.8\textwidth}
        \caption{$m_P^2/(m_{\rm val}+m_{\rm res})$ from combined 
            fits to NLO PQ$\chi$PT for $m_P^2$ and $f_P$ on the 
            (D, 0.72, 0.01/0.04) and (D, 0.72, 0.02/0.04) ensembles. 
            The dashed symbols are excluded from the fits.\label{fig:simultNLO_mpi}}
\end{center}
\end{figure}

\clearpage
\begin{figure}
\begin{center}
	\epsfig{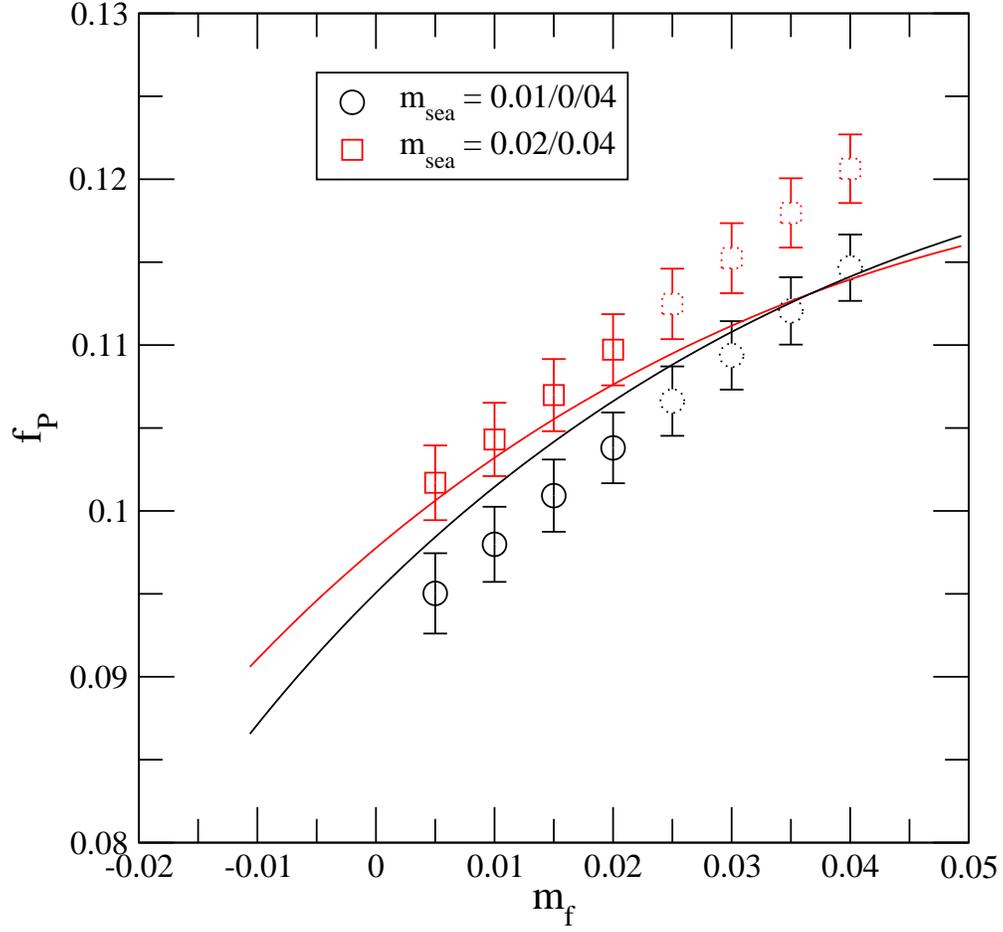}
        \caption{$f_P$ from combined 
            fits to NLO PQ$\chi$PT for $m_P^2$ and $f_P$ on the 
            (D, 0.72, 0.01/0.04) and (D, 0.72, 0.02/0.04) ensembles. 
            The dashed symbols are excluded from the fits.
            \label{fig:simultNLO_fpi}}
\end{center}
\end{figure}

\clearpage
\begin{figure}
\begin{center}
\epsfig{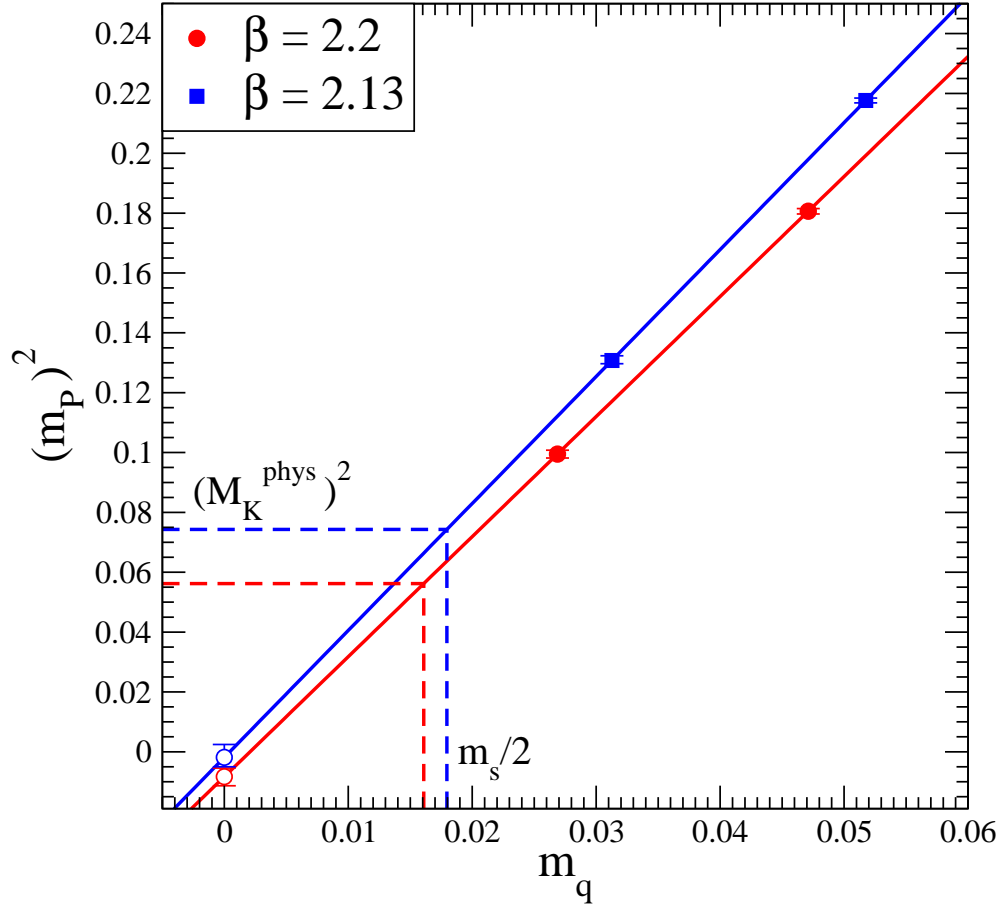}
\caption{The chiral extrapolation of the pseudoscalar meson mass
        squared versus quark mass for the two Iwasaki ensembles.
        The horizontal dashed lines show the physical kaon mass
        in lattice units as set by $r_0$.
\label{fig:IWmPchiral}}
\end{center}
\end{figure}

\begin{figure}
\begin{center}
\epsfig{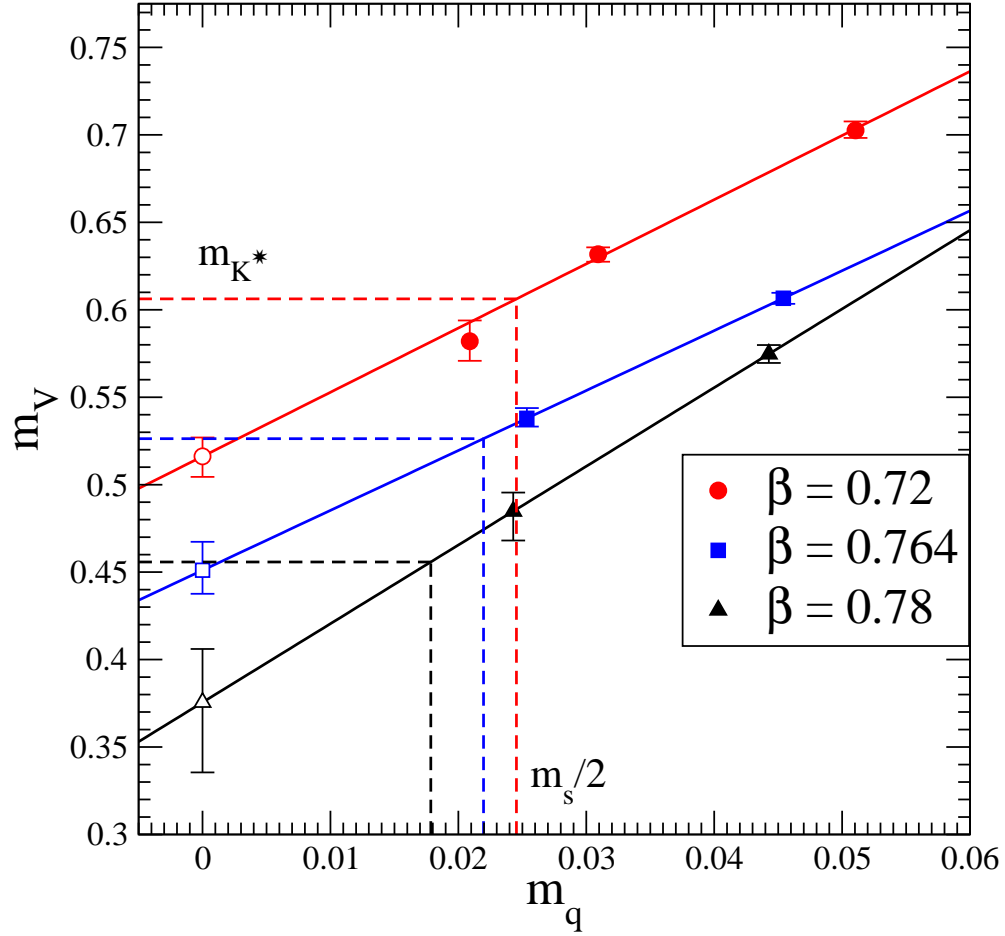}
\caption{The chiral extrapolation of the vector meson mass
        versus quark mass for the three DBW2 ensembles.
        The vertical dashed lines show half the strange 
        quark mass, enabling the $m_K^{\star}$ (horizontal lines)
        to be predicted from each ensemble.
\label{fig:DBW2mVchiral}}
\end{center}
\end{figure}

\begin{figure}
\begin{center}
	\epsfig{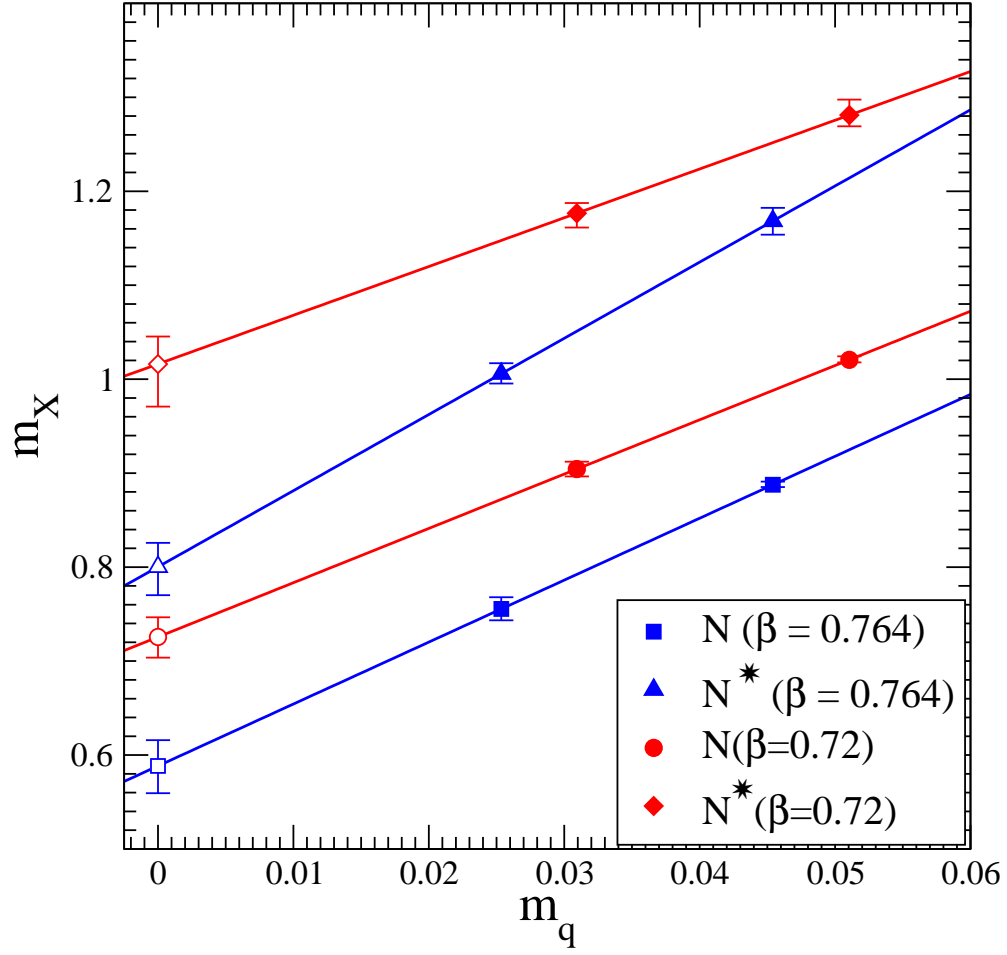}
\caption{The chiral extrapolation of the nucleon ($N$) and its 
         negative parity partner ($N^\star$) for the DBW2 $\beta=0.72$ 
         and $\beta = 0.764$ datasets. 
 \label{fig:DBW2mNchiral}}
\end{center}
\end{figure}

\begin{figure}
\begin{center}
	\epsfig{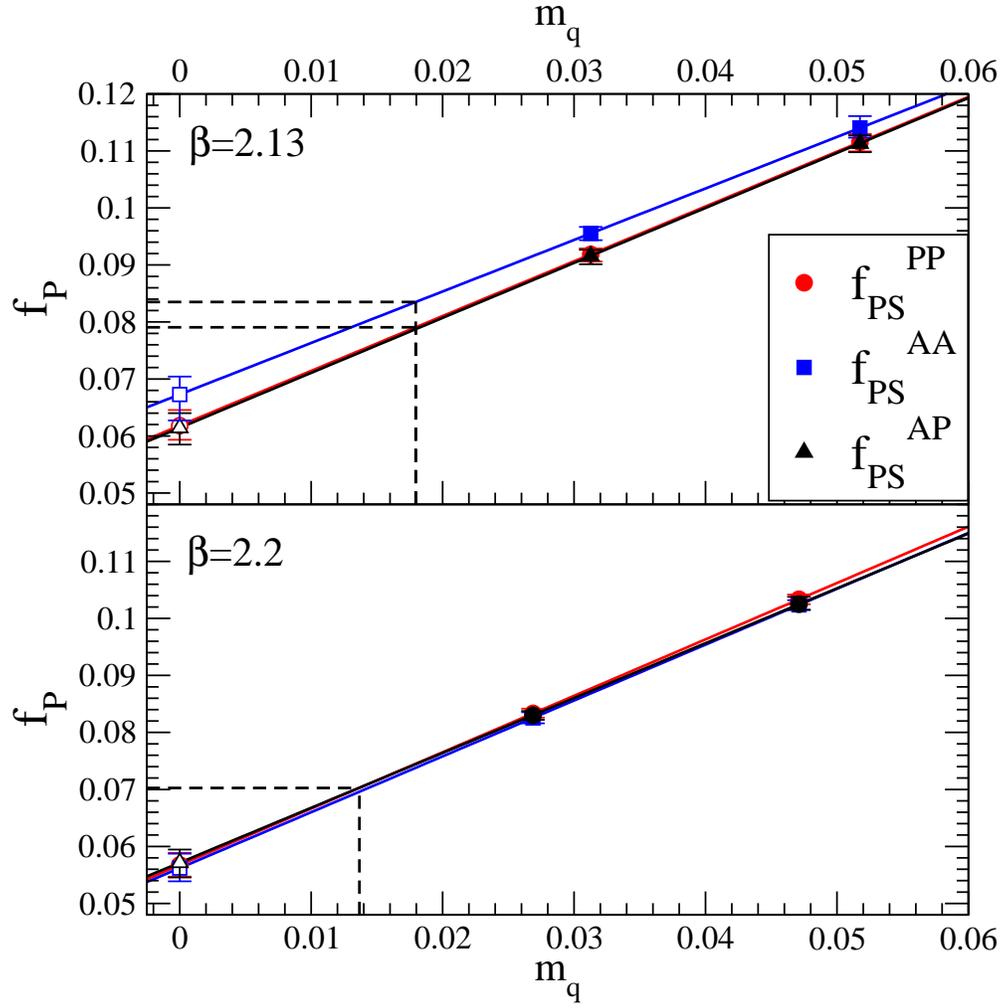}
\caption{The chiral extrapolation of the pseudoscalar decay constant
	 versus quark mass for the three methods. The upper panel
         shows the Iwasaki $\beta=2.13$ dataset and the lower
         $\beta=2.2$.
 \label{fig:DBW2fPchiral}}
\end{center}
\end{figure}

\clearpage
\begin{figure}
\begin{center}
	\epsfig{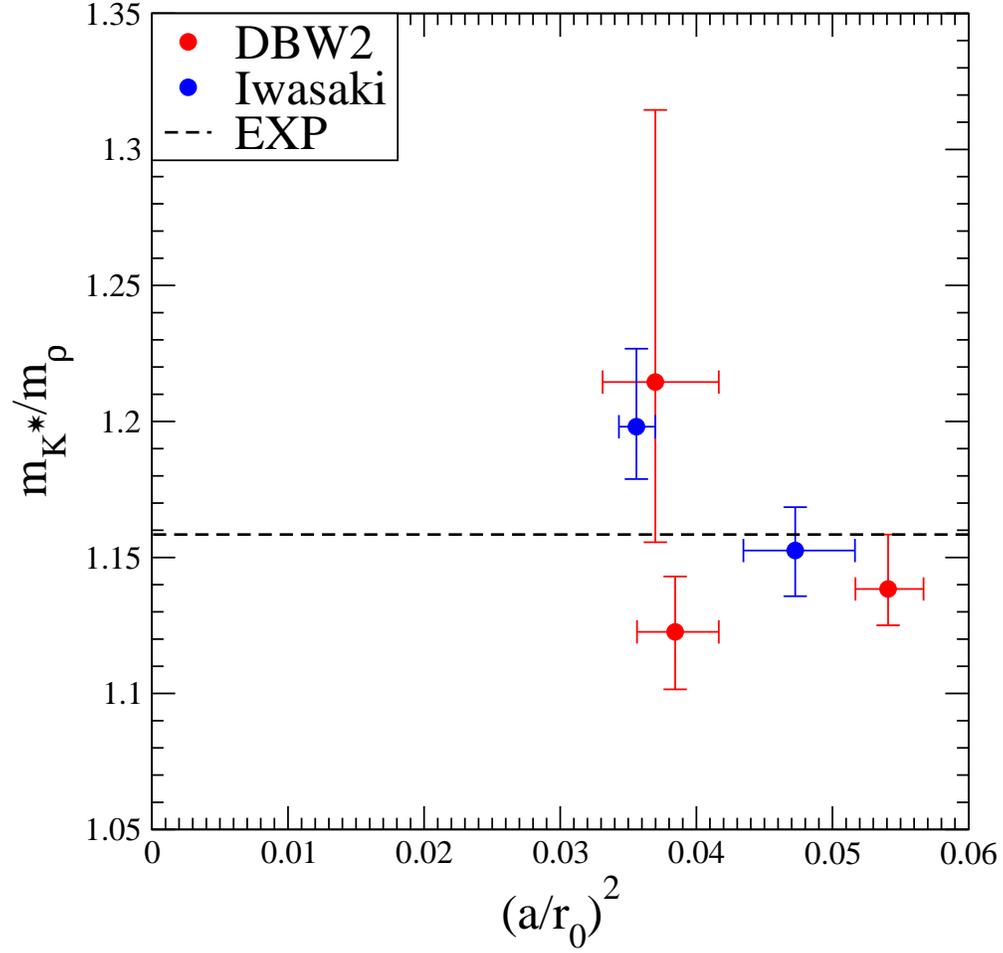}
\caption{Ratio of $m_{K^{*}}/m_{\rho}$ versus $(a/r_0)^{2}$ for all the datasets. The dotted lines are calculated from the ratio of the experimental values.} 
\label{fig:r0kstar}
\end{center}
\end{figure}

\clearpage
\begin{figure}
\begin{center}
	\epsfig{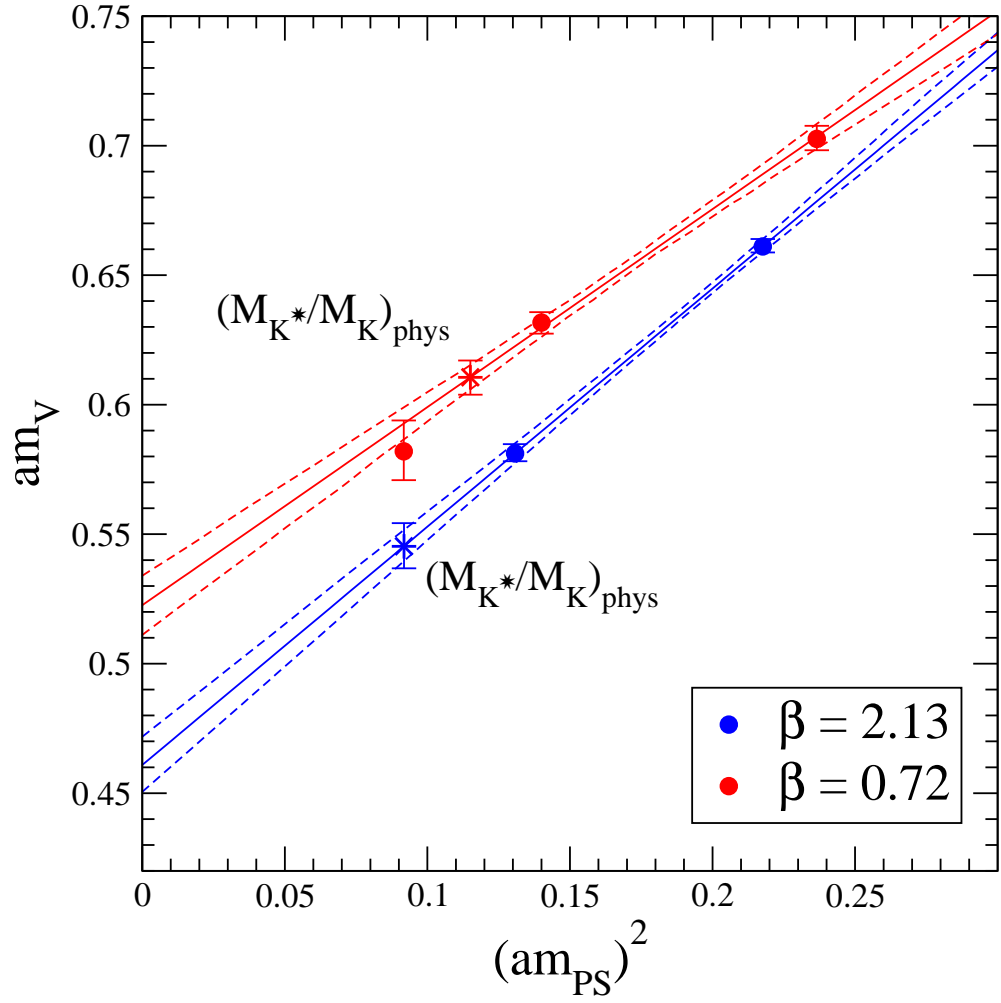}
\caption{Dependence of $m_{V}$ on $(m_{P})^2$.} 
\label{fig:J}
\end{center}
\end{figure}

\begin{figure}
\begin{center}
	\epsfig{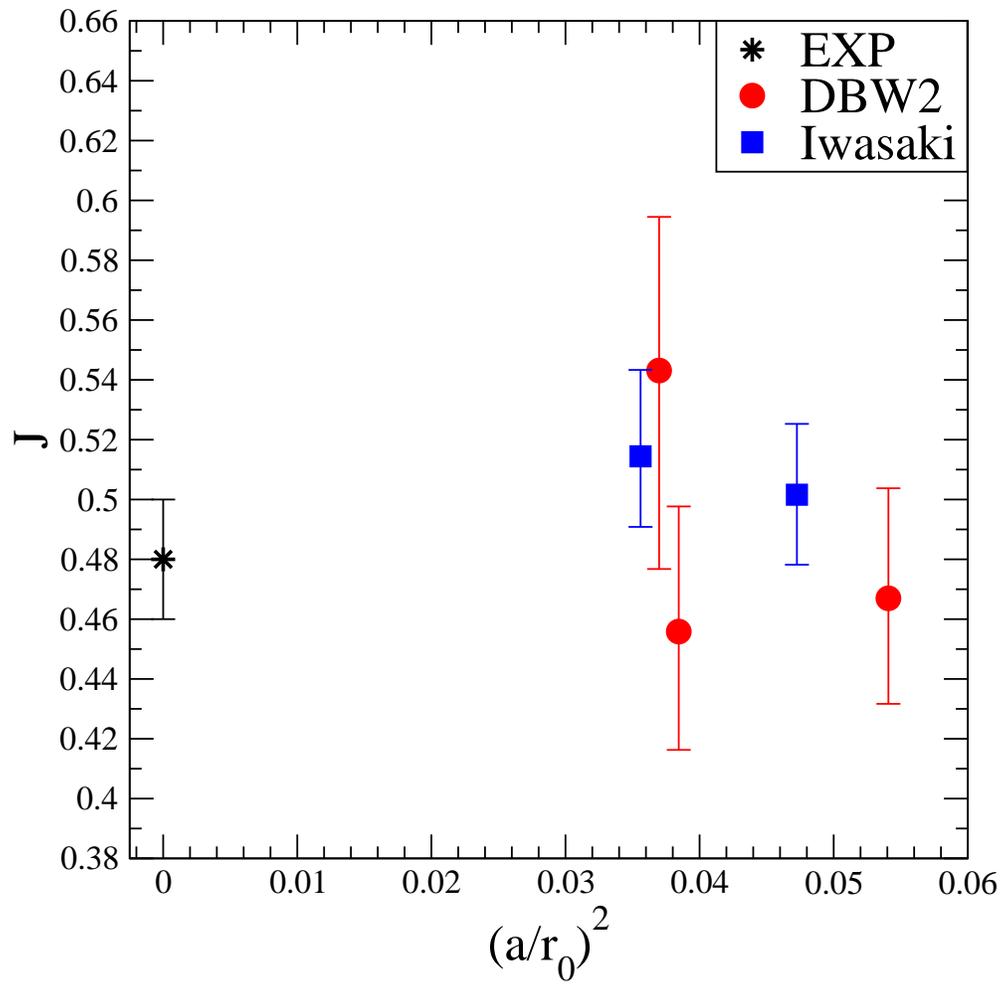}
\caption{The J parameter on all the datasets.} 
\label{fig:J-2}
\end{center}
\end{figure}

\clearpage
\begin{figure}
\begin{center}
	\epsfig{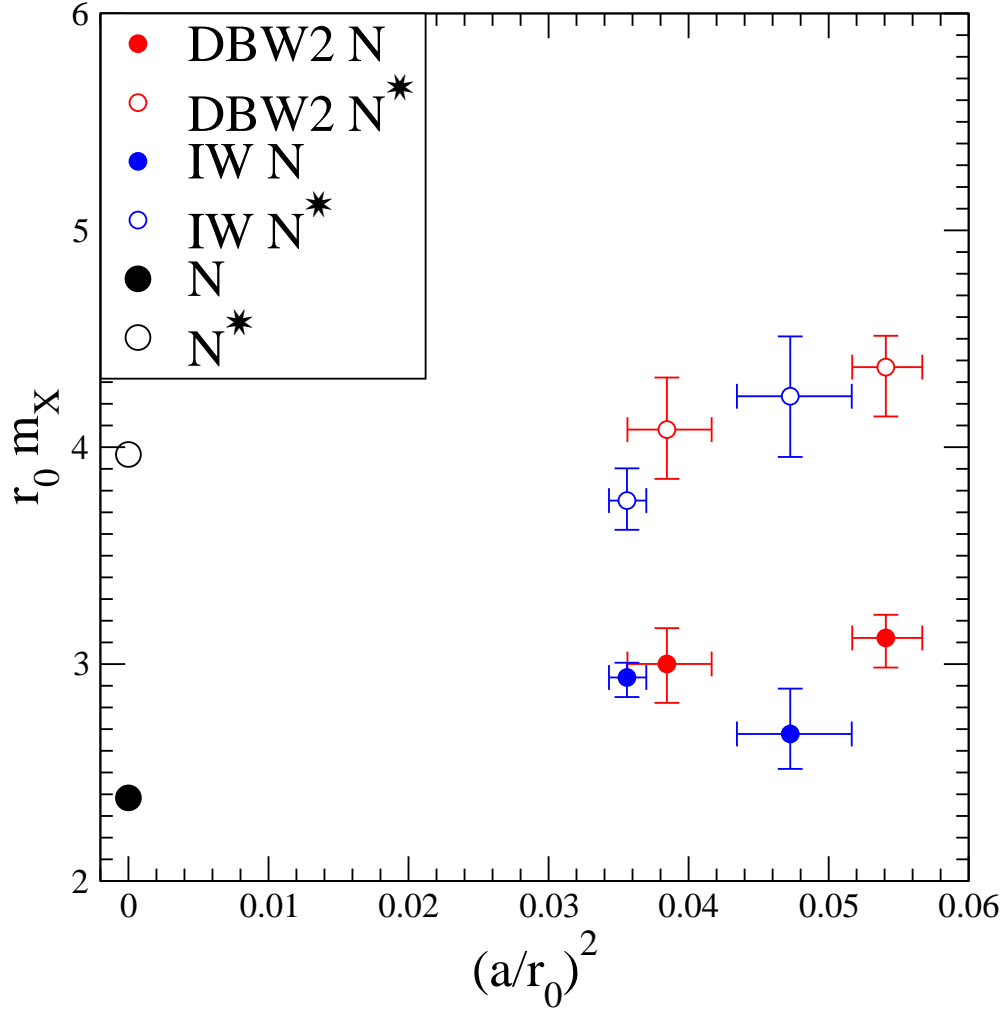}
\caption{Scaling of the baryon spectrum with lattice spacing
 squared. Closed circles denote the nucleon, $N$, and open circles the
 negative parity partner, $N^{\star}$ at the chiral limit.
Black symbols denote the experimental values scaled
by appropriate factors of $r_0$, red symbols
 the DBW2 ($\beta=0.72, 0.764$) ensembles and blue symbols
the Iwasaki ($\beta=2.13,2.2$) ensembles.
The value of $r_0=0.5$fm was chosen to
give an indication of the experimental spectrum in these units.}
\label{fig:scalenuc}
\end{center}
\end{figure}

\begin{figure}
\begin{center}
\epsfig{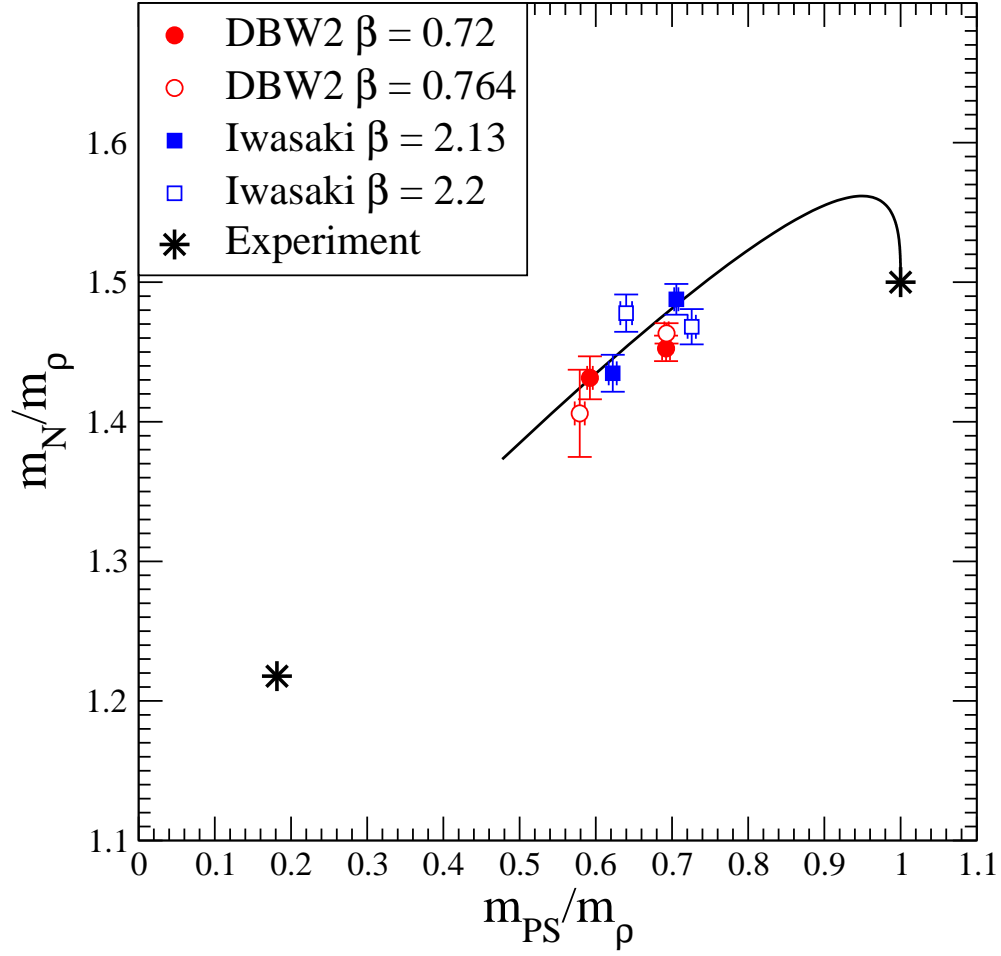}
\caption{The Edinburgh plot. Red symbols denote the DBW2 ensembles and
 blue symbols the Iwasaki ensembles.
The phenomenological curve derived from~\cite{Ono:1977ss}
has been shown to guide the eye. Experimental ratios and
the values obtained in the static quark limit,
where the hadron mass is equal to the sum of the valence quark masses,
are given by the starred points.}
\label{fig:edinburgh}
\end{center}
\end{figure}

\clearpage
\begin{figure}
\begin{center}
	\epsfig{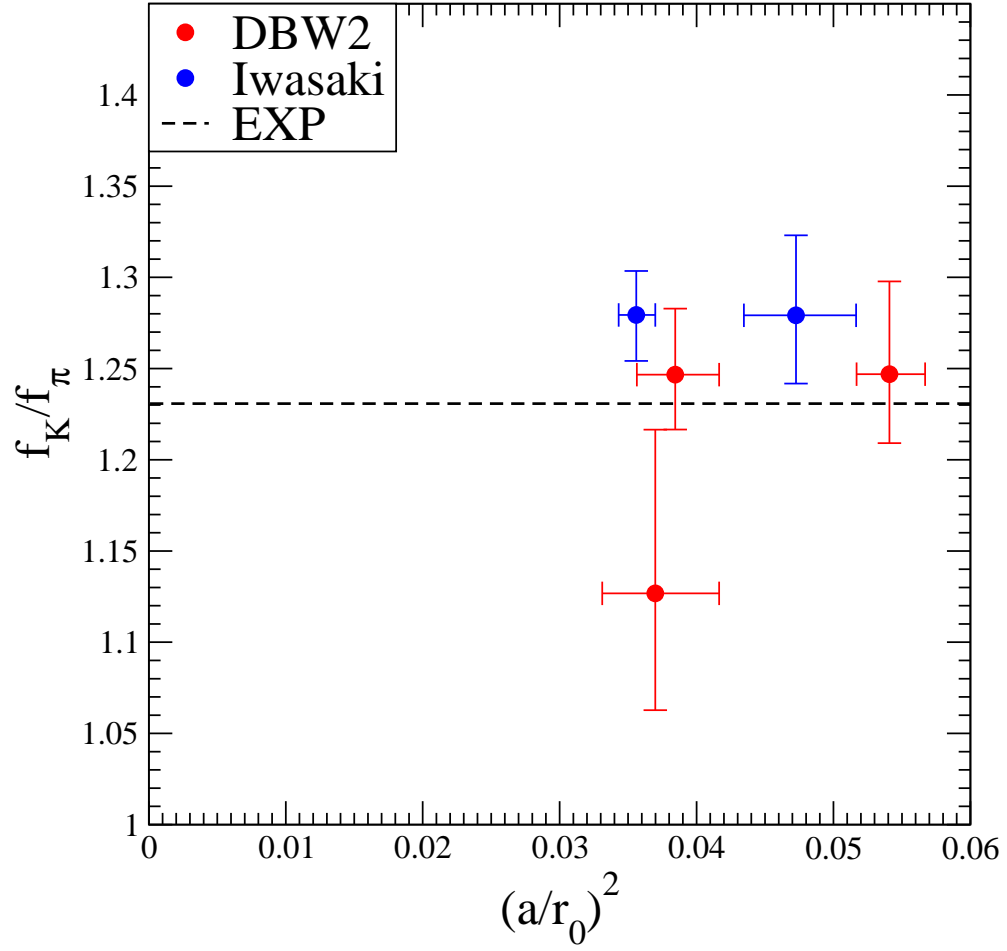}
\caption{Ratio of $f_{K}/f_\pi$ for plotted against $(a/r_0)^{2}$ for all the $\beta$ values. The dotted lines are calculated from the ratio of the experimental values.} 
\label{fig:r0fKfpi}
\end{center}
\end{figure}